\pdfoutput=1
\documentclass{article}
\usepackage{iclr2027_conference,times}
\iclrfinalcopy
\usepackage{wrapfig,needspace}
\usepackage[T1]{fontenc}
\usepackage[hyphens]{url}
\usepackage{graphicx,booktabs,amsmath,amssymb,amsthm,mathtools}
\usepackage{caption,array,placeins}
\usepackage{pgfplots}
\usepgfplotslibrary{groupplots}
\usetikzlibrary{calc}
\pgfplotsset{compat=1.18}
\usepackage{hyperref}
\hypersetup{hidelinks}

\newtheorem{theorem}{Theorem}
\newtheorem{proposition}{Proposition}

\newcommand{\cZ}{\mathcal Z}
\newcommand{\Law}{\mathcal L}
\newcommand{\E}{\mathbb E}
\newcommand{\Prob}{\mathbb P}
\newcommand{\Var}{\operatorname{Var}}
\newcommand{\1}{\mathbf 1}
\newcommand{\Up}{U_{\mathrm{prompt}}}
\newcommand{\Uw}{U_{\mathrm{within}}}
\newcommand{\Uc}{U_{\mathrm{call}}}
\newcommand{\Uo}{U_{\mathrm{order}}}
\newcommand{\Ui}{U_{\mathrm{int}}}
\newcommand{\Ut}{U_{\mathrm{total}}}
\newcommand{\Aw}{\widehat A_{\mathrm{within}}}
\newcommand{\Aa}{\widehat A_{\mathrm{across}}}
\newcommand{\sharedtablelabel}[1]{\label{#1}}
\newcommand{\maintextreproductionnote}{}
\makeatletter
\newcommand{\equationalias}[2]{\@ifundefined{r@#2}{}{\protected@write\@auxout{}{\string\newlabel{#1}{\csname r@#2\endcsname}}}}
\makeatother
\title{Second-Order Response Laws for LLM Judges:\\From Prompt Stability to Decision Quality}
\author{Pengbin Feng\textsuperscript{1,2*}, Chunlei Meng\textsuperscript{3}, Daozheng Qu\textsuperscript{4}, Zhilin Zhang\textsuperscript{2}, Haoran Liu\textsuperscript{5}, Jiekai Wu\textsuperscript{6,*}\kern-\tabcolsep\\
\textsuperscript{1}University of Southern California\\
\textsuperscript{2}Amazon\\
\textsuperscript{3}Fudan University\\
\textsuperscript{4}University of Portsmouth\\
\textsuperscript{5}The Chinese University of Hong Kong\\
\textsuperscript{6}Independent Researcher}
\begin{document}
\maketitle
\lhead{Under review as a conference paper at ICLR 2027}
{\renewcommand{\thefootnote}{*}\begin{NoHyper}\footnotetext{Corresponding authors: pengbinf@alumni.usc.edu and ketsu0612@gmail.com}\end{NoHyper}}
\raggedbottom
\begin{abstract}
LLM judges are used to compare models and select responses, yet a judge can change its verdict when a prompt is reworded. The change may come from the wording or from the randomness of another call. Confusing the two can distort model comparisons and response selection, but they are hard to separate with few repeated calls, and plug-in prompt-dispersion scores mix them. We develop a framework that connects measurements of judge variability to decision quality. It rests on second-order response laws, which describe how verdict distributions vary across prompts. Crossing prompts with candidate orders separates prompt, position, interaction, and call variation, and agreement between distinct calls removes the finite-call bias. We derive the corrected estimator's exact variance and matching bounds for detecting weak prompt effects under a fixed, balanced design. In over one million judgments from six judges on three benchmarks, the correction changes conclusions. On a matched panel, it reverses the estimated Qwen--Gemma-3 ranking by prompt sensitivity. Lowering the sampling temperature (cooling) makes Qwen's calls more consistent while its corrected prompt estimate rises nearly fivefold (call-level interval 3.3 to 8.5), a rise the plug-in estimate misses. Pre-specified studies on 320 unseen items and 24 unseen prompts confirm the increase. In votes balanced across candidate orders, cooling produces more ties in Qwen, and changing only the tie rule shifts the estimated accuracy effect of cooling by 9.84 percentage points. An analytical model shows how position dependence can push such votes toward ties and lower accuracy under abstention even when the judge favors the correct answer. The value of consistency therefore depends on alignment with correct answers, the voting rule, and the cost of abstention.
\end{abstract}
\section{Introduction}
\begin{wrapfigure}{r}{2.70in}
\vspace{-.5\baselineskip}\centering
\includegraphics[width=\linewidth]{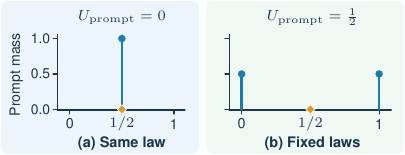}
\caption{The same marginal $\Pr(A)=1/2$, two mechanisms: (a) every prompt randomizes; (b) each prompt fixes a verdict, with half choosing each candidate. Horizontal: $\Pr(A\mid x)$; vertical: prompt mass.}
\label{fig:motivation}\vspace{-.3\baselineskip}
\end{wrapfigure}
Consider two LLM judges that each select candidate A half the time. For one, every prompt gives an equal chance of A or B. For the other, half the prompts always select A and half always select B. Their average judgments are the same. Repeating a prompt reveals randomness in the first judge; changing the prompt reveals systematic dependence in the second (Figure~\ref{fig:motivation}). Which kind of stability matters when a judge compares models or selects responses?

This question arises in instruction-following, preference, and correctness-based evaluation \citep{zheng2023judging,zeng2024llmbar,tan2024judgebench}. A changed verdict can reflect wording, position, or another random call. With few repeated calls, sampling noise inflates observed prompt dispersion and can distort comparisons of judges by prompt sensitivity.

We develop a framework that connects measurements of judge variability to decision quality. Its basis is the \emph{second-order response law}, which describes how verdict distributions vary across prompts. Crossing prompts with candidate orders separates prompt, position, interaction, and call variation. Agreement between distinct calls removes the inflation caused by limited repeats.

The correction reveals changes in prompt dependence that the plug-in score masks. Lowering the sampling temperature (cooling) improves repeatability in Qwen and Gemma while corrected prompt estimates increase. Qwen's prompt estimate rises 4.79-fold (call-level interval 3.3 to 8.5); the plug-in score rounds to .05456 before and after cooling. An independent replication, 32 further Gemma repeats per cell, and studies on 320 unseen items and 24 unseen prompts confirm the increase (Section~\ref{sec:temperature-story}).

\begin{figure}[!htbp]
\centering
\includegraphics[width=\linewidth]{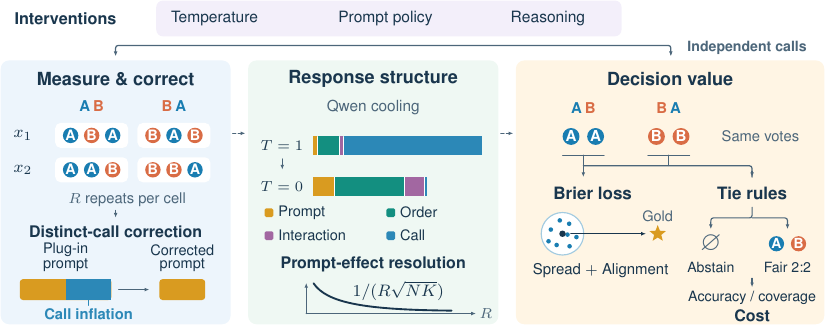}
\caption{From response structure to decision value.
Left: repeated calls for each prompt and order let us correct
finite-call inflation using agreement between distinct calls.
Middle: Qwen's corrected components under cooling
($N=40$ items, $K=48$ prompts, $R=4$ repeats), shown on a common
absolute scale, and the detection limit near a zero prompt effect
under a balanced design with fixed items and prompts.
Right: independent groups of balanced votes let us decompose Brier loss
and compare how tie rules affect accuracy, coverage, and cost on the same votes.
AABB illustrates one group; the cloud represents multiple deployment
vectors and their mean relative to gold.
Temperature, prompt policy, and reasoning are studied separately.
Colors track candidate identity; other symbols are schematic.
Dashed arrows connect results from separate call samples.}
\label{fig:methodology}
\end{figure}

We then collect independent groups of balanced votes to study how these changes affect decisions through gold alignment, voting rules, and abstention cost (Figure~\ref{fig:methodology}).

\paragraph{Relation to existing work.}
Prompt-sensitivity studies measure performance variation across formats and templates \citep{sclar2023formatspread,mizrahi2024state}, response-likelihood changes \citep{chatterjee2024posix}, within-instance correctness or quality-score differences \citep{zhuo2024prosa}, template-level performance distributions \citep{polo2024prompteval}, and repeat-adjusted verdict agreement under paraphrase \citep{bellibatlu2026judgesense}. Position-bias studies measure order effects \citep{wang2024fairevaluators,li2024splitmerge,shi2025positionbias}, which our order and interaction components estimate net of call noise. Reliability and bias audits report alignment, repeatability, bias, and tie behavior \citep{thakur2024judges,norman2026reliability,usami2026darkcurrent}, and CALM compares modification robustness with repeated-call consistency \citep{ye2025calm}. To improve the judgment itself, other methods infer judgment distributions \citep{wang2025judgmentdistribution}, calibrate A/B/tie aggregation \citep{dadkhahi2025distributioncalibrated}, aggregate label probabilities over orders \citep[TrustJudge;][]{wang2025trustjudge}, or denoise preference graphs from several evaluators \citep[PGED;][]{hu2024acyclic}. Our decomposition locates the instability that remains after such aggregation.

Crossed variance-component analysis \citep{cronbach1972dependability,brennan2001generalizability,searle1992variance} and second-order uncertainty measures \citep{sale2024variance} provide the basis for our decomposition; distinct-call correction uses within-cell quadratic U-statistics \citep{hoeffding1948class}. Related LLM work decomposes correctness variation \citep{romanou2026brittlebench}, separates sampling noise from system-prompt reward variation \citep{gao2026p1}, projects precision across evaluation configurations \citep{messing2026hidden}, and links variance to balanced judge allocation \citep{zhu2026cyclicjudge} or adaptive querying \citep{saha2026budget}. We decompose per-item verdict distributions, remove finite-call bias, and give finite-sample guarantees for the corrected measurement.

\paragraph{Contributions.}
Across six judge models and three benchmarks, over one million recorded judgments support three connected parts of the framework (Figure~\ref{fig:methodology}; Table~\ref{tab:reproduction-main-map}; Section~\ref{sec:output-cap-recovery}).

\noindent\emph{Measurement and correction.}
The plug-in prompt estimate has exact bias $(K-1)\Uc/(2KR)$; distinct-call correction removes this decoding-dependent term with two independent calls per cell. The estimator is the classical trace-mean-square correction; we add its finite-sample theory for categorical verdicts under a fixed crossed design: an exact variance, its unbiased estimate for $R\ge4$, a distribution-free test with a matching lower bound, exponential distribution-free intervals, and a sharp covariance constant (Table~\ref{tab:classical-map}). Identical-input controls, an independent reference with more repeats, and known-law simulations validate the measurement; on a matched panel, correction reverses the estimated Qwen--Gemma-3 ranking by prompt sensitivity (Section~\ref{sec:live-measurement-validation}).

\noindent\emph{Response structure.}
One analytical family shows how sharpening separates repeatability from prompt dispersion; another explains tie concentration at a fixed gold margin. In matched experiments, cooling separates repeatability from estimated prompt dependence, examples shift dependence toward position, and thinking increases call variation (Sections~\ref{sec:temperature-story}--\ref{sec:intervention-routes}).

\noindent\emph{Decision value.}
At a fixed positive gold margin, our position-dependence model shows that lower vector risk and Brier loss can come with lower strict-vote accuracy, so variance components alone do not determine decision quality. Relative to abstention, resolving pure-2:2 ties by a fair coin changes an accuracy contrast by half the change in pure ties; for Qwen's cooling contrast, the shift is 9.84 pp. A collision-moment criterion, based on polynomial unbiased estimation \citep{kimpara2023proper}, identifies when the risk difference between equal-budget designs can be estimated without bias from one call per cell (Propositions~\ref{prop:design-collisions}--\ref{prop:design-quadratic}). For a fixed target, a variance identity gives a continuous oracle allocation criterion for items and repeats (Proposition~\ref{prop:fixed-population-allocation}). On 120 unseen items, the strategy (one call or voting) chosen from measurement calls beat the other by 5.89 pp [4.48,7.30] on average over five pre-specified settings (Table~\ref{tab:discussion-e9}).

\section{Response Structure and Finite-Call Measurement}
\label{sec:iclr-methods}
We separate verdict variation by source and correct for finite-call bias. Fix an item and a judge configuration, including the checkpoint, rendered instructions, decoding, stopping, serving system, and parser. After restoring candidate identity across presentation orders, a completed verdict lies in
$\cZ=\{\textsc{candidate\_1},\textsc{candidate\_2},\textsc{tie},\bot\}$,
where $\bot$ is an invalid output. The target is therefore the full distribution of generated and parsed verdicts, including invalid outcomes. For prompt $X\sim Q$, write $P_X(c)=\Prob(Z=c\mid X)$ and $\bar p=\E_QP_X$. The second-order response law $\Law_Q(P_X)$ is the distribution of $P_X$ across prompts; judges with equal means $\bar p$ can differ in it (Figure~\ref{fig:motivation}). Define
\begin{equation}
\Up(Q)=\E_Q\|P_X-\bar p\|^2,\qquad
\Uw(Q)=\E_Q[1-\|P_X\|^2],\qquad
\Ut(Q)=1-\|\bar p\|^2.
\label{eq:u-prompt}
\end{equation}
\equationalias{eq:u-within}{eq:u-prompt}%
\equationalias{eq:u-total}{eq:u-prompt}%
The first quantity measures how verdict distributions vary across prompts; the second is the probability that two independent calls under the same prompt disagree.

\paragraph{Crossing prompts with candidate order.}
\leavevmode A pairwise verdict can also depend on which candidate is shown first, so we cross every prompt with both presentation orders. For a fixed set of $K$ equally weighted prompts and two equally weighted orders, let $P_{ko}=\Law(Z\mid x_k,o)$, $P_k^q=(P_{k,AB}+P_{k,BA})/2$, $P^o=K^{-1}\sum_kP_{ko}$, and $\bar p=K^{-1}\sum_kP_k^q$. Define
\begin{align}
\Up^{(q)}&=\frac1K\sum_k\|P_k^q-\bar p\|^2,
&\Uo&=\frac12\sum_o\|P^o-\bar p\|^2,\notag\\
\Ui&=\frac1{2K}\sum_{k,o}\|P_{ko}-P_k^q-P^o+\bar p\|^2,
&\Uc&=\frac1{2K}\sum_{k,o}(1-\|P_{ko}\|^2).
\label{eq:main-components}
\end{align}
\begin{proposition}[Disagreement decomposition]
\label{prop:categorical}\label{prop:crossed}
For any finite categorical alphabet,
\begin{equation}
\Ut(Q)=\Uw(Q)+\Up(Q),\qquad \Ut=\Up^{(q)}+\Uo+\Ui+\Uc,
\label{eq:categorical-decomposition}
\end{equation}
\equationalias{eq:crossed-decomposition}{eq:categorical-decomposition}%
with all four population components nonnegative.
\end{proposition}
Prompt dependence averaged over orders differs from prompt--order interaction: even when $\Up^{(q)}=0$, the effect of position can vary across prompts. Their sum is the average prompt dispersion within orders. The crossed profile shows how interventions redistribute variation among these sources.

\paragraph{Removing finite-call inflation.}
\leavevmode Each cell law is observed only through a few calls, and plug-in frequencies inflate the prompt component, because prompts with identical laws still produce different samples. Observe $R\ge2$ independent calls per cell, with independence across cells and items conditional on their fixed laws. For cell $a$, let $\widehat p_a$ be its empirical categorical vector. Estimate inner products by
\begin{equation}
\widehat A_{aa}=\frac{\sum_{r\ne r'}\1\{Z_{ar}=Z_{ar'}\}}{R(R-1)},
\qquad \widehat A_{ab}=\widehat p_a^\top\widehat p_b\quad(a\ne b).
\label{eq:main-cell-agreement}
\end{equation}
These are unbiased estimates of $P_a^\top P_b$. Substituting these agreements into the quadratic component formulas gives estimates using all distinct call pairs and preserves Equation~\eqref{eq:crossed-decomposition} for every sample. The corrected and plug-in prompt estimates satisfy
\begin{align}
\E\widehat U_{\rm plug,prompt}^{(q)}-\Up^{(q)}
 &=\frac{K-1}{2KR}\Uc,\label{eq:crossed-plugin-excess}\\
\widehat U_{\rm plug,prompt}^{(q)}-\widehat\Up^{(q)}
 &=\frac{K-1}{2KR}\widehat\Uc.
\label{eq:crossed-plugin-sample-identity}
\end{align}
At the same repeat budget, a larger call component produces greater finite-call inflation of the plug-in prompt estimate. We retain raw unbiased estimates, including negative values near zero. On this balanced design the estimator is algebraically equivalent to trace-mean-square correction (Section~\ref{sec:fixed-effects-equivalence}; Table~\ref{tab:classical-map}); complete estimators and proofs appear in the supplement.

\paragraph{Resolution of weak prompt dependence.}
The next result quantifies repeated-call uncertainty for fixed items and prompt laws. Let $\theta=N^{-1}\sum_i\Up^{(q)}(s_i)$ and let $\widehat\theta$ average the corrected estimates over $N$ items.
\begin{theorem}[Finite-call measurement and resolution]
\label{thm:main-finite-call}
For $N\ge1$, $K\ge2$, $R\ge2$, and mutually independent calls from arbitrary fixed categorical cell laws,
\begin{equation}
\E\widehat\theta=\theta,\qquad
\Var(\widehat\theta)\le\frac{\theta}{NKR}+
\frac{K-1}{8NK^2R(R-1)}.
\label{eq:story-envelope}
\end{equation}
Write $b=(K-1)/[8NK^2R(R-1)]$. For fixed $\alpha,\beta\in(0,1)$, the test
$\widehat\theta>\sqrt{(1-\alpha)b/\alpha}$ has type-I error at most $\alpha$ at $\theta=0$ and type-II error at most $\beta$ when $\theta\ge C_{\alpha,\beta}/(R\sqrt{NK})$. The constant depends only on the error levels. For even $K$ and $\alpha+\beta<3/4$, no test attains these levels uniformly at separations
$0<\theta\le\sqrt{\log(5/4)}/(4R\sqrt{NK})$.
Under the null, order-averaged prompt laws are equal within each item, and prompt--order interaction may be nonzero.
\end{theorem}
When prompt dispersion is zero, only the quadratic variance term remains. The matching bounds show how repeats and prompt coverage determine the smallest detectable effect. The exact variance, constants, test, and lower-bound construction are given in Theorems~\ref{thm:nn-variance}, \ref{thm:nn-cantelli}, and~\ref{thm:nn-lower}. The categorical covariance inequality $\operatorname{tr}(\Sigma_a\Sigma_b)\le1/4$, attained by identical fair binary laws, sharpens the variance envelope. For $(N,K,R)=(40,48,4)$, the explicit power bound guarantees 95\% power at 5\% type-I error when prompt dispersion is at least .0226; at $(152,6,16)$, the sufficient level is .00732. These are sufficient effect sizes for detecting prompt dependence within a single condition.

\section{Validating the Measurement}
\label{sec:iclr-empirical}\label{sec:live-measurement-validation}
We validate the correction against a zero-effect control and an independent reference, then compare judges. Throughout, $N$ counts items, $K$ prompts, and $R$ calls per item--prompt--order cell. Call-level intervals describe repeated-call uncertainty for fixed items and prompts; paired whole-item intervals describe variation in benchmark composition. Each result states its uncertainty target.

\begin{wrapfigure}{r}{2.80in}
\vspace{-.4\baselineskip}\centering\captionsetup{font=normalsize,skip=3pt}
\includegraphics[width=\linewidth]{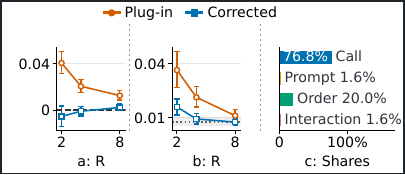}
\caption{Measurement validation: (a) zero-effect control; (b) MAE to an independent $R=16$ reference; (c) reference component shares. $R$: calls per cell. Gray: reference split-half disagreement. Bands summarize stratified item composition.}
\label{fig:qwen-depth}\vspace{-.4\baselineskip}
\end{wrapfigure}

\paragraph{A zero-effect control.}
All hidden prompt labels produce byte-identical model inputs, so true prompt dispersion is zero. At $R=2$, correction changes Qwen's estimate from .0406 to $-.0053$. Across $R=2,4,8$, corrected estimates remain near zero while the plug-in excess shrinks as calls are repeated, as Equation~\eqref{eq:crossed-plugin-excess} predicts (Figure~\ref{fig:qwen-depth}a).

\Needspace{16\baselineskip}
\paragraph{Correction improves agreement with the reference.}
Correction improves agreement with an independently collected $R=16$ reference at every tested repeat depth (Figure~\ref{fig:qwen-depth}b). At $R=2$, the mean absolute discrepancy across items falls from .03650 to .01599. Section~\ref{sec:recovery-reference-splits} reports reference split-half diagnostics.

\paragraph{Correction changes judge comparisons.}
On a matched four-configuration panel ($N=50$, $K=6$, $R=4$), correction reverses the estimated Qwen3.6--Gemma-3 ranking by prompt dependence: plug-in estimates of .02512 versus .01931 become corrected estimates of .00521 versus .01774. The plug-in ordering comes from a finite-call bias gap of $+.0183$ (Hoeffding interval $[.0126,.0241]$). The corrected difference is $-.0125$. At the call level, its exact-variance interval $[-.0207,-.0044]$ supports the reversal for these two judge configurations, and its item-paired interval is $[-.0256,.0005]$ (Section~\ref{sec:discussion-studies-ranking}). Call variation accounts for 76.3\% and 88.5\% of total disagreement for Qwen and GPT-OSS, whereas order has the largest point estimate for Gemma-3-27B-IT and Claude Haiku 4.5. Thus, at a fixed call budget, the bias depends on how each judge's variation is distributed across components.

\Needspace{17\baselineskip}
\noindent\begin{minipage}[t]{\dimexpr\linewidth-2.70in-12pt\relax}
\vspace{0pt}
The effect persists on a common five-model panel with 40 LLMBar and 40 JudgeBench items, 24 shared rewordings, both orders, and $R=2$. On LLMBar, correction reduces Qwen's estimate from .0718 to .0065; Figure~\ref{fig:five-model-correction} shows both datasets and the changed rankings. Claude Haiku's JudgeBench result is analyzed separately to examine its invalid outputs (Section~\ref{sec:semantic-five-model}).
\end{minipage}\hfill
\begin{minipage}[t]{2.70in}
\vspace{0pt}\centering\captionsetup{font=normalsize,skip=3pt,hypcap=false}
\includegraphics[width=\linewidth]{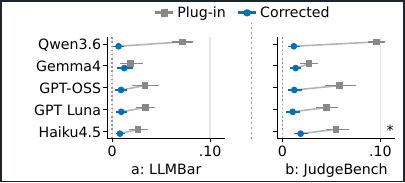}

\end{minipage}\par
{\captionsetup{font=normalsize,skip=3pt,hypcap=false}
\captionof{figure}{Prompt estimates on (a) LLMBar and (b) JudgeBench: gray plug-in, blue corrected. Each panel has 40 items, 24 prompts, two orders, $R=2$. Bars: 95\% item-composition intervals. Haiku* has 81.74\% invalid outputs and is a separate protocol case.}
\label{fig:five-model-correction}\par}
\smallskip

\begin{wrapfigure}{r}{3.30in}
\vspace{-.4\baselineskip}\centering\captionsetup{font=normalsize,skip=3pt}
\includegraphics[width=\linewidth]{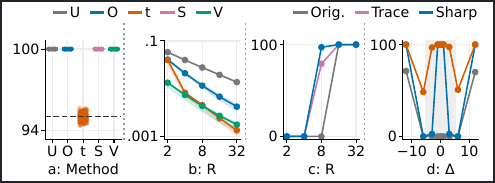}
\caption{Known-law validation: (a) coverage; (b) mean interval width across eight data-shaped laws (bands: range); (c) prompt-stability certification; (d) informative design comparisons at $R=32$ (gray: $\pm.005$). U/O/t/S/V: universal, observed variation, block-$t$, oracle S, oracle variance. $R$: calls/cell; $\Delta$: design contrast ($10^{-3}$). Section~\ref{sec:inference-validation-protocol} gives laws and Monte Carlo uncertainty.}
\label{fig:inference-validation}\vspace{-.4\baselineskip}
\end{wrapfigure}
\paragraph{Calibration under known laws.}
Known-law simulations (Figure~\ref{fig:inference-validation}) test coverage and certification of small prompt effects across 170 settings with 5,000 independent datasets each. With zero prompt dispersion, $N=152$, $K=6$, $R=8$ and interaction .08, the sharp covariance bound raises the rate of certifying prompt dispersion below .01 from 0\% (original) and 79.06\% (trace-refined) to 97.10\% (Monte Carlo interval [96.60\%,97.53\%]). At the true .01 boundary, no false certification occurs in the tested settings. The sharper bound can therefore certify that prompt dispersion is small without increasing the call budget in this setting.

Coverage and precision answer different questions in this experiment: approximate independent-block $t$ intervals cover 94.34--95.66\% on this grid, while categorical bounds provide finite-sample guarantees. Section~\ref{sec:temperature-panel-calibration} checks the separate two-panel temperature approximation under different patterns of variance concentration.
\WFclear

\section{Cooling Separates Repeatability from Prompt Dependence}
\label{sec:temperature-story}
With the measurement validated, we use cooling to test whether greater repeatability also means less prompt dependence. We evaluate Qwen3.6-27B and Gemma4-31B-IT on the same $N=40$ items (20 per LLMBar/JudgeBench), $K=48$ semantic prompts, both orders, and $T\in\{1,.2,0\}$. Both use native nonthinking generation, a 512-token cap, and unconstrained verdict output, retaining invalid outcomes. Components use $R=4$ calls per cell; independent four-call deployments use $J=8$ groups per item and temperature with matched prompt pairs. Each model contributes 49,920 analyzed calls (Section~\ref{sec:temperature-details}).

\begin{figure}[!htbp]
\centering
\includegraphics[width=\linewidth]{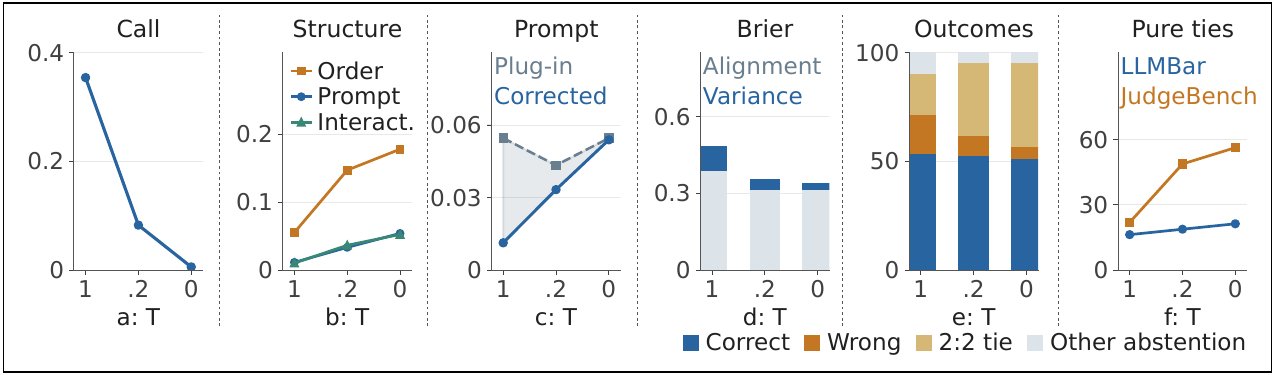}
\caption{Qwen across temperature $T$: (a) call variation; (b) prompt, order, and interaction components; (c) plug-in and corrected prompt estimates, with their gap shaded; (d) Brier loss decomposed into deployment variance and corrected mean-to-gold squared error; (e) correct, wrong, pure-tie, and other-abstention percentages; (f) pure candidate 2:2 ties by dataset. See Sections~\ref{sec:temperature-details} and~\ref{sec:temperature-full-repeat-inference}.}
\label{fig:temperature-response}
\end{figure}

\paragraph{Repeatability and prompt dependence diverge.}
Cooling makes repeated calls more consistent in both models while their corrected prompt estimates increase. From $T=1$ to $0$, Qwen's call estimate falls from .35386 to .00569, while its corrected prompt component rises from .01125 to .05387, a factor of 4.79 (call-level interval 3.3 to 8.5; the replication gives 5.5). The plug-in prompt estimate rounds to .05456 at both endpoints: the falling finite-call contribution offsets the growing prompt estimate (Equation~\eqref{eq:crossed-plugin-sample-identity}; Figure~\ref{fig:temperature-response}). Gemma's corrected prompt estimate rises by .01421.

\paragraph{Effect sizes and uncertainty.}
Table~\ref{tab:temperature-main} reports absolute contrasts with call-level and item-composition intervals. The upper-panel call-level intervals are worst-case bounds at $R=4$, which are wide for the prompt contrast. Calibrated intervals, a pre-specified replication (E1; Table~\ref{tab:discussion-studies-overview}), and 32 more Gemma repeats per cell (E6) place this contrast above zero (lower panel; Section~\ref{sec:discussion-studies-call-level}). So does an exponential distribution-free bound on the pooled calls at the four-contrast error level (post hoc; Proposition~\ref{prop:nn-exponential}). On 320 unseen items from three benchmarks and 24 unseen prompts (E2--E4, E9), the increases span $+.0262$ to $+.0421$ for Qwen and $+.0067$ to $+.0114$ for Gemma, each interval above zero (Section~\ref{sec:discussion-studies-items}). Sections~\ref{sec:temperature-full-repeat-inference} and~\ref{sec:temperature-panel-calibration} give full results and calibration.

\begin{table}[!htbp]
\centering\small
\setlength{\tabcolsep}{3pt}\renewcommand{\arraystretch}{1.0}
\caption{Endpoint contrasts ($T=0$ minus $T=1$). Upper panel: complete-pair $R=4$ estimates; call-level intervals have simultaneous coverage of at least 95\% for each model's four contrasts (two cooling levels, two components), and composition intervals are pointwise 95\% paired item-resampling summaries. Lower panel: call-level evidence for the prompt contrast on the same items and prompts. Calibrated exact-variance (EV) and item-paired (IP) intervals are Bonferroni-adjusted over the same four contrasts. The E1 and distribution-free (Cantelli) intervals jointly cover the prompt and call contrasts; those for all four use the exponential bound of Proposition~\ref{prop:nn-exponential} (post hoc).}
\label{tab:temperature-main}
\begin{tabular*}{\linewidth}{@{\extracolsep{\fill}}llrrr@{}}
\toprule
Model & Component & Estimate & Call-level interval & Composition interval\\
\midrule
Qwen & Call & $-.34818$ & $[-.37388,-.32247]$ & $[-.37860,-.32010]$\\
Qwen & Prompt & $+.04262$ & $[-.05699,.12589]$ & $[.03065,.05448]$\\
Gemma & Call & $-.04488$ & $[-.07059,-.01917]$ & $[-.06736,-.02526]$\\
Gemma & Prompt & $+.01421$ & $[-.07987,.09869]$ & $[.00731,.02261]$\\
\midrule
Qwen & Calibrated, original calls & $+.04262$ & \multicolumn{2}{r}{EV $[.0363,.0489]$, IP $[.0212,.0641]$}\\
Qwen & Distribution-free, pooled $R=8$ (post hoc) & $+.0440$ & \multicolumn{2}{r}{$[.008,.080]$; four contrasts $[.0173,.0701]$}\\
Qwen & Replication E1, EV & $+.0454$ & $[.0398,.0510]$ & \\
Gemma & Calibrated, original calls & $+.01421$ & \multicolumn{2}{r}{EV $[.0120,.0164]$, IP $[.0012,.0272]$}\\
Gemma & Distribution-free, pooled $R=40$ (E6) & $+.0133$ & \multicolumn{2}{r}{$[.0026,.0240]$; four contrasts $[.0046,.0219]$}\\
Gemma & Replication E1, EV & $+.0119$ & $[.0100,.0138]$ & \\
\bottomrule
\end{tabular*}
\end{table}

\paragraph{Common metrics miss the increase.}
Metrics that compare single calls across prompts also count call disagreement. On the same Qwen calls, across-prompt agreement rises from .57 to .71 and the ProSA-style binary correctness gap falls from .32 to .10 \citep{zhuo2024prosa}, while the corrected prompt estimate rises from .011 to .054. Section~\ref{sec:stored-label-metrics} defines these metrics, and Table~\ref{tab:discussion-metrics} shows the same pattern on the unseen items.

\paragraph{A sharpening mechanism.}
\leavevmode A two-prompt model shows why cooling can raise prompt dispersion. Take two equally weighted binary prompts, identical across orders, with fixed opposite logit gaps $\pm a$, $a>0$. For $T>0$, let $p_\pm(T)=\sigma(\pm a/T)$, where $\sigma$ is the logistic function. Then
\begin{equation}
\Up^{(q)}(T)=\tfrac12\tanh^2\!\left(\frac{a}{2T}\right),\qquad
\Uc(T)=\frac{1}{2\cosh^2(a/(2T))}.
\label{eq:temperature-sharpening}
\end{equation}
Their sum remains $1/2$: cooling converts within-prompt randomness into between-prompt dispersion while keeping the underlying preferences fixed. Prompts that share a preferred candidate instead converge toward the same verdict as $T\to0$. Qwen and Gemma show the same opposing changes in call and prompt estimates when measured after generation, stopping, and verdict parsing.

\paragraph{The increase holds within each prompt family.}
The 48 prompts fall into four predefined families: lexical, syntactic, discourse, and clause-order variation. In all four, call estimates fall and corrected prompt estimates rise in both models; each family alone ($K=12$) and each set of three families ($K=36$) gives an increase of $+.0346$ to $+.0436$ for Qwen and $+.0127$ to $+.0148$ for Gemma, with every exact-variance interval above zero (Section~\ref{sec:discussion-studies-items}). The clause-order subset uses the same five sentences in 12 orders. Section~\ref{sec:temperature-prompt-families} reports this exploratory reanalysis.

\section{From Response Structure to Final Judgments}
\label{sec:decision-geometry}\label{sec:temperature-rescoring}
Cooling trades call variation for prompt dependence. What this means for a final judgment depends on gold alignment and the voting rule, which we examine using independent four-call deployments. Design D draws two distinct prompts, presents each in both orders, and forms a verdict-frequency vector $V$. The original (strict) rule selects a candidate only if it is the unique most frequent category. It abstains if categories share the maximum count or if the top category is tie or invalid.

\paragraph{Gold alignment.}
For $\E V=\bar p$ and gold candidate $y$, squared loss separates into
\begin{equation}
\E\|V-e_y\|^2=\underbrace{\E\|V-\bar p\|^2}_{\text{vector risk}}+
\underbrace{\|\bar p-e_y\|^2}_{\text{gold alignment error}}.
\label{eq:story-brier}
\end{equation}
\equationalias{eq:design-brier}{eq:story-brier}%
The first term measures how tightly vote vectors cluster around the judge's mean; the second measures how close that mean is to the gold answer. Cooling lowers Qwen's vector-risk estimate from .09643 to .02924 and Brier estimate from .48516 to .34141. In Gemma, estimated vector risk also falls, but Brier loss changes little because alignment responds differently. These estimates use the same independent votes; Table~\ref{tab:temperature-loss-precision} reports Qwen's paired item-composition intervals.

\paragraph{A fixed-margin position mechanism.}
\leavevmode Does lower loss mean better decisions? Cooling also raises Qwen's order component (Figure~\ref{fig:temperature-response}b), so we vary position dependence in balanced voting at a fixed mean gold preference. Consider a binary family in which every prompt chooses the gold candidate with probability $\pi+\delta$ in AB order and $\pi-\delta$ in BA order. Here $1/2\le\pi<1$ and $0\le\delta\le1-\pi$. Its mean gold margin $m=\bar p_{\rm gold}-\bar p_{\rm other}=2\pi-1$ stays fixed, $\Up^{(q)}=\Ui=0$, and $\Uo=2\delta^2$. Independent balanced votes have vector risk $[\pi(1-\pi)-\delta^2]/2$. Their pure-2:2 probability $T_{22}$ and strict accuracy follow from the sum of independent $\operatorname{Bin}(2,\pi+\delta)$ and $\operatorname{Bin}(2,\pi-\delta)$ variables.

\begin{proposition}[Position dependence and the tie rule]
\label{prop:position-margin}
For $1/2<\pi<2/3$, increasing $\delta^2$ decreases vector risk, gold Brier loss, and strict accuracy, while increasing pure ties and fair-coin accuracy. The mean gold margin remains positive and fixed.
\end{proposition}
\noindent\begin{minipage}[t]{\dimexpr\linewidth-3.30in-12pt\relax}
\vspace{0pt}
For example, at $\pi=.6$, increasing $\delta$ from 0 to .4 reduces vector risk from .12 to .04 while strict accuracy falls from .4752 to .36. Figure~\ref{fig:position-theory} shows the accompanying loss and tie-rule curves. The proof and zero-margin endpoint are in the supplement.
\end{minipage}\hfill
\begin{minipage}[t]{3.30in}
\vspace{0pt}\centering\captionsetup{font=normalsize,skip=3pt,hypcap=false}
\includegraphics[width=\linewidth]{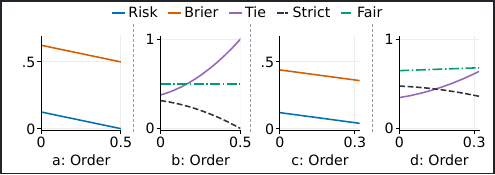}

\end{minipage}\par
{\captionsetup{font=normalsize,skip=3pt,hypcap=false}
\captionof{figure}{Exact binary families: (a,b) $m=0$; (c,d) $m=.2$. Horizontal: order component $\Uo$. Loss panels show vector risk and gold Brier loss; decision panels show pure ties and strict/fair-coin accuracy.}
\label{fig:position-theory}\par}

\paragraph{Changing the rule on the same independent votes.}
\leavevmode The recorded votes show this dependence on the tie rule. A fair pure-2:2 rule chooses either candidate with probability $1/2$ only at a pure candidate tie, leaving all other outcomes unchanged. We compute its expected score exactly from the stored vote groups. If $A$ is accuracy over all deployments,
\begin{equation}
A_{\rm coin22}=A_{\rm strict}+\tfrac12T_{22},\qquad
\Gamma_t:=\Delta_t A_{\rm coin22}-\Delta_t A_{\rm strict}
=\tfrac12\Delta_t T_{22}.
\label{eq:story-coin22}
\end{equation}
Here $\Delta_t$ compares $T=t$ with $T=1$. Qwen's pure ties increase from 19.06\% to 38.75\% at $t=0$, producing a direct rule interaction of 9.84 pp [1.68,18.01]. The interval is simultaneous across both cooling contrasts and both models, conditional on the fixed 40 items and assigned prompt pairs. Each contrast uses 320 paired independent vote groups. Table~\ref{tab:main-decisions} reports both rule-specific accuracy contrasts. Gemma's smaller tie change, 6.88\% to 7.81\%, corresponds to an estimated interaction of .47 pp [$-1.54$,2.48]. Qwen's rule interaction replicates on fresh calls, on 200 unseen items, and in E9. Qwen's estimated accuracy gain from cooling under the fair 2:2 rule is positive in every collection, +2.88 pp [1.02,4.73] pooled post hoc over 240 items (Section~\ref{sec:discussion-studies-decisions}).

\begin{table}[!htbp]\centering\small
\setlength{\tabcolsep}{3pt}\renewcommand{\arraystretch}{1.0}
\caption{Cooling contrasts ($T=0$ minus $T=1$, pp). Decision intervals jointly cover twelve endpoints within each model; rule-interaction intervals cover four interactions across both models. All are finite-sample 95\% intervals.}
\label{tab:main-decisions}
\begin{tabular*}{\linewidth}{@{\extracolsep{\fill}}llrr@{}}\toprule
Model & Endpoint / rule & Estimate & Simultaneous interval\\\midrule
Qwen & Accuracy / original & $-1.88$ & $[-14.43,10.68]$\\
Qwen & Accuracy / fair 2:2 & $+7.97$ & $[-.87,16.81]$\\
Qwen & Wrong decision / original & $-12.81$ & $[-23.07,-2.56]$\\
Qwen & Accuracy rule interaction & $+9.84$ & $[1.68,18.01]$\\
Gemma & Accuracy / original & $-.31$ & $[-5.26,4.63]$\\
Gemma & Accuracy / fair 2:2 & $+.16$ & $[-4.52,4.83]$\\
Gemma & Accuracy rule interaction & $+.47$ & $[-1.54,2.48]$\\\bottomrule
\end{tabular*}
\end{table}

The item-level results link position dependence and baseline preferences to ties. The estimated order component rises from .055 to .178 in Qwen, versus .058 to .063 in Gemma. In Qwen, order-component changes correlate with changes in independently sampled pure ties ($r=.75$ centered within dataset), and the link replicates on fresh calls, unseen items, and RewardBench~2 ($r=.81$, .76, and .79 without centering; Section~\ref{sec:discussion-studies-position}). At $T=0$, all four calls chose the first-shown candidate in 905 of Qwen's 929 pure ties in five collections (post hoc; Table~\ref{tab:further-ties}). A clause favoring the second position moved Qwen's ties toward it (E10 at $T=1$, pre-specified; Table~\ref{tab:further-e10}); an earlier intervention, E7, did not meet its joint criterion (Section~\ref{sec:discussion-studies-position}). Cooling raises Qwen's ties by a mean 36.8 pp on items with a $T=1$ gold margin $|m|\le.25$, versus .7 pp when $m>.25$ (Section~\ref{sec:temperature-details}).

\paragraph{What remains after aggregation.}
On 120 unseen items (E8), TrustJudge \citep{wang2025trustjudge} and PGED (\citealp{hu2024acyclic}; applied post hoc) raise accuracy and remove order conflict (Table~\ref{tab:aggregation-baselines}). Our decomposition shows that PGED's remaining disagreement between prompts is mostly call noise and falls from 18.3\% to 11.0\% when calls per verdict double, whereas TrustJudge's is mostly systematic (2.37 pp [.75,3.98] for Qwen, 4.18 pp [1.80,6.56] for Gemma) and barely changes.

\paragraph{The cost of abstention.}
When abstentions increase, both wrong decisions and overall accuracy can decrease. With cost one for a wrong candidate, zero for a correct one, and $c\in[0,1]$ for abstention,
\begin{equation}
L_c=\Prob(\text{wrong})+c\Prob(\text{abstain}).
\label{eq:position-abstention-cost}
\end{equation}
\noindent\begin{minipage}[t]{\dimexpr\linewidth-2.70in-12pt\relax}
\vspace{0pt}
For Qwen at $T=0$, the original rule's estimated loss change is $-.128125+.146875c$: fewer mistakes offset extra abstentions when abstention is cheap (Figure~\ref{fig:temperature-decision-rules}). The estimated losses are equal at $c=.872$. A finite-sample band covering both rules, both cooling contrasts, and every $c\in[0,1]$ simultaneously supports lower original-rule loss for $c<.165$. Deployment choice thus depends on the rule and abstention cost; the supplement gives both models' outcomes and interval constructions.
\end{minipage}\hfill
\begin{minipage}[t]{2.70in}
\vspace{0pt}\centering\captionsetup{font=normalsize,skip=3pt,hypcap=false}
\includegraphics[width=\linewidth]{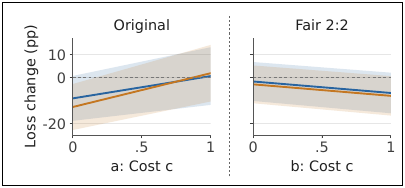}
\captionof{figure}{Qwen loss change (pp) relative to $T=1$: (a) original rule; (b) fair 2:2. Blue: $T=.2$; orange: $T=0$. Bands jointly cover both rules, both temperatures, and all costs at 95\%, given items and prompt pairs.}
\label{fig:temperature-decision-rules}
\end{minipage}\par

\section{Prompting, Reasoning, and Deployment Design}
\label{sec:intervention-routes}
Like cooling, prompt policies and reasoning change response variation and final decisions.

\begin{wrapfigure}{r}{2.70in}
\vspace{-.4\baselineskip}\centering\captionsetup{font=normalsize,skip=3pt}
\includegraphics[width=\linewidth]{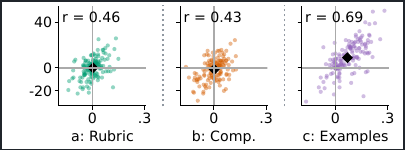}
\caption{Changes from paraphrasing on 152 items. Horizontal: corrected order component ($R=16$); vertical: independent pure-2:2 ties (pp; D/E mean, 64 groups/item/design). Diamonds: paired means; $r$: descriptive correlation.}
\label{fig:policy-item-link}\vspace{-.4\baselineskip}
\end{wrapfigure}
\paragraph{Examples redirect dependence toward position.}
In an exploratory comparison of Qwen's four six-prompt policies, replacing paraphrasing with examples lowers the prompt estimate from .00339 to .00014 and raises the order estimate from .09784 to .16677. Under D, pure candidate ties rise from 27.88\% to 36.85\%. Fair pure-2:2 rescoring changes the original-rule accuracy contrast of $-6.36$ pp [$-8.99,-3.73$] to $-1.88$ pp [$-4.51,.76$], showing that tie handling accounts for much of the observed gap (Section~\ref{sec:policy-decisions}).

Changes in position dependence correlate with changes in independently sampled ties across all three policies: $r=.69$ for examples, .46 for rubrics, and .43 for comparison (Figure~\ref{fig:policy-item-link}). When order is reversed, a preferred position corresponds to the opposite candidate, so balanced votes can cluster near a tie. Section~\ref{sec:recovery-response-decision-figures} gives the full profiles.
\WFclear
\Needspace{21\baselineskip}
\begin{wrapfigure}{r}{3.20in}
\vspace{-.4\baselineskip}\centering\captionsetup{font=normalsize,skip=3pt}
\includegraphics[width=\linewidth]{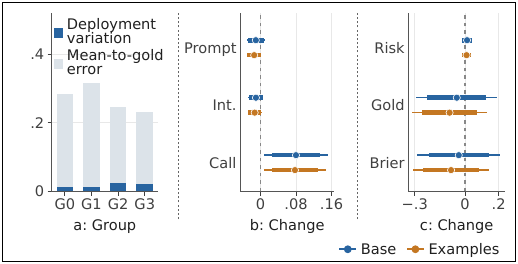}
\caption{Gemma: (a) Brier decomposition; (b,c) thinking-minus-nonthinking contrasts. G0/G1: nonthinking base/examples; G2/G3: thinking base/examples. Int./Risk/Gold: interaction, deployment variance, squared mean-to-gold error. Thick/thin intervals: 95\%/99.583\% centered whole-item bootstrap intervals, paired across arms and stratified by domain; the latter target approximate 95\% coverage over twelve endpoints. These quantify item composition conditional on recorded calls.}
\label{fig:intervention-alignment}\vspace{-.4\baselineskip}
\end{wrapfigure}

\paragraph{Reasoning increases call variation.}
On 32 RewardBench~2 items, a Gemma4-31B-IT study combines base/example instructions with either nonthinking generation or a 1,024-token thinking budget \citep{malik2025rewardbench2}. Each arm uses 48 prompts, two orders, $R=2$, eight independent D deployments per item, and the same constrained A/B/TIE final decoder. Thinking increases call variation under both instructions, from .01302 to .09277 for the base instruction, while Brier point estimates also fall. Figure~\ref{fig:intervention-alignment} separates the lower Brier point estimates into increased deployment variation and a larger estimated reduction in mean-to-gold error. Adjusted composition intervals support both call increases in the exploratory twelve-contrast analysis.

On 80 unseen items, thinking raises call variation by $.0531$ $[.0222,.0840]$ (Section~\ref{sec:discussion-studies-items}). Gemma's paired component and loss contrasts appear in Figure~\ref{fig:intervention-alignment}; Section~\ref{sec:gemma-fourarm-details} gives the pre-specified tests and exploratory analyses. Qwen's policy intervals are approximate simultaneous 95\% item-composition intervals.
\WFclear

\paragraph{Response components also determine sampling tradeoffs.}
Design D uniformly samples two distinct prompts and queries both orders; E uniformly samples four prompts and randomly assigns two to each order. For $K\ge4$ equally weighted prompts, both estimate the same response mean. Their exact squared-error risk contrast is
\begin{equation}
\mathcal R_E-\mathcal R_D=\frac{K}{4(K-1)}\bigl(\Ui-\Up^{(q)}\bigr).
\label{eq:main-design-tradeoff}
\end{equation}
At the same four-call budget, E has lower population vector risk precisely when prompt dispersion exceeds prompt--order interaction. Both designs cancel the order main effect. Because the same-cell terms also cancel in their risk difference, it can be estimated without bias from one measurement call per cell. More generally, this comparison is possible for designs with the same budget and target weights exactly when their within-cell collision moments are equal (Propositions~\ref{prop:design-collisions}--\ref{prop:design-quadratic}). For a fixed benchmark target, an exact variance identity gives a continuous oracle solution for dividing the budget between items and repeats (Proposition~\ref{prop:fixed-population-allocation}). Table~\ref{tab:semantic-calibration} compares predicted and observed risk differences across three models and two benchmarks, using separate measurement and deployment calls on the same test pool. Table~\ref{tab:observed-variation-intervals} reports simultaneous 95\% call-level intervals for eight Qwen conditions; Section~\ref{sec:recovery-estimation-checks} compares raw, fixed- and held-out-shrinkage, and zero predictors.

\paragraph{Choosing a design and a strategy.}
Before deployment, Equation~\eqref{eq:main-design-tradeoff} bounds the design gain in squared-error risk by $K/(4(K-1))$ times the within-order prompt dispersion. In ten settings, the design chosen in development is within 1.77 pp of either fixed design in test accuracy. The choice between one call and voting (D) matters more. In a pre-specified study on 120 unseen items (E9), strategies chosen from measurement calls on the Section~\ref{sec:temperature-story} items gained 5.89 pp [4.48,7.30] in mean accuracy over the alternative in five settings (judge, temperature, tie rule). Post hoc, they beat both fixed strategies on average over all eight (Table~\ref{tab:discussion-e9}). D never has higher risk than one call, yet on the E9 items it is less accurate under the original rule, so we report risk and decision loss separately (Section~\ref{sec:discussion-studies-selection}). Choosing the strategy at each abstention cost from 0 to 1 under that rule gives 3.02 pp [2.39,3.65] less mean loss on these items than always using one call (post hoc; Section~\ref{sec:discussion-studies-further}).

\begin{table}[!htbp]\centering\small
\setlength{\tabcolsep}{3pt}\renewcommand{\arraystretch}{1.05}
\caption{Accuracy gain (pp) on the 120 unseen E9 items, with item-paired 95\% intervals.}
\label{tab:discussion-e9}
\begin{tabular*}{\linewidth}{@{\extracolsep{\fill}}lll@{}}\toprule
Chosen strategy minus & Mean gain & Largest gain\\\midrule
the other strategy, five pre-specified settings & $+5.89$ $[4.48,7.30]$ & $+11.49$ $[7.81,15.17]$\\
always one call, all eight settings (post hoc) & $+.88$ $[.54,1.22]$ & $+5.37$ $[3.30,7.44]$\\
always D, all eight settings (post hoc) & $+3.36$ $[2.03,4.69]$ & $+11.49$ $[7.81,15.17]$\\\bottomrule
\end{tabular*}
\end{table}

\section{Conclusion}
Repeated calls and reworded prompts test different kinds of stability, and a judge can gain one while losing the other. Finite-call correction separates prompt dependence from call randomness, and matching bounds show how small a prompt effect is detectable. Analytical models and matched interventions show how repeatability, prompt dependence, and position effects change together. The value of consistency for final judgments depends on gold alignment, voting rules, and abstention cost; the response components also give exact risk comparisons between deployment designs and guide how to divide a call budget between items and repeats. Pre-specified studies confirm the cooling result for both judges on fresh calls and on unseen items and prompts. On 120 unseen items, the strategy (one call or voting) chosen from measurement calls beat the other.

\paragraph{Limitations and future work.}
Our guarantees apply to independent calls that return categorical verdicts for a fixed set of prompts. Future work can study generated prompts, dependent calls, and continuous scores. We also aim to test more judges and use human review to check that prompts preserve meaning. Other directions are to estimate intervention and design differences more precisely and allocate calls based on the observed variation.

\label{page:iclr-main-end}
\clearpage
\subsection*{AI use statement}
AI tools helped polish the writing, analyze experimental results, and check and correct mathematical formulations. The authors led the formulation and description of the problem, designed and validated the experiments, carried out the theoretical analysis, and prepared the initial and final manuscripts. They take responsibility for the complete work.
\subsection*{Ethics statement}
This work studies LLM evaluation using public benchmarks and model-generated judgments, without human-subject experiments. The framework supports transparent analysis of judge variability, including effects of prompt wording, candidate order, and decision rules that can influence model comparison and response selection.
\subsection*{Reproducibility statement}
The main paper and appendix give the estimators, proofs, and experimental protocols. A compact data package provides recorded categorical judgments, prompt and item metadata, gold labels, independent vote groups, and analysis scripts for reproducing the main numerical results. Section~\ref{sec:discussion-studies} reports the pre-specified replication and extension studies; we will release their plans, code, and other materials with the paper.
\bibliographystyle{iclr2027_conference}
\bibliography{rev_references}
\clearpage
\makeatletter
\providecommand{\equationalias}[2]{%
  \@ifundefined{r@#2}{}{%
    \protected@write\@auxout{}{\string\newlabel{#1}{\csname r@#2\endcsname}}}}
\makeatother
\appendix
\setcounter{section}{0}
\setcounter{secnumdepth}{2}
\setcounter{equation}{0}
\setcounter{table}{0}
\setcounter{figure}{0}
\setcounter{theorem}{0}
\setcounter{proposition}{0}
\renewcommand{\theequation}{S\arabic{equation}}
\renewcommand{\theHequation}{supp.\arabic{equation}}
\renewcommand{\thetable}{S\arabic{table}}
\renewcommand{\thefigure}{S\arabic{figure}}
\renewcommand{\thetheorem}{S\arabic{theorem}}
\renewcommand{\theproposition}{S\arabic{proposition}}
\renewcommand{\theHtheorem}{supp.\arabic{theorem}}
\renewcommand{\theHproposition}{supp.\arabic{proposition}}
\renewcommand{\theremark}{S\arabic{remark}}
\renewcommand{\thedefinition}{S\arabic{definition}}
\renewcommand{\sharedtablelabel}[1]{\label{#1-supp}}
\renewcommand{\maintextreproductionnote}{}
\section*{Supplementary Material}
\section{Response Laws and Complete-Pair Estimation}
\label{sec:s-proofs}\label{sec:s-crossed}\label{sec:iclr-estimator-background}
The primary analysis gives equal weight to all prompts in the declared census. Calls are independent conditional on the fixed item--prompt--order laws, with iid repetitions within each cell. Before analysis, we map verdicts back to the original candidate identities and assign invalid completed outputs to a fourth category $\bot$. We report raw estimates, including negative values when a component is near zero.

\subsection{Population decomposition}
Let $e_Z$ be the one-hot vector of the categorical response and
$P_X=\E[e_Z\mid X]$. The conditional-mean residual is orthogonal to every
function of $X$, hence
\begin{equation}
\E\|e_Z-\bar p\|_2^2
=\E\|e_Z-P_X\|_2^2+\E\|P_X-\bar p\|_2^2.
\label{eq:s-total-variance}
\end{equation}
Because $\|e_Z\|_2^2=1$,
\begin{align}
\E[\|e_Z-P_X\|_2^2\mid X]&=1-\|P_X\|_2^2,\\
\E\|e_Z-\bar p\|_2^2&=1-\|\bar p\|_2^2.
\end{align}
Substitution proves
$\Ut=\Uw+\Up$. The same proof applies when the categorical alphabet includes invalid
completed outputs $\bot$. Conditioning on validity defines a separate
response law, as specified below.

Let $X\sim Q$ and $O\sim q$ be independent by design. Put
\[
\begin{aligned}
P_{XO}&=\Law(Z\mid X,O),&
P_X^q&=\E[P_{XO}\mid X],\\
P^O&=\E[P_{XO}\mid O],&
\bar p&=\E P_{XO},
\end{aligned}
\]
and
\begin{align}
H_{XO}&=\E_{X,O}\|P_{XO}\|_2^2,
&H_X&=\E_X\|P_X^q\|_2^2,\\
H_O&=\E_O\|P^O\|_2^2,
&H_0&=\|\bar p\|_2^2.
\end{align}
The components are
\begin{align}
\Uc&=1-H_{XO},\\
\Up^{(q)}&=H_X-H_0,\\
\Uo&=H_O-H_0,\\
\Ui&=H_{XO}-H_X-H_O+H_0.
\end{align}

Write
\[
P_{XO}-\bar p
=(P_X^q-\bar p)+(P^O-\bar p)
+(P_{XO}-P_X^q-P^O+\bar p).
\]
The three centered terms are mutually orthogonal in
$L^2(Q\otimes q)$: the interaction has conditional mean zero given $X$ or
$O$, and the main effects are orthogonal by independence. Therefore
\[
\begin{aligned}
H_{XO}-H_0
&=(H_X-H_0)+(H_O-H_0)\\
&\quad+(H_{XO}-H_X-H_O+H_0).
\end{aligned}
\]
Adding $1-H_{XO}$ proves
\[
\Ut=\Uc+\Up^{(q)}+\Uo+\Ui.
\]
More explicitly,
\[
\begin{aligned}
\Up^{(q)}
&=\E_X\|P_X^q-\bar p\|_2^2,\\
\Uo
&=\E_O\|P^O-\bar p\|_2^2,\\
\Ui
&=\E_{X,O}
\left\|P_{XO}-P_X^q-P^O+\bar p\right\|_2^2.
\end{aligned}
\]
Moreover,
\[
\Uc=\E_{X,O}\!\left[1-\|P_{XO}\|_2^2\right].
\]
Hence every population component is nonnegative.

\subsection{Unbiased complete-pair estimator for a fixed census}

Let $Q_K=K^{-1}\sum_k\delta_{x_k}$ and let every positive-weight
prompt--order cell contain $R_{ko}\ge2$ calls. Assume all distinct calls are
mutually independent conditional on the fixed array $\{P_{ko}\}$. Define
the same-cell agreement
\begin{equation}
\widehat A_{ko,ko}=
\frac{1}{R_{ko}(R_{ko}-1)}
\sum_{r\ne r'}\1\{Z_{kor}=Z_{kor'}\}.
\label{eq:s-cell-agreement-same}
\end{equation}
For distinct cells, define
\begin{equation}
\widehat A_{ko,\ell o'}
=\widehat p_{ko}^{\top}\widehat p_{\ell o'},
\qquad (k,o)\ne(\ell,o').
\label{eq:s-cell-agreement}
\end{equation}
Set
\begin{align}
\widehat H_{XO}
&=\frac1K\sum_k\sum_oq_o\widehat A_{ko,ko},\\
\widehat H_X
&=\frac1K\sum_k\sum_{o,o'}q_oq_{o'}
\widehat A_{ko,ko'},\\
\widehat H_O
&=\sum_oq_o\frac1{K^2}\sum_{k,\ell}
\widehat A_{ko,\ell o},\\
\widehat H_0
&=\frac1{K^2}\sum_{k,\ell}\sum_{o,o'}q_oq_{o'}
\widehat A_{ko,\ell o'}.
\end{align}
Each $\widehat H$ is unbiased for its fixed-census counterpart. The component
estimators
\begin{align}
\widehat\Uc&=1-\widehat H_{XO},\\
\widehat\Up^{(q)}&=\widehat H_X-\widehat H_0,\\
\widehat\Uo&=\widehat H_O-\widehat H_0,\\
\widehat\Ui
&=\widehat H_{XO}-\widehat H_X-\widehat H_O+\widehat H_0,\\
\widehat\Ut&=1-\widehat H_0
\label{eq:s-crossed-estimators}
\end{align}
are therefore unbiased and satisfy the exact decomposition in every sample.

To express the census correction, define
\begin{align}
\widehat A_X
&=\frac1K\sum_k\sum_{o,o'}q_oq_{o'}\widehat A_{ko,ko'},\\
\widehat A_{\ne X}
&=\frac1{K(K-1)}\sum_{k\ne\ell}\sum_{o,o'}q_oq_{o'}
\widehat A_{ko,\ell o'}.
\end{align}
Then
\begin{equation}
\widehat\Up^{(q)}
=\frac{K-1}{K}(\widehat A_X-\widehat A_{\ne X}).
\end{equation}
The inclusive $K^{-2}$ sums target the realized fixed census.

\subsection{Exact order mixture and plug-in excess}

For one prompt with order laws $P^o$, the declared mixture is
$P^q=\sum_oq_oP^o$, so
\begin{equation}
\|P^q\|_2^2
=\sum_oq_o^2\|P^o\|_2^2
+\sum_{o\ne o'}q_oq_{o'}(P^o)^\top P^{o'}.
\end{equation}
Agreement between distinct calls in the same order estimates the diagonal terms;
agreement between calls in different orders estimates the off-diagonal terms. Pooling equal numbers of AB and BA
calls into one unordered sample introduces finite-$R$ bias in mixture
agreement. If $a=P^{AB}$, $b=P^{BA}$, and there are $n$ calls in each
order, ordinary pooled distinct-pair agreement has expectation
\begin{equation}
\E\widehat A_{\rm pool}
=\left\|\frac{a+b}{2}\right\|_2^2
-\frac{\|a-b\|_2^2}{4(2n-1)}.
\label{eq:s-pooled-order-bias}
\end{equation}

For the plug-in prompt estimate in the fixed crossed design,
\[
\widehat U_{\rm plug,prompt}^{(q)}
=\frac1K\sum_k
\|\widehat P_k^q-\bar{\widehat P}^{\,q}\|_2^2,
\qquad
\widehat P_k^q=\sum_oq_o\widehat p_{ko},
\]
with $R_o$ calls in each prompt cell of order $o$,
\begin{equation}
\begin{aligned}
\E\widehat U_{\rm plug,prompt}^{(q)}
&=\Up^{(q)}+\frac{K-1}{K^2}\\
&\quad{}\times
\sum_k\sum_o\frac{q_o^2}{R_o}
(1-\|P_{ko}\|_2^2).
\end{aligned}
\label{eq:s-crossed-excess-general}
\end{equation}
For two orders with $q_o=1/2$ and equal $R$,
\begin{equation}
\E\widehat U_{\rm plug,prompt}^{(q)}
=\Up^{(q)}+\frac{K-1}{2KR}\Uc.
\label{eq:s-crossed-excess-balanced}
\end{equation}
The balanced two-order design gives this simplified expression.
At the cell level, the empirical plug-in concentration and the
distinct-call agreement satisfy
\[
\|\widehat p_{ko}\|_2^2-\widehat A_{ko,ko}
=
\frac{1-\widehat A_{ko,ko}}{R}.
\]
This follows directly by writing both quantities in terms of the categorical
cell counts. Averaging this identity over the two equally weighted orders and
the \(K\) prompts, and then applying the fixed-census factor
\((K-1)/K\), gives the exact sample identity
\begin{equation}
\widehat U_{\rm plug,prompt}^{(q)}-\widehat\Up^{(q)}
=\frac{K-1}{2KR}\widehat\Uc .
\label{eq:s-crossed-excess-sample}
\end{equation}
At $K=6$, the $R$-scaled excess equals $(5/12)\widehat\Uc$ in every
sample. The simulation with known cell laws, the control with byte-identical inputs,
and the independent reference with more repeated calls test the estimator
against separately known or measured targets.

\paragraph{Deterministic and near-deterministic runtimes.}
If every fixed prompt--order cell is deterministic, each $P_{ko}$
is a point mass, $\Uc=0$, and the finite-repeat excess vanishes for every
$R$. More generally, Equation~\eqref{eq:s-crossed-excess-balanced} shows that
the excess shrinks in proportion to $\Uc/R$. At temperature zero, any
remaining variability of the configured serving system contributes to
$P_{ko}$ and is measured by the call component.

\paragraph{Fixed census versus a prompt generator.}
For a single-order response law, write $\widehat a_k$ for within-cell distinct-call agreement and $\widehat b_{k\ell}$ for cross-cell agreement, and average them as $\Aw=K^{-1}\sum_k\widehat a_k$ and $\Aa=[K(K-1)]^{-1}\sum_{k\ne\ell}\widehat b_{k\ell}$. The fixed-census estimators are $\widehat U_w=1-\Aw$, $\widehat U_p=(K-1)(\Aw-\Aa)/K$, and $\widehat U_t=1-\Aw/K-(K-1)\Aa/K$. Their expectations follow from $\E\widehat a_k=\|P_k\|^2$ and $\E\widehat b_{k\ell}=P_k^\top P_\ell$. If prompts are instead iid draws from a generator, $\Aw-\Aa$ targets its prompt variance, and the usual sample-variance plug-in has expectation $U_p+U_w/R$. A declared census uses denominator $K$ and inflation $(K-1)U_w/(KR)$. These are different sampling targets; the model studies use fixed censuses.
\subsection{What the classical mean-square baseline estimates}
\label{sec:fixed-effects-equivalence}
The classical fixed-effects method estimates the same quadratic dispersion
over the fixed prompt census. Let $Y_{kor}$ be the one-hot outcome for one item,
$\widehat P_{ko}$ its cell mean, and $\widehat P_k$ the order-averaged
mean. With $O=2$, define trace sums of squares
\begin{align}
SS_p&=OR\sum_k\|\widehat P_k-\widehat{\bar P}\|^2,\\
SS_{\rm int}&=R\sum_{k,o}\|
\widehat P_{ko}-\widehat P_k-\widehat P^o
+\widehat{\bar P}\|^2,\\
MS_e&=\frac{\sum_{k,o,r}\|Y_{kor}-\widehat P_{ko}\|^2}
 {KO(R-1)}=\widehat U_{\rm call}.
\end{align}
The numerator degrees of freedom are $d_p=K-1$ and
$d_{\rm int}=(K-1)(O-1)$. The corresponding corrections for the finite census satisfy
\begin{align}
\frac{d_p}{KOR}\left(\frac{SS_p}{d_p}-MS_e\right)
 &=\widehat U_{\rm plug,prompt}^{(q)}
 -\frac{K-1}{2KR}\widehat U_{\rm call}
 =\widehat\Up^{(q)},\\
\frac{d_{\rm int}}{KOR}
 \left(\frac{SS_{\rm int}}{d_{\rm int}}-MS_e\right)
 &=\widehat U_{\rm plug,int}
 -\frac{K-1}{2KR}\widehat U_{\rm call}
 =\widehat\Ui.
\end{align}
Thus the fixed-effects trace-mean-square correction and the complete-pair
estimator coincide in every sample in this balanced design, even
when categorical cell variances differ. Subtracting the two
component estimates cancels their common correction, so the plug-in contrast
also equals the unbiased contrast. These identities hold for the
fixed-census target. Random-effects REML instead estimates a quantity defined
by a hierarchical population model. Comparing it with the fixed-census
estimate requires specifying that model and the relation between the two
targets. We treat methods that agree in every sample as the same estimator.

\section{Fixed-Census Inference Near Zero}
\label{sec:nn-main}

Fix $N\ge1$ benchmark items, a uniform census of $K\ge2$ prompts, a
categorical alphabet with $|\cZ|\ge2$, and order weights $q_o>0$ with
$\sum_o q_o=1$. Each item--prompt--order cell
contains $R\ge2$ iid categorical calls, and all calls are independent
conditional on the fixed cell laws. Define
\[
\theta_i=\Up^{(q)}(s_i),\qquad
\theta=\frac1N\sum_i\theta_i,\qquad
\widehat\theta=\frac1N\sum_i\widehat\Up^{(q)}(s_i),
\]
where the estimate uses every eligible pair of distinct calls. The following
results quantify uncertainty from repeated calls while holding the items and
prompt census fixed.

Index a prompt--order cell by $a=(k,o)$, write its law on item $i$ as
$p_{ia}$, and put
\begin{align}
 W_{(k,o),(\ell,o')}
 &=\left(\frac{\1\{k=\ell\}}K-\frac1{K^2}\right)q_oq_{o'},
 \label{eq:nn-W}\\
 \Sigma_{ia}&=\operatorname{diag}(p_{ia})-p_{ia}p_{ia}^{\top},
 &d_{ia}&=2\sum_bW_{ab}p_{ib}.
 \label{eq:nn-covariance}
\end{align}
Thus $\theta_i=\sum_{a,b}W_{ab}p_{ia}^{\top}p_{ib}$.

\begin{theorem}[Exact fixed-census variance and a uniform bound]
\label{thm:nn-variance}
Under the stated fixed-cell design, $\E\widehat\theta=\theta$ and
$\Var(\widehat\theta)=N^{-2}\sum_iV_i$, where
\begin{equation}
\begin{aligned}
V_i={}&\frac1R\sum_a d_{ia}^{\top}\Sigma_{ia}d_{ia}
+\frac{2}{R(R-1)}\sum_a W_{aa}^2
   \operatorname{tr}(\Sigma_{ia}^2)\\
&+\frac4{R^2}\sum_{a<b}W_{ab}^2
   \operatorname{tr}(\Sigma_{ia}\Sigma_{ib}).
\end{aligned}
\label{eq:nn-exact-variance}
\end{equation}
Here $a<b$ uses any fixed ordering of the cells. With $C=|\cZ|$ and $s_q=\sum_o q_o^2$,
\begin{align}
 \Var(\widehat\theta)&\le a_{N,R}\theta+b_{N,R},
 \label{eq:nn-variance-envelope}\\
 a_{N,R}&=\frac{2s_q}{NKR},&
 b_{N,R}&=\frac{(K-1)s_q^2}{2NK^2R(R-1)}.
 \label{eq:nn-envelope-constants}
\end{align}
For six prompts, two balanced orders, and four verdict categories,
$a_{N,R}=1/(6NR)$ and $b_{N,R}=5/[288NR(R-1)]$.
\end{theorem}

Write $P_{ik}^{q}=\sum_o q_op_{i,(k,o)}$ and
$\bar p_i=K^{-1}\sum_kP_{ik}^{q}$.
The first term in Equation~\eqref{eq:nn-exact-variance} vanishes when
$\theta=0$: indeed,
$d_{i,ko}=2q_o(P_{ik}^{q}-\bar p_i)/K$.
Under this null, the order-averaged prompt laws are the same within each
item, while prompt--order interaction is allowed. The remaining two terms
retain the call variation within each cell.

\begin{theorem}[Finite-sample test and upper confidence bound]
\label{thm:nn-cantelli}
Let $0<\alpha<1$, $\lambda_\alpha=(1-\alpha)/\alpha$, and abbreviate
$a=a_{N,R}$, $b=b_{N,R}$. For any prespecified
$0\le\tau\le1-1/|\cZ|$, the test
\begin{equation}
 \text{reject }H_0:\theta\le\tau
 \quad\text{if}\quad
 \widehat\theta>\tau+\sqrt{\lambda_\alpha(a\tau+b)}
 \label{eq:nn-test}
\end{equation}
has type-I error at most $\alpha$. Define $x_+=\max\{\widehat\theta,0\}$
and
\begin{equation}
\begin{aligned}
\mathcal U_\alpha
=\min\Bigg\{1-\frac1{|\cZ|},\;&x_++\frac{\lambda_\alpha a}{2}\\
&+\sqrt{\lambda_\alpha(ax_++b)
              +\frac{\lambda_\alpha^2a^2}{4}}\Bigg\}.
\end{aligned}
\label{eq:nn-upper-bound}
\end{equation}
Then $\Prob(\theta\le\mathcal U_\alpha)\ge1-\alpha$ for every fixed array
of cell laws in this design.
\end{theorem}

The categorical covariance inequalities
$\|\Sigma\|_{\rm op}\le1/2$ and
$\operatorname{tr}(\Sigma_a\Sigma_b)\le1/4$ control the linear and
quadratic terms, respectively. The latter constant is attained by two
identical laws assigning probability $1/2$ to each of two categories,
regardless of the alphabet size. For $C=4$, the resulting bound on the quadratic terms is one third
smaller than the bound obtained from
$\operatorname{tr}(\Sigma_a\Sigma_b)\le(1-1/C)/2$.
Combining this sharp covariance bound with the inequalities for the
coefficients and repeat counts gives the variance envelope. Cantelli's
inequality converts it into a distribution-free bound at a prespecified
sample size.
Equation~\eqref{eq:nn-upper-bound} uses the positive part of the estimate
to construct the bound; we report the raw point estimate.
An upper bound below a prespecified tolerance certifies that the target is
below that tolerance at the stated confidence level. Supplementary Section~\ref{sec:nn-proofs}
also gives an explicit power bound.

\begin{theorem}[A near-zero testing lower bound]
\label{thm:nn-lower}
Consider the same balanced observation design with two equally weighted
orders, an even $K\ge2$, $R\ge2$, $N\ge1$, and $|\cZ|\ge2$.
Let $\mathcal F_0$ and $\mathcal F_\delta$ be the classes of fixed cell-law
arrays with $\theta=0$ and $\theta=\delta$, respectively. If
\begin{equation}
 0<\delta\le
 \delta_{N,K,R}:=\frac{\sqrt{\log(5/4)}}{4R\sqrt{NK}},
 \label{eq:nn-lower-separation}
\end{equation}
every test $\phi\in[0,1]$ based on all the calls satisfies
\begin{equation}
 \sup_{P\in\mathcal F_0}\E_P\phi
 +\sup_{P\in\mathcal F_\delta}\E_P(1-\phi)
 \ge\frac34.
 \label{eq:nn-lower-risk}
\end{equation}
Here $\delta$ is the prompt-dispersion target itself. For $K=6$,
$\delta_{N,6,R}=\sqrt{\log(5/4)}/(4R\sqrt{6N})$.
\end{theorem}

The construction permits different response contrasts across items and
prompt pairs. Together with the power bound for
Equation~\eqref{eq:nn-test}, it gives the near-zero separation order
$1/(R\sqrt{NK})$ for fixed error probabilities whose sum is below
$3/4$ in this balanced design.
\subsection{Fixed-Census Inference Near Zero: Proofs}
\label{sec:nn-proofs}

Throughout this section, the benchmark items and all prompt--order cell
laws are fixed. The randomness comes from the categorical calls, which
are mutually independent, with iid repetitions within each cell.

\subsubsection{Orthogonal expansion and exact variance}
\label{sec:nn-exact-proof}

We first prove a slightly more general identity allowing different fixed
repeat counts $R_a\ge2$ across cells of one item. Suppress the item index,
and let $Y_{ar}=e_{Z_{ar}}$, $\xi_{ar}=Y_{ar}-p_a$, and
\[
 \bar\xi_a=\frac1{R_a}\sum_r\xi_{ar},\qquad
 \E\xi_{ar}=0,\qquad
 \E\xi_{ar}\xi_{ar}^{\top}=\Sigma_a.
\]
For a symmetric deterministic matrix $W$, define
\[
 T_W=\sum_{a,b}W_{ab}p_a^{\top}p_b.
\]
Estimate a diagonal product with same-cell distinct-call agreement and an
off-diagonal product with the dot product of the two empirical
distributions. This gives the complete-pair estimator $\widehat T_W$.
Both estimates are unbiased for the corresponding population products.

For a diagonal product, direct expansion gives
\begin{equation}
 \widehat A_{aa}
 =\|p_a\|^2+2p_a^{\top}\bar\xi_a+D_a,
 \qquad
 D_a=\frac1{R_a(R_a-1)}
       \sum_{r\ne s}\xi_{ar}^{\top}\xi_{as}.
 \label{eq:nn-diagonal-expansion}
\end{equation}
For $a\ne b$,
\begin{equation}
 \widehat A_{ab}
 =p_a^{\top}p_b+p_b^{\top}\bar\xi_a
   +p_a^{\top}\bar\xi_b+\bar\xi_a^{\top}\bar\xi_b.
 \label{eq:nn-offdiagonal-expansion}
\end{equation}
Set $d_a=2\sum_bW_{ab}p_b$ and $Q_{ab}=\bar\xi_a^{\top}\bar\xi_b$.
Symmetry of $W$ yields
\begin{equation}
 \widehat T_W-T_W
 =\sum_a d_a^{\top}\bar\xi_a
  +\sum_aW_{aa}D_a
  +2\sum_{a<b}W_{ab}Q_{ab}.
 \label{eq:nn-orthogonal-expansion}
\end{equation}

All terms in this expansion have mean zero and are pairwise uncorrelated.
The linear summands are independent across cells. A linear term and a
$D_a$ term have zero covariance because every product contributing to the
covariance contains a centered observation occurring only once; conditioning
on the other observations gives zero. The same argument applies to a linear
term and $Q_{ab}$. Distinct $D_a$ terms are independent. A $D_a$ term and
any $Q_{bc}$ are uncorrelated, since either their cells are disjoint or the
other cell of $Q_{bc}$ has conditional mean zero. Finally, distinct
$Q_{ab}$ terms are independent if their cell pairs are disjoint; if their
pairs share one cell, conditioning on that cell leaves the two other
centered cell means independent with expectation zero.

The linear variance is
\[
 \Var(d_a^{\top}\bar\xi_a)=\frac{d_a^{\top}\Sigma_a d_a}{R_a}.
\]
In $\E D_a^2$, a term formed from two ordered repeat pairs contributes only
when the second pair is identical to the first or reverses its order. There are
$2R_a(R_a-1)$ such terms. Since two distinct repeats are independent,
\[
 \E(\xi_{ar}^{\top}\xi_{as})^2
 =\operatorname{tr}(\Sigma_a^2),
 \qquad r\ne s,
\]
and hence
\[
 \Var(D_a)=\frac{2\operatorname{tr}(\Sigma_a^2)}{R_a(R_a-1)}.
\]
Independence of the two cell means also gives
\[
 \Var(Q_{ab})
 =\frac{\operatorname{tr}(\Sigma_a\Sigma_b)}{R_aR_b}.
\]
Taking variances in Equation~\eqref{eq:nn-orthogonal-expansion} proves
\begin{equation}
\begin{aligned}
\Var(\widehat T_W)
={}&\sum_a\frac{d_a^{\top}\Sigma_a d_a}{R_a}
+2\sum_a\frac{W_{aa}^2\operatorname{tr}(\Sigma_a^2)}{R_a(R_a-1)}\\
&+4\sum_{a<b}\frac{W_{ab}^2
             \operatorname{tr}(\Sigma_a\Sigma_b)}{R_aR_b}.
\end{aligned}
\label{eq:nn-general-exact-variance}
\end{equation}
For the prompt matrix in Equation~\eqref{eq:nn-W}, $T_W=\theta_i$.
Setting all $R_a=R$ gives Equation~\eqref{eq:nn-exact-variance}.
Independence across fixed items then gives
$\Var(\widehat\theta)=N^{-2}\sum_iV_i$ and
$\E\widehat\theta=\theta$.

\subsubsection{A uniform variance envelope}
\label{sec:nn-envelope-proof}

For any categorical probability vector $p$,
$\Sigma=\operatorname{diag}(p)-pp^{\top}$ is positive semidefinite,
$\operatorname{tr}(\Sigma)=1-\|p\|^2\le1-1/C$, where $C=|\cZ|$, and
$\|\Sigma\|_{\mathrm{op}}\le1/2$.
For the last bound, take any vector $v$ and note that
$v^{\top}\Sigma v=\Var(v_Z)$. A real random variable in $[m,M]$ has
variance at most $(M-m)^2/4$: indeed,
$\E[(X-m)(M-X)]\ge0$ implies
$\Var(X)\le(M-\E X)(\E X-m)\le(M-m)^2/4$.
Apply this with $m=\min_c v_c$ and $M=\max_c v_c$, and use
$(M-m)^2\le2\|v\|^2$. To control the quadratic terms, write
$s_2=\sum_c p_c^2$ and $s_3=\sum_c p_c^3$. Cauchy--Schwarz gives
$s_2^2\le(\sum_c p_c^3)(\sum_c p_c)=s_3$, and hence
\[
\operatorname{tr}(\Sigma^2)=s_2-2s_3+s_2^2
\le s_2(1-s_2)\le\frac14.
\]
For any two categorical covariance matrices, Frobenius
Cauchy--Schwarz therefore implies
\begin{equation}
 \operatorname{tr}(\Sigma_a^2)\le\frac14,
 \qquad
 0\le\operatorname{tr}(\Sigma_a\Sigma_b)\le\frac14.
 \label{eq:nn-covariance-bounds}
\end{equation}
Both upper bounds are attained when the two laws equal
$(1/2,1/2,0,\ldots,0)$.

For the prompt target, let
$\Delta_{ik}=P_{ik}^{q}-\bar p_i$. Summing the coefficients in
Equation~\eqref{eq:nn-W} gives
\[
 d_{i,ko}=\frac{2q_o}{K}\Delta_{ik},\qquad
 \theta_i=\frac1K\sum_k\|\Delta_{ik}\|^2.
\]
Therefore the first variance term obeys
\begin{equation}
 \frac1R\sum_{k,o}d_{i,ko}^{\top}\Sigma_{i,ko}d_{i,ko}
 \le\frac{2s_q}{KR}\theta_i.
 \label{eq:nn-linear-bound}
\end{equation}
Write $H_K=I_K-\mathbf 1\mathbf 1^{\top}/K$. The coefficient matrix is
$W=(H_K/K)\otimes(qq^{\top})$. Since $H_K$ is a rank-$K-1$ orthogonal
projection,
\begin{equation}
 \|W\|_F^2=\frac{K-1}{K^2}\left(\sum_o q_o^2\right)^2
          =\frac{K-1}{K^2}s_q^2.
 \label{eq:nn-W-frobenius}
\end{equation}
Using Equation~\eqref{eq:nn-covariance-bounds} and
$R^{-2}\le[R(R-1)]^{-1}$, the two quadratic terms in
Equation~\eqref{eq:nn-exact-variance} are at most
\begin{equation}
\begin{aligned}
 &\frac{1}{2R(R-1)}\sum_aW_{aa}^2
   +\frac{1}{R^2}\sum_{a<b}W_{ab}^2\\
 &\hspace{1cm}\le\frac{\|W\|_F^2}{2R(R-1)}
 =\frac{(K-1)s_q^2}{2K^2R(R-1)}.
\end{aligned}
\label{eq:nn-quadratic-bound}
\end{equation}
Equations~\eqref{eq:nn-linear-bound}--\eqref{eq:nn-quadratic-bound},
summed over items and divided by $N^2$, prove
Equation~\eqref{eq:nn-variance-envelope} and complete the proof of
Theorem~\ref{thm:nn-variance}.

Because every $\theta_i$ is nonnegative, $\theta=0$ implies
$\Delta_{ik}=0$ for all $i,k$. The individual $P_{iko}$ may vary across
prompts at each order, so the null bound covers arrays with nonzero
prompt--order interaction.
For two orders, $s_q=1/2$ gives the constants stated for $K=6$.

\subsubsection{Cantelli test, upper bound, and power}
\label{sec:nn-cantelli-proof}

For a centered real random variable $Y$ with variance $v$, Cantelli's
inequality states
\begin{equation}
 \Prob(Y\ge t)\le\frac{v}{v+t^2},\qquad t>0.
 \label{eq:nn-cantelli-inequality}
\end{equation}
Applying the inequality to $-Y$ gives the same bound for the lower tail.

Let $v_\tau=a\tau+b$ and
$h_\tau=\sqrt{\lambda_\alpha v_\tau}$.
Under $\theta\le\tau$, rejection in Equation~\eqref{eq:nn-test} implies
$\widehat\theta-\theta>h_\tau$, and
$\Var(\widehat\theta)\le a\theta+b\le v_\tau$.
The function $v/(v+h_\tau^2)$ increases in $v$, so Cantelli gives
\[
 \Prob_\theta(\text{reject})
 \le\frac{v_\tau}{v_\tau+h_\tau^2}
 =\alpha.
\]
This proves the test guarantee for the full composite null.

To prove the upper confidence bound, let $x=\widehat\theta$,
$y=x_+$, and define the untruncated bound
\[
 u(y)=y+\frac{\lambda_\alpha a}{2}
       +\sqrt{\lambda_\alpha(ay+b)
                       +\frac{\lambda_\alpha^2a^2}{4}}.
\]
It is the larger root of
$(u-y)^2=\lambda_\alpha(au+b)$ and exceeds $y$. If
$\theta>u(y)$, then
\[
 \theta-x\ge\theta-y>
       \sqrt{\lambda_\alpha(a\theta+b)}.
\]
The strict inequality follows by considering the upward-opening quadratic
$(t-y)^2-\lambda_\alpha(at+b)$ beyond its larger root.
Cantelli's lower-tail bound and
$\Var(\widehat\theta)\le a\theta+b$ imply
$\Prob\{\theta>u(x_+)\}\le\alpha$.
Finally,
\[
 0\le\theta_i
 =\frac1K\sum_k\|P_{ik}^{q}\|^2-\|\bar p_i\|^2
 \le1-\frac1{|\cZ|},
\]
so intersecting the bound with this known maximum cannot remove a valid
parameter value. This proves Equation~\eqref{eq:nn-upper-bound} and
Theorem~\ref{thm:nn-cantelli}. In particular, if a tolerance $\tau$ is
prespecified and the true $\theta>\tau$, the event
$\mathcal U_\alpha\le\tau$ has probability at most $\alpha$.

\begin{proposition}[Explicit sufficient separation for power]
\label{prop:nn-power}
For the test in Equation~\eqref{eq:nn-test}, fix a desired type-II error
$0<\beta<1$, set $\lambda_\beta=(1-\beta)/\beta$, and retain
$v_\tau=a\tau+b$ and $h_\tau=\sqrt{\lambda_\alpha v_\tau}$.
Define
\begin{equation}
\begin{aligned}
 D_{\alpha,\beta}(\tau)
 =h_\tau+\frac{\lambda_\beta a}{2}
 +\sqrt{\lambda_\beta v_\tau
       +\lambda_\beta a h_\tau
       +\frac{\lambda_\beta^2a^2}{4}}.
\end{aligned}
\label{eq:nn-power-separation}
\end{equation}
Every alternative with
$\theta\ge\tau+D_{\alpha,\beta}(\tau)$ has rejection probability at
least $1-\beta$.
\end{proposition}

\paragraph{Proof.}
Write $\theta=\tau+\delta$. Failure to reject implies
$\widehat\theta-\theta\le-(\delta-h_\tau)$. The number
$D_{\alpha,\beta}(\tau)$ is the larger root of
\[
 (\delta-h_\tau)^2=\lambda_\beta(v_\tau+a\delta),
\]
and exceeds $h_\tau$. For $\delta$ at least this root, the lower-tail
Cantelli bound is at most
\[
 \frac{v_\tau+a\delta}
      {v_\tau+a\delta+(\delta-h_\tau)^2}
 \le\frac1{1+\lambda_\beta}=\beta.
\]
This proves the proposition. If the sufficient separation exceeds the
parameter range, no alternative satisfies the power condition.

For two equally weighted orders, let $L=NK$. Since $K\ge2$ and $R\ge2$,
\[
 a=\frac1{LR},\qquad
 b=\frac{K-1}{8NK^2R(R-1)}\le\frac1{4LR^2}.
\]
At $\tau=0$, write $h_0=\sqrt{\lambda_\alpha b}$ and use
$\sqrt{u+v+w}\le\sqrt u+\sqrt v+\sqrt w$ and
$\sqrt{\lambda_\beta a h_0}\le(\lambda_\beta a+h_0)/2$ in
Equation~\eqref{eq:nn-power-separation}. This gives
\begin{equation}
\begin{aligned}
 D_{\alpha,\beta}(0)
 &\le\left(\frac32\sqrt{\lambda_\alpha}
              +\sqrt{\lambda_\beta}\right)\sqrt b
        +\frac32\lambda_\beta a\\
 &\le\frac{C_{\alpha,\beta}}{R\sqrt{NK}},\qquad
 C_{\alpha,\beta}
 =\frac{\tfrac32\sqrt{\lambda_\alpha}+\sqrt{\lambda_\beta}}{2}
   +\frac32\lambda_\beta.
\end{aligned}
\label{eq:nn-upper-rate}
\end{equation}
The last step uses $a=1/(LR)\le1/(R\sqrt L)$. This explicit bound
establishes the stated separation order. The exact expression
\eqref{eq:nn-power-separation} is smaller and is preferable for numerical
planning.

\subsubsection{Heterogeneous-item testing lower bound}
\label{sec:nn-lower-proof}

We prove Theorem~\ref{thm:nn-lower} directly for every even $K\ge2$.
Use two categories, embedded in the full alphabet by assigning zero
probability to the other categories. Let $P_0$ be the fixed-law array
with success probability $1/2$ in every cell. Pair the prompts as
$(1,2),(3,4),\ldots,(K-1,K)$.

For every fixed sign array
$\boldsymbol\sigma=(\sigma_{ij})\in\{-1,1\}^{N\times K/2}$,
define an alternative $P_{\boldsymbol\sigma,\varepsilon}$ by
\begin{align}
 \Prob(Z=1\mid i,2j-1,o)&=\tfrac12+\sigma_{ij}\varepsilon,\\
 \Prob(Z=1\mid i,2j,o)&=\tfrac12-\sigma_{ij}\varepsilon,
 \qquad o\in\{AB,BA\}.
 \label{eq:nn-lower-alternatives}
\end{align}
For each fixed sign array, observations are independent calls from
fixed cell laws. The two orders have identical
laws, the mean prompt law is $(1/2,1/2)$, and each prompt contributes
$2\varepsilon^2$ to its squared distance from that mean. Thus
\begin{equation}
 \theta_i=\theta=2\varepsilon^2=\delta
 \label{eq:nn-delta-not-epsilon}
\end{equation}
for every alternative. Each item has $K/2$ independent prompt-pair
signs; opposite signs within each pair keep the mean prompt law fixed exactly.

For the lower-bound argument only, put the uniform prior on all sign
arrays, so the $NK/2$ signs are independent Rademacher variables.
Denote the resulting mixture of observation laws by
$\overline P_\varepsilon$. The prior used in this proof averages observation laws over fixed parameter arrays.

Consider one item and one prompt pair. It contributes $4R$ Bernoulli
observations: two prompts, two orders, and $R$ repetitions. Under $P_0$,
write each Bernoulli observation as a symmetric variable $Y\in\{-1,1\}$.
If the prompt's fixed orientation within its pair is
$t\in\{-1,1\}$ and the pair sign is $\sigma$, the one-observation
likelihood ratio is $1+2\sigma t\varepsilon Y$.
For two candidate signs $\sigma,\sigma'$, independence of observations
therefore gives
\[
 \E_0[L_\sigma L_{\sigma'}]
 =(1+4\sigma\sigma'\varepsilon^2)^{4R}.
\]
Averaging over the two independent signs in the mixture's second moment
yields the exact identity
\begin{equation}
 1+\chi^2(\overline P_{\varepsilon,\mathrm{pair}},P_{0,\mathrm{pair}})
 =\frac{(1+4\varepsilon^2)^{4R}
              +(1-4\varepsilon^2)^{4R}}2.
 \label{eq:nn-pair-chisquare}
\end{equation}
For $0\le4\varepsilon^2<1$, using $1+x\le e^x$ and
$1-x\le e^{-x}$ bounds the right side by
$\cosh(16R\varepsilon^2)$. The inequality
$\cosh u\le e^{u^2/2}$ follows from
$\log\cosh u=\int_0^u\tanh v\,\mathrm dv\le u^2/2$ for $u\ge0$,
and symmetry for $u<0$. Thus the one-pair second moment is at most
$\exp(128R^2\varepsilon^4)$.

The $NK/2$ item--pair groups are independent under both the prior and
the null observation law, so their likelihood-ratio second moments multiply.
Substituting $\varepsilon^2=\delta/2$ from
Equation~\eqref{eq:nn-delta-not-epsilon} gives
\begin{equation}
 \chi^2(\overline P_\varepsilon,P_0)
 \le\exp(64NKR^2\varepsilon^4)-1
 =\exp(16NKR^2\delta^2)-1.
 \label{eq:nn-product-chisquare}
\end{equation}
Under Equation~\eqref{eq:nn-lower-separation}, the final expression is
at most $1/4$. The proposed $\delta$ also satisfies $\delta<1/2$ for
$N\ge1$, $K\ge2$, and $R\ge2$, so
$4\varepsilon^2=2\delta<1$ and all Bernoulli probabilities used above
are strictly between zero and one.

If $L=\mathrm d\overline P_\varepsilon/\mathrm dP_0$, Cauchy--Schwarz
gives
\[
 \|\overline P_\varepsilon-P_0\|_{\mathrm{TV}}
 =\tfrac12\E_0|L-1|
 \le\tfrac12\sqrt{\E_0(L-1)^2}\le\tfrac14.
\]
For a possibly randomized test $0\le\phi\le1$,
\begin{align*}
 \E_0\phi+\E_{\overline P_\varepsilon}(1-\phi)
 &=1-\bigl(\E_{\overline P_\varepsilon}\phi-\E_0\phi\bigr)\\
 &\ge1-\|\overline P_\varepsilon-P_0\|_{\mathrm{TV}}
 \ge\tfrac34.
\end{align*}
The supremum over $\mathcal F_0$ is at least the error at $P_0$, and the
supremum over $\mathcal F_\delta$ is at least the prior-average type-II
error, because every sign-array alternative has target $\delta$.
This proves Equation~\eqref{eq:nn-lower-risk}.

The lower-bound subclass has zero order effect and interaction. A
lower bound on this subclass applies to the full fixed-cell class; the
upper bound also covers arrays with zero prompt dispersion and nonzero interaction.

For the balanced fixed-repeat design with even $K$, the upper and lower
bounds therefore match at order $1/(R\sqrt{NK})$ for fixed error
probabilities whose sum is below $3/4$.

\section{Acquisition, Uncertainty, and Reproducibility}
\label{sec:experimental-materials}\label{sec:reproduction-index}\label{sec:inference-navigation}
\subsection{Observation units and study design}
For measurement, we use $N$ items, a fixed census of $K$ prompts, two candidate orders, and $R$ repeated calls in each cell. We collect separate calls to form $J$ independent four-vote deployments per item and condition. D selects two distinct prompts and queries both orders for each; E selects four distinct prompts and assigns two to each order. Measurement and deployment use separate sets of calls. Results give equal weight to each item, except for conditional rates with explicitly stated denominators.

\begin{table}[!htbp]\centering\small
\caption{Distinct study populations and acquired observations. Each study retains its own benchmark population; shared calls are counted once in the controlled-formatting total.}
\label{tab:reproduction-main-map}\label{tab:notation-navigation}
\setlength{\tabcolsep}{3pt}\begin{tabular*}{\linewidth}{@{\extracolsep{\fill}}>{\raggedright\arraybackslash}p{.22\linewidth}>{\raggedright\arraybackslash}p{.44\linewidth}>{\raggedright\arraybackslash}p{.27\linewidth}@{}}\toprule
Study & Measurement / independent deployment & Retained calls\\\midrule
Controlled formatting & 50 matched items; $K=6$; null/evaluation $R=8$, reference $R=16$; 369 further items at $R=4$ & 44,112 unique calls across Qwen and the four-configuration panel\\
Semantic comparison & Common 80 items, $K=24,R=2$; open models additionally use 120 test items and unseen-prompt deployments & 162,816 calls including development/canaries\\
Matched temperatures & Same 40 items, $K=48,R=4$, $T\in\{1,.2,0\}$; $J=8$ D groups & 49,920 calls per model\\
Qwen policies & 152 test items, $K=6,R=16$; $J=64$ groups per D/E design and policy & 428,032 test calls\\
Gemma reasoning & 32 RewardBench~2 pairs, $K=48,R=2$; $J=8$ D groups in each of four arms & 28,672 formal judgments\\
Independent JudgeBench & 100 development, 200 test items; $K=6,R=8$ test measurement; $J=16$ per D/E & 49,600 formal calls\\\bottomrule
\end{tabular*}\end{table}

\paragraph{Uncertainty targets.}
Theorems in Section~\ref{sec:nn-main} and the temperature bounds in Section~\ref{sec:temperature-full-repeat-inference} concern fresh independent calls conditional on fixed items and prompt assignments. Whole-item bootstrap intervals describe sensitivity to which questions appear in the sample: all prompts, orders, repeats, conditions and independent groups for a selected item are resampled together. The families of temperature decision comparisons condition on the assigned prompt pairs. Deployment trace variance measures variation across deployments. Block-$t$ and bootstrap intervals use approximate calibration; the finite-call bounds have distribution-free guarantees.

\subsection{Rendering, outcome parsing, and model configurations}
\label{sec:s-prompts}\label{sec:s-configurations}\label{sec:extension-configurations}
All pairwise calls preserve the candidate texts. After reversing their order, we map verdicts back to the original candidate identities. The common parser reads A, B or TIE from the first nonempty final line; other completed outputs map to $\bot$. Completed invalid or truncated outputs are retained. Only infrastructure failures without a returned completion are retried under the original conditions. Conditioning within each cell on valid outputs defines a different response law; pooling surviving calls additionally changes cell weights. Requiring at least two valid calls per cell can change which items enter the panel. The primary analysis uses the full four-category law.

The six formatting wrappers use the same judging criterion and required output format while changing headings or delimiters. The shared criterion judges instruction adherence and correctness, forbids rewarding length, style or confidence on their own, and allows TIE. Exact strings, separators and candidate fields are given in Section~\ref{sec:recovery-prompt-strings}. The semantic bank has 48 variants, 12 each emphasizing lexical, syntactic, clause-order or discourse edits; six formatting controls are retained. A set of 216 automated checks, including controls, documents their intended equivalence. The four procedural policy banks apply the criterion through distinct judging procedures.

\begin{table*}[t]\setlength{\tabcolsep}{2.5pt}\renewcommand{\arraystretch}{1.0}
\centering
\footnotesize
\setlength{\tabcolsep}{2.5pt}
\caption{Judge configurations in the matched comparison. Each configuration
includes its model, renderer, decoder, stopping rule, and categorical parser.
All item--prompt--order cells are shared across configurations.}
\label{tab:judge-configurations}
\begin{tabular*}{\linewidth}{@{\extracolsep{\fill}}>{\raggedright\arraybackslash}p{.19\textwidth}>{\raggedright\arraybackslash}p{.26\textwidth}>{\raggedright\arraybackslash}p{.18\textwidth}>{\raggedright\arraybackslash}p{.25\textwidth}@{}}
\toprule
Configuration & Model & Renderer & Decoding and terminal rule \\
\midrule
Qwen3.6-27B &
\texttt{Qwen/Qwen3.6-27B} &
Qwen chat renderer &
$T=1$, top-$p=1$, 8-token cap; categorical parser \\
GPT-OSS-120B &
\texttt{openai/gpt-oss-120b} &
Harmony renderer &
$T=1$, top-$p=1$, 512-token cap; final channel only \\
Gemma-3-27B-IT &
\texttt{google/\allowbreak gemma-3-27b-it} &
official chat template &
$T=1$, top-$p=1$, top-$k=-1$, 32-token cap; newline stop \\
Claude Haiku 4.5 &
\texttt{claude-haiku-4-5-}\newline
\texttt{20251001} &
Messages request &
$T=1$; top-$p$ omitted (provider default); no provider seed; 512-token cap;
provider \texttt{end\_turn} \\
\bottomrule
\end{tabular*}
\end{table*}
\begin{table}[!htbp]\setlength{\tabcolsep}{2.5pt}\renewcommand{\arraystretch}{1.0}
\centering\small
\setlength{\tabcolsep}{2.5pt}
\caption{Fixed configurations in the five-model extension. Output caps
specify token limits; computational budgets vary across models.
Hosted GPT sampling uses its supported reasoning-effort setting;
requests omit temperature and top-$p$.}
\label{tab:extension-configurations}
\begin{tabular*}{\linewidth}{@{\extracolsep{\fill}}>{\raggedright\arraybackslash}p{.17\linewidth}>{\raggedright\arraybackslash}p{.40\linewidth}r>{\raggedright\arraybackslash}p{.20\linewidth}@{}}
\toprule
Configuration & Recorded model ID & Cap & Generation settings\\
\midrule
Qwen3.6 & \texttt{Qwen/Qwen3.6-27B} & 512 & Thinking off; $T=1$\\
Gemma4 & \texttt{google/gemma-4-31B-it} & 512 & Thinking off; $T=1$\\
GPT-OSS & \texttt{openai/gpt-oss-120b} & 4096 & Low effort; $T=1$\\
GPT Luna & \texttt{gpt-5.6-luna} & 4096 & Low effort\\
Haiku4.5 & \texttt{claude-haiku-4-5-20251001} & 512 & $T=1$\\
\bottomrule
\end{tabular*}
\end{table}

The original four-configuration panel uses the same 50 LLMBar items, six wrappers and $R=4$ (2,400 calls/configuration); Gemma-3 and the later Gemma4 studies are distinct configurations. Generation settings are recorded in Table~\ref{tab:judge-configurations}. The analysis assumes independent calls under stationary cell laws across runtime periods. Each configuration is defined by a fixed model ID and fixed decoding and rendering settings. The cross-model comparison therefore includes both model weights and runtime settings.

\subsection{Data and analysis materials}
The accompanying data package provides recorded categorical judgments, prompt and item metadata, gold labels, model settings, independent vote groups, and analysis code for the main numerical results. Fixed probability arrays and trial summaries support the known-law simulations. The analyses retain all conditions, invalid outputs and negative component estimates.

\section{Measurement Validation and Cross-Model Evidence}
\label{sec:extended-measurement}\label{sec:live-validation-details}
\subsection{Identical-input null and independent reference}
The controlled study uses 50 LLMBar items, ten per stratum. The Qwen corpus contains 4,800 null-control calls with byte-identical inputs, 4,800 matched $R=8$ evaluation calls, a disjoint 9,600-call $R=16$ reference, and 17,712 calls on the remaining 369 items at $R=4$. A matched four-configuration panel reuses 2,400 Qwen calls and adds 7,200 calls, giving 44,112 unique outcomes. Within each item--order--repeat unit, the six null labels share exactly the same model-visible input but have distinct generation seeds. The evaluation and reference samples share items, prompts, orders, checkpoint and parser but use disjoint calls and seeds. Analyses at lower repeat depths use fixed prefixes of the recorded calls.

\begin{table*}[!htbp]\setlength{\tabcolsep}{2.5pt}\renewcommand{\arraystretch}{1.0}
\centering
\small
\setlength{\tabcolsep}{2.5pt}
\caption{Prompt-instability estimates in the live experiments. Entries are means across items, with 95\% percentile bands for benchmark composition from 20,000
stratified whole-item bootstrap draws. Each resample retains all six hidden
labels or templates and both answer orders. The hidden-factor null has an
exactly zero target because its labels induce byte-identical inputs.
The other rows concern the fixed six-prompt census.}

\label{tab:real_prompt_estimates}
\begin{tabular*}{\linewidth}{@{\extracolsep{\fill}}llcc@{}}
\toprule
Setting & $R$ & Plug-in & Debiased all-$R$ \\
\midrule
Byte-identical hidden-factor null & 2 & 0.0406 [0.0313, 0.0505] & -0.0053 [-0.0139, 0.0038] \\
 & 4 & 0.0206 [0.0153, 0.0263] & -0.0008 [-0.0050, 0.0033] \\
 & 8 & 0.0126 [0.0089, 0.0168] & 0.0024 [-0.0002, 0.0054] \\
Matched evaluation & 2 & 0.0392 [0.0289, 0.0501] & 0.0007 [-0.0056, 0.0072] \\
 & 4 & 0.0251 [0.0191, 0.0313] & 0.0052 [0.0011, 0.0096] \\
 & 8 & 0.0143 [0.0107, 0.0179] & 0.0039 [0.0013, 0.0066] \\
Broader item sample & 4 & 0.0245 [0.0223, 0.0268] & 0.0026 [0.0010, 0.0044] \\
\bottomrule
\end{tabular*}
\end{table*}

\begin{table}[!htbp]
\centering\small
\setlength{\tabcolsep}{3pt}\renewcommand{\arraystretch}{1.0}
\caption{Itemwise MAE against the independent $R=16$ Qwen reference on the same 50 items, six prompts and two orders. The evaluation and reference samples use disjoint calls; corrected estimates use all distinct-call pairs. Reduction is the paired difference $\mathrm{MAE}_{\rm plug}-\mathrm{MAE}_{\rm corr}$, with a pointwise 95\% composition interval from 20,000 stratified whole-item resamples.}
\label{tab:matched-reference-precision}
\begin{tabular*}{\linewidth}{@{\extracolsep{\fill}}rrrr@{}}
\toprule
Evaluation $R$ & Plug-in MAE & Corrected MAE & Reduction [95\% interval]\\
\midrule
2 & 0.0365 & 0.0160 & 0.0205 [0.0131, 0.0288]\\
4 & 0.0214 & 0.0093 & 0.0120 [0.0077, 0.0165]\\
8 & 0.0112 & 0.0076 & 0.0036 [0.0009, 0.0064]\\
\bottomrule
\end{tabular*}
\end{table}

All bands in this subsection use 20,000 stratified whole-item resamples to describe variation in item composition. Compared with the independently acquired unbiased $R=16$ reference, complete-pair correction lowers mean absolute discrepancy at $R=2,4,8$. After acquisition, contiguous and interleaved splits of the reference into two independent $R=8$ halves give MAEs of .00758 and .00819 (correlations .150 and .197). For evaluation estimates $A_i,C_i$ and independent unbiased reference $T_i$, $\E[(A_i-T_i)^2-(C_i-T_i)^2]=\E(A_i-\theta_i)^2-\E(C_i-\theta_i)^2$. Observed squared-loss improvements are .002311, .000716 and .000179. The choice to report complete all-$R$ estimates was made after acquisition. Consecutive-two-repeat estimates give an estimated reference-MAE reduction of $-.0005$ at $R=8$, with composition band [$-.0041$,.0029]; complete pairs use additional cross-repeat information.

The independent reference gives mean components of .1957 for call variation, .0040 for prompts, .0511 for order and .0040 for interaction. The disjoint 369-item extension also has small corrected prompt dispersion (.0026, composition band [.0010,.0044]), with component means of .2098 for call variation, .0521 for order and .0028 for interaction. This extends the profile within LLMBar under the same wrappers and configuration. The independent $R=16$ reference covers the initial 50-item panel.

\subsection{Matched model configurations and invalid outputs}
\label{sec:semantic-five-model}
\begin{table*}[t]\setlength{\tabcolsep}{2.5pt}\renewcommand{\arraystretch}{1.0}
\centering
\small
\setlength{\tabcolsep}{2.5pt}
\caption{Raw, unclipped complete-pair decomposition for the matched \(R=4\)
panel. Values are macroaverages over the same 50 items. Every row obeys the
sample decomposition identity up to display rounding.}
\label{tab:s-cross-model-components}
\begin{tabular*}{\linewidth}{@{\extracolsep{\fill}}lrrrrr@{}}
\toprule
Judge configuration & Call & Prompt & Answer order & Interaction & Total \\
\midrule
Qwen3.6-27B & $.19111$ & $.00521$ & $.04694$ & $.00719$ & $.25046$ \\
GPT-OSS-120B & $.08028$ & $-.00123$ & $.00935$ & $.00228$ & $.09068$ \\
Gemma-3-27B-IT & $.01500$ & $.01774$ & $.11017$ & $.01733$ & $.16024$ \\
Claude Haiku 4.5 & $.05917$ & $.00986$ & $.08826$ & $.00799$ & $.16528$ \\
\bottomrule
\end{tabular*}
\end{table*}

The separate semantic comparison uses the same 80 test items (40/dataset), 24 prompts, both orders and $R=2$ across five configurations. Each open model's larger acquisition includes 120 test items, 40 development items, four canaries and a four-item identical-input diagnostic, with 48 semantic prompts plus six controls on the test sample. Closed models use the nested common panel. The complete extension has 162,816 outcomes. Unseen-prompt deployments for the open models are analyzed separately in Section~\ref{sec:sampling-designs}.
\begin{table}[!htbp]\setlength{\tabcolsep}{2.5pt}\renewcommand{\arraystretch}{1.0}
\centering\small
\setlength{\tabcolsep}{2.5pt}
\caption{Common semantic-prompt panel. P is the plug-in prompt component; $\widehat U_p$, $\widehat U_o$, and $\widehat U_{po}$ are corrected prompt, order, and interaction components. Accuracy, $\bot$, and truncation are percentages of all completed calls. Values are configuration-specific point estimates. The Haiku JudgeBench condition is reported separately as an output-protocol case: invalid outputs dominate its response profile. The remaining rows support comparisons of judge prompt sensitivity.}
\label{tab:semantic-components}
\begin{tabular*}{\linewidth}{@{\extracolsep{\fill}}llrrrrrrr@{}}
\toprule
Model & Dataset & P & $\widehat U_p$ & $\widehat U_o$ & $\widehat U_{po}$ & Acc. & $\bot$ & Trunc.\\
\midrule
Qwen3.6 & LLMBar & 0.0718 & 0.0065 & 0.0224 & 0.0116 & 72.27 & 0.36 & 0.05\\
Qwen3.6 & JudgeBench & 0.0965 & 0.0118 & 0.1102 & 0.0108 & 53.12 & 11.02 & 8.70\\
Gemma4 & LLMBar & 0.0190 & 0.0123 & 0.0410 & 0.0143 & 82.60 & 0.03 & 0.21\\
Gemma4 & JudgeBench & 0.0272 & 0.0136 & 0.1110 & 0.0140 & 64.11 & 1.04 & 1.72\\
GPT-OSS & LLMBar & 0.0336 & 0.0090 & 0.0271 & 0.0027 & 79.35 & 0.00 & 0.00\\
GPT-OSS & JudgeBench & 0.0587 & 0.0123 & 0.0070 & 0.0024 & 63.46 & 0.05 & 0.03\\
GPT Luna & LLMBar & 0.0338 & 0.0092 & 0.0512 & 0.0118 & 75.03 & 0.00 & 0.00\\
GPT Luna & JudgeBench & 0.0454 & 0.0106 & 0.0174 & 0.0132 & 71.46 & 0.00 & 0.00\\
Haiku4.5 & LLMBar & 0.0265 & 0.0079 & 0.0652 & 0.0064 & 69.71 & 0.94 & 0.00\\
\midrule
\multicolumn{9}{l}{\textit{Output-protocol case: 81.74\% invalid completed outputs}}\\
Haiku4.5 & JudgeBench & 0.0549 & 0.0187 & 0.0304 & 0.0092 & 11.15 & 81.74 & 56.35\\
\bottomrule
\end{tabular*}
\end{table}

Claude's order-minus-call point difference in the matched formatting panel is .02910, with a composition band of [$-.01493$,.07514]. The matched formatting panel contains 10 Qwen and 64 GPT-OSS invalid outputs and none for Gemma-3 or Haiku. Recoding $\bot$ as TIE changes the panel's prompt means by at most .00049 at $R=2$ and .00036 at $R=4$. Haiku's semantic JudgeBench condition has 81.74\% invalid completions and 56.35\% cap stops. Its response profile therefore primarily measures output usability. Retaining $\bot$ keeps these configuration-dependent outcomes in the measured law.

\paragraph{Dependence checks.}\label{sec:s-diagnostics}
We assess monotone drift in the seven original acquisition streams using agreement slopes against rank gap and elapsed time. Whole-item resampling preserves the dependence among agreement pairs that share calls. All outputs remain included; complete-case filtering at $R=2$ would retain 42/50 null, 45/50 matched and 329/369 broader items.

\subsection{Known-law validation}
\label{sec:inference-validation-protocol}\label{sec:inference-validation}\label{sec:new-simulation-results}
The basic bias experiment crosses $K\in\{5,6\}$, $R\in\{2,4,8\}$, two balanced orders and prompt targets zero or .03, with 200,000 independent items per setting. The base law is $(.4,.3,.2,.1)$; the alternative moves mass along $(1,-1,0,0)/\sqrt2$ with a centered, unit-variance census score. At $K=6$, the null plug-in biases of .1459, .0730 and .0365 match the $1/R$ identity. Corrected bias lies within 1.44 Monte Carlo standard errors of zero across all 12 settings. Negative null estimates (52.7--56.1\% itemwise) are retained.

A separate validation uses 170 fixed settings with 5,000 independent datasets each: 13 analytical families at $N=32$ and 152, plus eight data-shaped arrays at $N=152$, crossed with $R=2,4,8,16,32$, $K=6$ and two balanced orders. The analytical grid includes homogeneous, heterogeneous and interaction-only prompt nulls, a stability-tolerance boundary, and signed D/E contrasts. Data-shaped laws use categorical probabilities $(n_c+.5)/18$ from 16-repeat measurements. Treating these arrays as fixed synthetic laws gives the simulations exact known targets. Stored arrays, seeds and trial summaries permit exact reconstruction of each scenario.

The universal and observed-variation design intervals cover their targets in all 5,000 datasets in every setting. The Wilson 95\% Monte Carlo lower bound of 99.923\% quantifies uncertainty in each observed coverage rate. The approximate independent-block $t$ comparison has coverage of 94.34--95.66\% on this grid. Section~\ref{sec:temperature-panel-calibration} separately evaluates the two-panel temperature approximation.
At the interaction-only prompt null with $U_{\rm int}=.08$, $N=152$ and $R=8$, the fraction of datasets that certify the fixed tolerance $\theta<.01$ is 0\%, 79.06\% and 97.10\% for the original, trace-refined and sharp categorical covariance bounds. No tested boundary case $\theta=.01$ falsely certifies the tolerance (per-setting Wilson upper bound .077\%). In the lower-bound family, four designs $(N,K,R)=(16,4,4),(64,4,4),(16,8,16),(64,8,16)$ use $\delta=c/(R\sqrt{NK})$ and 5,000 datasets/setting. Power is .04--.12\% at $c=.5$, 56.92--65.16\% at $c=2$, and 100\% at $c=8$ for the trace-envelope test. These power simulations show how the proven separation rate translates into finite-call resolution.

\subsection{Explicit laws and matched estimator comparisons}
\label{sec:simulation-evidence}
The 170-setting study tests inference across heterogeneous fixed laws. A separate ten-law study isolates estimator efficiency and equal-budget design selection. Both studies generate independent categorical calls with known targets. Section~\ref{sec:temperature-panel-calibration} reports the separate temperature-panel calibration.

\paragraph{The 170-setting probability arrays.}
For $k=1,\ldots,6$, define $h=(-1,-1,-1,1,1,1)$, $g=\sqrt{3/2}(-1,0,1,-1,0,1)$ and $u=(1,-1,0,0)/\sqrt2$. The prompt directions are centered, orthogonal and have unit mean square. For item $i=0,\ldots,N-1$, let $b_i=(.44+t_i,.36-t_i,.15,.05)$, where $t_i=.04\sin(1.61803398875\,i)$. Nine families use
\begin{equation}
\begin{aligned}
P_{ik,AB}&=b_i+\sqrt S\,h_ku,\\
P_{ik,BA}&=b_i+\sqrt S(\rho h_k+\sqrt{1-\rho^2}g_k)u,
\end{aligned}
\qquad S=.04,\quad \rho=-\frac{\overline\Delta}{\gamma S},\quad \gamma=\frac6{20},
\label{eq:simulation-regular-laws}
\end{equation}
with $\overline\Delta\in\{-.012,-.006,-.003,-.001,0,.001,.003,.006,.012\}$. Hence
\[
\theta=\frac{S-\overline\Delta/\gamma}{2},\qquad
\overline U_{\rm int}=\frac{S+\overline\Delta/\gamma}{2},
\]
so $\overline\Delta=.006$ puts prompt dispersion at the tolerance boundary $\theta=.01$. Four further families have $\theta=0$: the constant binary law $(.5,.5,0,0)$; laws $b_i\pm(.1+.03\cos i)u$ constant across prompts; and interaction-only laws $b_i\pm\sqrt S h_ku$ for $S=.005,.08$, with signs denoting AB/BA. Each of these 13 families uses $N=32$ and 152.

The other eight arrays use the four output-cap and four policy conditions, each with $N=152$. A cell's 16 observed category counts define $P_c=(n_c+.5)/18$. We treat these probabilities as fixed synthetic laws, retaining the dispersion present in the finite source sample. Combining the 34 arrays with $R=2,4,8,16,32$ gives 170 settings, each with 5,000 independent datasets and its own random stream. Methods share calls within each dataset; random streams are independent across settings.

Figure~\ref{fig:inference-validation}a reports coverage for each setting. Panel (b) averages full interval widths over the eight data-shaped laws at $N=152$, with shading from the minimum to the maximum law-specific mean. Panel (c) uses the interaction-only null with $S=.08,N=152$. Panel (d) uses the nine contrast families at $N=152,R=32$; gray marks the practical tolerance $\pm.005$. Rate bars are 95\% Wilson intervals over 5,000 datasets per setting.

\paragraph{Point estimators compared.}
Write $\widehat\theta=N^{-1}\sum_i\widehat U_{p,i}^{(q)}$ for the complete-pair macroaverage. The five prompt estimates are the plug-in macroaverage, $\widehat\theta$, $\max(0,\widehat\theta)$, $\widehat\theta/2$, and zero, all scored by $\E(\widehat t-\theta)^2$. The classical trace-mean-square estimate equals $\widehat\theta$ sample by sample (Section~\ref{sec:fixed-effects-equivalence}); the maximum numerical discrepancy was $2.78\times10^{-16}$. For $\overline\Delta$, we compare the pooled unbiased estimate, fixed half shrinkage, zero and the independent-panel mean. The shrinkage factor is fixed before generating test draws.

At the interaction-only null with $U_{\rm int}=.08$, $N=152,R=8$, prompt MSE is $8.82\times10^{-4}$ for plug-in, $1.07\times10^{-6}$ for complete pairs, and $5.28\times10^{-7}$ for the macro positive part; zero has zero error at this null. For $S=.04$, $\overline\Delta=.001$ and $N=152,R=8$, the unbiased and half-shrunk contrast estimates have MSEs of $5.57/3.86\times10^{-7}$, respectively, versus $10^{-6}$ for zero. At $R=32$ they become $1.03/2.78\times10^{-7}$; at $\overline\Delta=.006,R=8$ they are $5.64\times10^{-7}$ and $9.09\times10^{-6}$. These comparisons quantify how signal size and precision affect estimation error.

\paragraph{Ten-law efficiency and allocation study.}
Within each scenario, all items share the same law. Set $a_k=(k-3.5)/\sqrt{35/12}$ and $z_k=(a_k^2-1)/\{6^{-1}\sum_\ell(a_\ell^2-1)^2\}^{1/2}$. The centered, orthogonal directions have unit mean square. With the same $u$ as above,
\[
P_{k,AB}=p_0+A a_ku,\qquad
P_{k,BA}=p_0+(B a_k+C z_k)u.
\]
Table~\ref{tab:ten-law-generators} specifies all ten arrays. The prompt module uses $N=50$, $R=2,4,8$ and 5,000 independent outer datasets per law. Repeat depths use nested prefixes within a dataset. Disjoint-panel estimates average the complete-pair estimates from consecutive two-repeat blocks, whereas all-$R$ estimates also use cross-block pairs. Monte Carlo uncertainty is calculated across independent outer datasets.

\begin{table}[!htbp]
\centering\small
\setlength{\tabcolsep}{5pt}
\caption{Complete ten-law generator. The default $p_0=(.50,.30,.15,.05)$ lists probabilities for candidate 1, candidate 2, TIE and $\bot$, in that order. Only the two indicated bases differ. All probabilities are valid.}
\label{tab:ten-law-generators}
\begin{tabular*}{\linewidth}{@{\extracolsep{\fill}}lrrrl@{}}
\toprule
Family & $A$ & $B$ & $C$ & Alternative $p_0$\\
\midrule
No crossed effects & 0 & 0 & 0 & ---\\
Deterministic null & 0 & 0 & 0 & $(1,0,0,0)$\\
Small prompt & .025 & .025 & 0 & ---\\
Prompt boundary & .10 & .10 & 0 & ---\\
Pure prompt & .14 & .14 & 0 & ---\\
Pure interaction & .14 & $-.14$ & 0 & ---\\
Regular design tie & .14 & 0 & .14 & ---\\
Near tie, E favored & .14 & .01 & .14 & ---\\
Near tie, D favored & .14 & $-.01$ & .14 & ---\\
High invalid probability & .14 & .14 & 0 & $(.38,.27,.10,.25)$\\
\bottomrule
\end{tabular*}
\end{table}

At the pure-interaction prompt null ($U_{\rm int}=.0196$), $R=8$ aggregate MSEs are $3.96\times10^{-6}$ for all-$R$, $2.13\times10^{-5}$ for disjoint panels and $1.03\times10^{-3}$ for plug-in. Itemwise clipping, $N^{-1}\sum_i\max(0,\widehat U_{p,i}^{(q)})$, reduces mean itemwise MSE from $1.99\times10^{-4}$ to $1.23\times10^{-4}$. Its positive bias raises aggregate MSE to $3.22\times10^{-5}$. Clipping \emph{after} averaging projects $\widehat\theta$ onto $[0,\infty)$ and weakly reduces squared error for every nonnegative macro target. Across 27 positive-variance law/depth settings, empirical/exact variance ratios are .963--1.051 and oracle-normal coverage is 94.3--95.6\%; these intervals use variances calculated from the known laws.

\begin{figure}[!htbp]
\centering
\includegraphics[width=\linewidth]{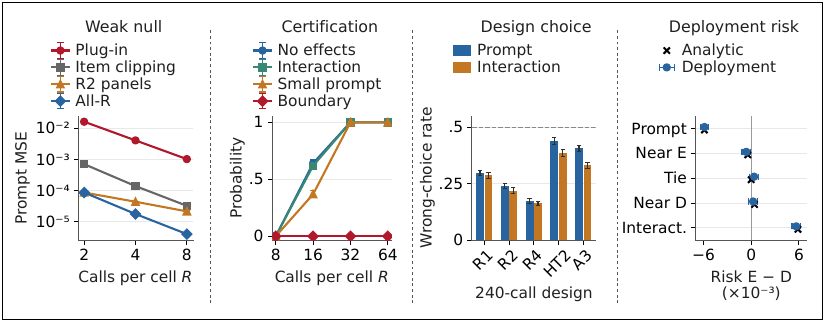}
\caption{Distinct checks under controlled categorical laws. From left to right: weak-null prompt MSE versus repeat depth; certification of $\theta<.01$ versus depth; wrong sign-based design choices at 240 measurement calls; independently observed versus analytic D/E risk contrasts. MSE and risk bars are $\pm1.96$ Monte Carlo standard errors over 5,000 outer datasets. Rate bars are 95\% Wilson intervals, using 1,000 datasets/law for the separate depth diagnostic and 5,000 for design choice. R1/R2/R4 denote census repeat depths; HT2 (Pair2) and A3 denote the two-call pair and three-call anchor. Near E/D are the corresponding nonzero near-tie laws. The bars quantify Monte Carlo uncertainty under fixed known laws.}
\label{fig:simulation-efficiency-allocation}
\end{figure}

The separate depth diagnostic uses four fixed laws (no effects, pure interaction, small prompt and boundary), $N=50$, 1,000 datasets/law, and $R=8,16,32,64$. It uses the original Cantelli constant $b^{\rm orig}=(K-1)/[4NK^2R(R-1)]$, whose minimum 95\% upper endpoint is .01981 at $R=8$. Pure-interaction certification rates for $\theta<.01$ are 0/61.9/100/100\%; the small-prompt law reaches 36.9\% at $R=16$ and 100\% at $R=32$. No boundary case falsely certifies at any depth (per-setting Wilson upper endpoint .38\%). The fixed grid measures the original bound's resolution at prespecified budgets.

For equal-budget selection, 240 calls permit census $R=1,2,4$ on 20, 10 and 5 items, or 120 independent Pair2 units or 80 Anchor3 units (Equations~\ref{eq:design-pair}--\ref{eq:design-anchor}). All methods estimate the same $\Delta$ under the homogeneous scenario law. Forced sign choice selects E for a negative estimate and D otherwise. Under pure prompt variation, wrong-choice rates are 29.8/23.9/17.5\% for the census designs, 44.0\% for Pair2 and 40.8\% for Anchor3; census $R=4$ has wrong-choice rates of 46.4--46.8\% near a regular tie. The universal-bound commitment rule abstains throughout the tested 120- and 240-call configurations. Measurement draws are independent across methods; all methods share a separate 100-item deployment evaluation within each outer dataset. Observed $E-D$ vector-risk differences are $-.00587$ and $.00567$ under pure prompt and pure interaction, respectively (MCSE .00027 each), versus analytic $-.00588$ and $.00588$. These homogeneous laws quantify the precision gain from additional repeats; Section~\ref{sec:heterogeneous-validation} evaluates allocation under item heterogeneity.

\FloatBarrier

\section{Matched Temperature Interventions}
\label{sec:temperature-details}\label{sec:gemma-temperature-details}
\subsection{Shared acquisition and outcome definitions}
Both Qwen/Qwen3.6-27B and \texttt{google/gemma-4-31B-it} use the same fixed 40 items (20 LLMBar, 20 JudgeBench), all 48 prompts, two orders, and temperatures $1,.2,0$. Four repeats give 46,080 measurement calls per model. Eight independent D deployment groups per item and temperature add 3,840 calls, for 49,920 formal calls per model. Prompt-pair assignments are matched across temperatures and models; measurement and deployment calls are generated separately. All invalid and capped completions enter the four-category law.

Both configurations use bfloat16, native nonthinking, a 512-token output cap, top-$p=1$, top-$k$ disabled, unconstrained final generation and the fixed first-nonempty-line parser. Qwen runs on one H100 PCIe with tensor parallelism one; Gemma runs on one H200 NVL with a 16,384-token context. The serving stack is vLLM 0.25.1, Torch 2.11.0+cu130, Transformers 5.14.1 and tokenizers 0.22.2. Temperature conditions are interleaved within logical units with distinct recorded generation seeds. Gemma's 681 technical calls are separate from the formal sample.

Under the original rule, a deployment returns the unique plurality winner among candidate 1, candidate 2, TIE and $\bot$; a tied maximum is TIE. A unique invalid winner remains invalid. Fair resolution changes only a pure candidate 2:2 vote vector, assigning exact expected accuracy $1/2$, wrong-candidate frequency $1/2$ and coverage one. All other outcomes retain their original verdict; candidate selection is independent of the gold label. Thus $A_f=A_o+T_{22}/2$, $W_f=W_o+T_{22}/2$, $C_f=C_o+T_{22}$, using exact expectations. Every deployment is included in the denominator of each decision rate.

\paragraph{Analysis timing and endpoint families.}
The four component contrasts compare call and prompt variation between each cooling temperature and $T=1$. Complete-pair $R=4$ and complete-group $J=8$ analyses use all acquired observations. Their finite-sample constructions were selected after collection. Qwen rule rescoring was developed after inspecting descriptive outcomes; Gemma's rule and cost families were specified before its collection. The direct rule interaction is a secondary analysis of the stored groups. Each coverage guarantee applies to its stated model, reporting pool and endpoint family.
\subsection{Using all four repeats}
\label{sec:temperature-full-repeat-inference}
For each temperature, the complete-pair estimator uses all six distinct-call pairs within each cell. Write $x_t$ and $y_t$ for its prompt and call estimates. Conditional on the fixed items and prompts, Theorem~\ref{thm:nn-variance} gives
\[
\operatorname{Var}(x_t)\le a\theta_t+b,\qquad a=\frac1{4NK},\qquad b=\frac{K-1}{96NK^2},
\]
where $\theta_t$ is the population prompt component for the four-category response law. Using all four repeats leaves the linear coefficient unchanged and reduces the quadratic coefficient by a factor of three relative to averaging two $R=2$ panels.

Set $\eta=.025/6$, $\lambda=(1-\eta)/\eta$, and $x_+=\max(x,0)$. Inverting the two one-sided Cantelli inequalities gives
\begin{align*}
U(x)&=\min\left\{\frac34,x_++\frac{\lambda a}{2}+\sqrt{\lambda(ax_++b)+\frac{\lambda^2a^2}{4}}\right\},\\
L(x)&=\begin{cases}0,&x\le\sqrt{\lambda b},\\
\displaystyle\max\left\{0,x+\frac{\lambda a}{2}-\sqrt{\lambda(ax+b)+\frac{\lambda^2a^2}{4}}\right\},&x>\sqrt{\lambda b}.
\end{cases}
\end{align*}
Each limit has one-sided error at most $\eta$. The prompt-contrast interval for $t\in\{.2,0\}$ is $[L(x_t)-U(x_1),U(x_t)-L(x_1)]$.

For call variation, partitioning the four repeats into two disjoint pairs produces $M=4NK$ independent mismatch indicators at each temperature. The complete-pair estimate is the conditional mean of this partitioned estimate given the cell counts. Jensen's inequality gives the same Hoeffding bound for the complete-pair estimate:
\[
\Pr\!\left(\left|(y_t-y_1)-(U_{\mathrm{call},t}-U_{\mathrm{call},1})\right|\ge h\right)\le2e^{-Mh^2}.
\]
We use $h=\sqrt{\log(2/.0125)/M}$ for each call contrast. The six prompt limits and two call intervals have total error $6\eta+2(.0125)=.05$, giving simultaneous coverage at least 95\% for the four contrasts within each model under independent calls with fixed categorical laws. Table~\ref{tab:temperature-primary} reports these intervals using pooled $R=4$ point estimates throughout. Paired item-composition intervals use the same estimates and 20,000 resamples that preserve dataset and stratum; they describe changes in the questions represented by the panel.

\begin{table}[!htbp]\setlength{\tabcolsep}{2.5pt}\renewcommand{\arraystretch}{1.0}
\centering\small\setlength{\tabcolsep}{2.5pt}
\caption{Full-repeat finite-sample intervals in all ten calibration settings, with 5,000 simulated datasets per setting. Coverage concerns all four contrasts jointly. FWER counts exclusion of zero for any true-null contrast. Width is the mean width of the two prompt intervals; the split-panel column uses finite-sample bounds on the same calls.}
\label{tab:temperature-full-r4-calibration}
\begin{tabular*}{\linewidth}{@{\extracolsep{\fill}}lrrrr@{}}\toprule
Setting & Coverage (\%) & FWER (\%) & Full $R=4$ width & Split-panel width \\\midrule
Homogeneous binary null & 100.00 & 0.00 & .1113 & .1619 \\
Interaction-only weak null & 100.00 & 0.00 & .1107 & .1610 \\
Heterogeneous null & 100.00 & 0.00 & .1190 & .1678 \\
Four-category null & 100.00 & 0.00 & .1210 & .1695 \\
Near-deterministic null & 100.00 & 0.00 & .1089 & .1584 \\
Temperature-shaped signal & 100.00 & -- & .1689 & .2218 \\
Prompt null; call signal & 100.00 & 0.00 & .1108 & .1611 \\
10 stochastic items & 100.00 & 0.00 & .1101 & .1601 \\
2 stochastic items & 100.00 & 0.00 & .1094 & .1592 \\
1 stochastic item & 100.00 & 0.00 & .1093 & .1590 \\\bottomrule
\end{tabular*}\end{table}
All intervals were available in every setting. The Wilson 95\% Monte Carlo interval for each 100\% coverage estimate is [99.923\%,100\%]; for each zero FWER estimate it is [0\%,.077\%].
\begin{table}[!htbp]\setlength{\tabcolsep}{2.5pt}\renewcommand{\arraystretch}{1.0}
\centering\scriptsize
\setlength{\tabcolsep}{2.5pt}
\caption{Temperature contrasts using all four repeats. Call-level intervals have simultaneous coverage of at least 95\% for the four contrasts of each model under independent fixed cell laws. Composition intervals are separate pointwise 95\% paired resampling summaries. Both interval columns use the same complete-pair $R=4$ point estimate.}
\label{tab:temperature-primary}
\label{tab:gemma-temperature-primary}
\label{tab:temperature-full-r4}
\begin{tabular*}{\linewidth}{@{\extracolsep{\fill}}lllrrr@{}}
\toprule
Model & Contrast & Component & Estimate & Call-level interval & Composition interval\\
\midrule
Qwen & $.2-1$ & Call & $-.27135$ & $[-.29706,-.24565]$ & $[-.29228,-.25265]$\\
Qwen & $.2-1$ & Prompt & $.02200$ & $[-.06996,.09928]$ & $[.01514,.02901]$\\
Qwen & $0-1$ & Call & $-.34818$ & $[-.37388,-.32247]$ & $[-.37860,-.32010]$\\
Qwen & $0-1$ & Prompt & $.04262$ & $[-.05699,.12589]$ & $[.03065,.05448]$\\
Gemma & $.2-1$ & Call & $-.03659$ & $[-.06230,-.01088]$ & $[-.05482,-.02036]$\\
Gemma & $.2-1$ & Prompt & $.00999$ & $[-.07987,.09314]$ & $[.00488,.01592]$\\
Gemma & $0-1$ & Call & $-.04488$ & $[-.07059,-.01917]$ & $[-.06736,-.02526]$\\
Gemma & $0-1$ & Prompt & $.01421$ & $[-.07987,.09869]$ & $[.00731,.02261]$\\
\bottomrule
\end{tabular*}
\end{table}
\subsection{Finite-sample intervals for independent deployments}
\label{sec:temperature-group-inference}\label{sec:decision-finite-intervals}
The decision contrasts use all eight independently acquired groups per item. We condition on the item census and assigned prompt pairs, averaging over fresh generation at those assignments. Groups may have different response laws; the two temperatures within a group are paired. Write $D_j\in[-1,1]$ for a paired difference in accuracy, wrong-decision frequency, or coverage, and $n=8N$. For $\lambda=1/2$ and $\psi=\log 2-1/2$, define
\begin{equation}
\widehat\Delta=\frac1n\sum_{j=1}^nD_j,\qquad
h_M=\frac{\log(2M/.05)+\psi\sum_{j=1}^nD_j^2}{\lambda n}.
\label{eq:decision-finite-radius}
\end{equation}
The intervals $[\widehat\Delta-h_M,\widehat\Delta+h_M]\cap[-1,1]$ have simultaneous coverage at least 95\% for a family of $M$ endpoints. This fixed-time empirical-Bernstein construction uses a zero predictable center for each contrast \citep{howard2021confidence}. The value of $\lambda$ was fixed for the post-acquisition analysis and used across all endpoints, models and reporting pools. The interval has positive width even when the observed second moment is zero.

For $x\ge-1$, $\exp(\lambda x-\psi x^2)\le1+\lambda x$. With $\mu_j=\E D_j$,
\[
\E\exp\{\lambda(D_j-\mu_j)-\psi D_j^2\}\le e^{-\lambda\mu_j}(1+\lambda\mu_j)\le1.
\]
Independence across groups lets us multiply these bounds. Markov's inequality for both signs and a union bound over the $M$ endpoints prove the claim with unequal group means and variances. This fixed-$\lambda$ construction yields the reported intervals.

The two-rule family has $M=12$: two cooling contrasts, two rules, and three scores. Adding candidate-only majority gives an $M=18$ family; its complete bounds are included in the accompanying numerical results. Each family is defined within one model and reporting pool. Item-composition intervals describe a separate source of variation.

\paragraph{Direct interaction between temperature and the tie rule.}
For each cooling level $t\in\{.2,0\}$, write $\Delta_t$ for the contrast with $T=1$. The rule interaction is
\begin{equation}
\Gamma_t=\Delta_t A_{\rm coin22}-\Delta_t A_{\rm strict}
=\tfrac12\Delta_t T_{22}.
\label{eq:temperature-rule-interaction}
\end{equation}
This direct contrast was specified for a secondary analysis of the stored groups. For paired group $j$, let $X_j=\1\{\text{pure tie at }t\}-\1\{\text{pure tie at }1\}\in[-1,1]$. Apply Equation~\eqref{eq:decision-finite-radius} to $X_j$ with the same $\lambda=1/2$, then divide the estimate and radius by two. The family has $M=4$: both cooling contrasts in both models. Thus $\widehat\Gamma_t=\sum_jX_j/(2n)$ and the half-width is
\[
\frac{\log(8/.05)+(\log2-1/2)\sum_jX_j^2}{n},\qquad n=320.
\]
The intervals are clipped to $[-1/2,1/2]$ and condition on the fixed items and assigned prompt pairs. They quantify the change between rules directly, retaining the pairing of the same votes. Their simultaneous guarantee covers these four interactions as a separate family.

\begin{table}[!htbp]\setlength{\tabcolsep}{2.5pt}\renewcommand{\arraystretch}{1.0}\centering\small
\caption{Temperature-by-rule interactions, in percentage points. A positive value means fair resolution of pure 2:2 ties increases the accuracy effect of cooling relative to the original rule. The finite-sample 95\% intervals are simultaneous across all four rows. Each row uses 320 paired independent deployment groups.}
\label{tab:temperature-rule-interaction}
\begin{tabular*}{\linewidth}{@{\extracolsep{\fill}}llrr@{}}\toprule
Model & Temperature contrast & $\widehat\Gamma_t$ & Simultaneous interval\\\midrule
Qwen & $.2-1$ & $7.34$ & $[-.34,15.03]$\\
Qwen & $0-1$ & $9.84$ & $[1.68,18.01]$\\
Gemma & $.2-1$ & $.63$ & $[-1.44,2.69]$\\
Gemma & $0-1$ & $.47$ & $[-1.54,2.48]$\\\bottomrule
\end{tabular*}
\end{table}
The groupwise identity in Equation~\eqref{eq:temperature-rule-interaction} holds for all 256 categorical four-vote patterns with either candidate as gold, and for all 1,920 stored groups. Tables~\ref{tab:temperature-rescoring-fixed} and~\ref{tab:gemma-temperature-rules} report the accuracy contrasts under each rule.

For abstention costs, $\Delta L(c)=(1-c)\Delta W+c(-\Delta A)$. Applying Equation~\eqref{eq:decision-finite-radius} jointly to the eight endpoints of the four cooling/rule curves gives simultaneous bands for every $c\in[0,1]$ by linear interpolation. Crossover sets are obtained by inverting these bands, retaining full, empty, and boundary-truncated sets.

\begin{table}[!htbp]\setlength{\tabcolsep}{2.5pt}\renewcommand{\arraystretch}{1.0}\centering\scriptsize\setlength{\tabcolsep}{2.5pt}
\caption{Decision-interval calibration with 5,000 datasets per setting. Coverage requires all 12 intervals to be available and cover their targets. Unavailable intervals under the two-panel approximation are reported separately. Width is the mean full interval width in percentage points; other entries are percentages.}
\label{tab:decision-finite-calibration}
\begin{tabular*}{\linewidth}{@{\extracolsep{\fill}}lrrrrr@{}}\toprule
Scenario & Finite coverage & Width & Panel coverage & Panel FWER & Unavailable\\\midrule
Distributed, null & 100.00 & 37.07 & 96.46 & 3.54 & 0.00\\
Distributed, shift & 100.00 & 35.59 & 96.88 & 0.32 & 0.00\\
Heterogeneous, null & 99.98 & 23.05 & 0.00 & 2.48 & 100.00\\
Heterogeneous, shift & 100.00 & 23.12 & 0.00 & 0.58 & 100.00\\
10 stochastic items & 100.00 & 15.06 & 89.12 & 10.88 & 0.00\\
2 stochastic items & 100.00 & 9.19 & 41.74 & 29.68 & 42.68\\
1 stochastic item & 100.00 & 8.45 & 6.92 & 21.82 & 91.68\\
Rare events, null & 100.00 & 8.74 & 0.00 & 0.22 & 100.00\\
Rare events, shift & 100.00 & 8.84 & 0.00 & 1.30 & 100.00\\
Shared randomness, shift & 99.94 & 16.09 & 0.00 & 0.00 & 100.00\\\bottomrule
\end{tabular*}\end{table}
All finite-sample intervals were available. The separate eight-endpoint cost family had joint coverage of 99.96--100\% across the same settings. The Wilson 95\% Monte Carlo interval for 5,000 successes is [99.923\%,100\%]. The coverage guarantee follows from Equation~\eqref{eq:decision-finite-radius}.

\subsection{Calibration under concentrated call variation}
\label{sec:temperature-panel-calibration}
An approximate alternative partitions $R=4$ into two independent $R=2$ panels and compares their studentized difference with a $t_{39}$ reference distribution. Calibration uses the exact $N=40,K=48$, two-order design and 5,000 independent datasets in each of ten fixed-law settings. The settings include interaction-only weak nulls, invalid-output laws, near-deterministic laws, temperature signals and variance concentrated in 10, 2 or 1 items. All calls are independent conditional on the fixed laws; the scenarios were specified before simulation.
\begin{table}[!htbp]\setlength{\tabcolsep}{2.5pt}\renewcommand{\arraystretch}{1.0}
\centering\small
\caption{Calibration of the two-panel procedure. Simultaneous coverage counts a dataset as successful only when all four intervals are available and cover their targets. Parentheses give Wilson 95\% Monte Carlo intervals. FWER is the probability that Holm rejects at least one true null. All settings have 5,000 draws.}
\label{tab:temperature-panel-calibration}
\begin{tabular*}{\linewidth}{@{\extracolsep{\fill}}lrrr@{}}
\toprule
Setting & Four-interval coverage (\%) & FWER (\%) & Availability (\%) \\
\midrule
Homogeneous binary null & 96.26 (95.70, 96.75) & 3.74 & 100.00 \\
Interaction-only weak null & 95.40 (94.78, 95.95) & 4.60 & 100.00 \\
Heterogeneous null & 96.18 (95.61, 96.68) & 3.82 & 100.00 \\
Four-category null & 96.36 (95.80, 96.84) & 3.64 & 100.00 \\
Near-deterministic null & 97.00 (96.49, 97.44) & 3.00 & 100.00 \\
Temperature-shaped signal & 96.02 (95.44, 96.53) & -- & 100.00 \\
Prompt null; call signal & 96.12 (95.55, 96.62) & 4.58 & 100.00 \\
10 stochastic items & 92.22 (91.44, 92.93) & 7.78 & 100.00 \\
2 stochastic items & 65.70 (64.37, 67.00) & 34.26 & 99.92 \\
1 stochastic item & 42.48 (41.12, 43.86) & 55.82 & 95.86 \\
\bottomrule
\end{tabular*}
\end{table}

Coverage is 92.22\%, 65.70\% and 42.48\% in the concentrated cases. With one stochastic item, 4.14\% of datasets have at least one unavailable interval because the empirical panel variance is zero; availability is counted in simultaneous coverage. The descriptive ranges for effective item counts are 11.35--16.06 for Qwen and 4.23--5.17 for Gemma, with largest contributions 17.4\% and 43.5\%. These quantities summarize realized variance concentration. Table~\ref{tab:temperature-full-r4-calibration} reports the complete-pair bounds' coverage and width under concentrated variance.

\subsection{Complete response and decision results}
\label{sec:temperature-rescoring-details}
The primary component table reports fixed-call intervals and separate pointwise composition intervals for the same $R=4$ estimates. The realized prompt inflation is $47\widehat U_c/384$ at $K=48,R=4$.
\begin{table}[!htbp]\setlength{\tabcolsep}{2.5pt}\renewcommand{\arraystretch}{1.0}
\centering\footnotesize
\caption{Qwen temperature component profiles. Entries are pooled $R=4$ complete-pair estimates, averaged equally over items; plug-in prompt dispersion is shown alongside the corrected components. Each dataset has 20 items.}
\label{tab:temperature-components}
\begin{tabular*}{\linewidth}{@{\extracolsep{\fill}}llrrrrrr@{}}
\toprule
Panel & $T$ & Prompt & Order & Interaction & Call & Total & Plug-in \\
\midrule
All 40 & 1 & 0.01125 & 0.05516 & 0.01018 & 0.35386 & 0.43045 & 0.05456 \\
All 40 & 0.2 & 0.03325 & 0.14688 & 0.03637 & 0.08251 & 0.29902 & 0.04335 \\
All 40 & 0 & 0.05387 & 0.17773 & 0.05231 & 0.00569 & 0.28959 & 0.05456 \\
\addlinespace
LLMBar & 1 & 0.00694 & 0.02216 & 0.00916 & 0.29366 & 0.33193 & 0.04288 \\
LLMBar & 0.2 & 0.01715 & 0.07345 & 0.02439 & 0.06753 & 0.18252 & 0.02541 \\
LLMBar & 0 & 0.03520 & 0.10318 & 0.03615 & 0.00208 & 0.17661 & 0.03546 \\
\addlinespace
JudgeBench & 1 & 0.01556 & 0.08816 & 0.01119 & 0.41406 & 0.52897 & 0.06624 \\
JudgeBench & 0.2 & 0.04936 & 0.22032 & 0.04835 & 0.09748 & 0.41551 & 0.06129 \\
JudgeBench & 0 & 0.07253 & 0.25228 & 0.06847 & 0.00929 & 0.40256 & 0.07367 \\
\bottomrule
\end{tabular*}
\end{table}
\begin{table}[!htbp]\setlength{\tabcolsep}{2.5pt}\renewcommand{\arraystretch}{1.0}
\centering\small
\setlength{\tabcolsep}{2.5pt}
\caption{Gemma response components, pooled $R=4$ descriptive means over all 40 items. The four corrected components sum to corrected total disagreement.}
\label{tab:gemma-temperature-components}
\begin{tabular*}{\linewidth}{@{\extracolsep{\fill}}lrrr@{}}
\toprule
Component & $T=1$ & $T=.2$ & $T=0$ \\
\midrule
Prompt & $.01859$ & $.02858$ & $.03280$ \\
Order & $.05810$ & $.06369$ & $.06280$ \\
Interaction & $.01734$ & $.02690$ & $.03062$ \\
Call & $.04501$ & $.00842$ & $.00013$ \\
Total disagreement & $.13904$ & $.12759$ & $.12636$ \\
Plug-in prompt & $.02410$ & $.02961$ & $.03282$ \\
Finite-call prompt excess & $.00551$ & $.00103$ & $.00002$ \\
\bottomrule
\end{tabular*}
\end{table}

We estimate deployment vector risk using the unbiased trace sample variance across the eight independent groups. Gold Brier loss is the average squared distance of the four-category vote vector to the gold one-hot vector. Subtracting trace variance from Brier loss estimates squared mean-to-gold distance; we retain the raw estimates. In Qwen, cooling decreases both terms. Gemma's vector risk changes from .02494 to .01959 while mean-to-gold error changes from .35592 to .35932, so Brier loss changes from .38086 to .37891. Gold loss combines vote variation and mean-to-gold alignment.

\paragraph{Gold-loss decomposition contrasts.}
Table~\ref{tab:temperature-loss-precision} gives the paired Qwen loss changes on all 40 items. We estimate vector risk with the unbiased trace sample variance of the eight independent deployment vectors. Subtracting vector risk from Brier loss estimates squared mean-to-gold distance. Both terms decrease under cooling in these composition summaries.

\begin{table}[!htbp]
\centering\small
\setlength{\tabcolsep}{3pt}\renewcommand{\arraystretch}{1.0}
\caption{Qwen loss-decomposition contrasts using all $J=8$ independent four-vote groups per item and temperature. Entries are changes with pointwise 95\% paired item-composition intervals from 20,000 resamples that preserve dataset and stratum. Units are squared one-hot-vector distance; these intervals describe item composition conditional on the recorded calls.}
\label{tab:temperature-loss-precision}
\begin{tabular*}{\linewidth}{@{\extracolsep{\fill}}lrr@{}}
\toprule
Metric & $\Delta_{.2-1}$ & $\Delta_{0-1}$\\
\midrule
Vector risk & $-.05709\;[-.07042,-.04364]$ & $-.06719\;[-.07958,-.05458]$\\
Gold Brier loss & $-.13125\;[-.16641,-.09766]$ & $-.14375\;[-.18086,-.10547]$\\
Squared mean-to-gold & $-.07416\;[-.10246,-.04632]$ & $-.07656\;[-.10865,-.04425]$\\
\bottomrule
\end{tabular*}
\end{table}

Tables~\ref{tab:temperature-rescoring-fixed} and~\ref{tab:gemma-temperature-rules} report all twelve endpoints spanning the two rules and cooling levels, in the pooled sample and each dataset. Each pool has its own simultaneous endpoint family. Qwen's original-rule wrong-decision change at $T=0$ is $-12.81$ pp [$-23.07$,$-2.56$]. Gemma's pure-tie increase is .94 pp (6.88\% to 7.81\%), and its original-rule coverage is 83.13\% at both endpoints.
\begin{table}[!htbp]\setlength{\tabcolsep}{2.5pt}\renewcommand{\arraystretch}{1.0}\centering\scriptsize\setlength{\tabcolsep}{2.5pt}
\caption{Qwen cooling contrasts relative to $T=1$ (percentage points). Brackets are finite-sample 95\% intervals simultaneous over the twelve endpoints within each reporting pool. All eight groups per item are used; items and assigned prompt pairs are fixed.}
\label{tab:decision-finite-qwen-12}\label{tab:temperature-rescoring-fixed}
\begin{tabular*}{\linewidth}{@{\extracolsep{\fill}}lllccc@{}}\toprule
Pool & $T$ & Rule & Accuracy & Wrong & Coverage\\\midrule
All 40 & .2 & Original & $-.63$\,$[-13.42,12.17]$ & $-9.06$\,$[-19.08,.95]$ & $-9.69$\,$[-25.50,6.12]$\\
All 40 & .2 & Fair 2:2 & $6.72$\,$[-2.36,15.80]$ & $-1.72$\,$[-10.44,7.00]$ & $5.00$\,$[-.79,10.79]$\\
All 40 & 0 & Original & $-1.88$\,$[-14.43,10.68]$ & $-12.81$\,$[-23.07,-2.56]$ & $-14.69$\,$[-31.70,2.33]$\\
All 40 & 0 & Fair 2:2 & $7.97$\,$[-.87,16.81]$ & $-2.97$\,$[-11.44,5.51]$ & $5.00$\,$[-1.03,11.03]$\\
\addlinespace
LLMBar & .2 & Original & $5.63$\,$[-10.06,21.31]$ & $-5.63$\,$[-17.45,6.20]$ & $.00$\,$[-16.89,16.89]$\\
LLMBar & .2 & Fair 2:2 & $6.88$\,$[-5.19,18.94]$ & $-4.38$\,$[-15.71,6.96]$ & $2.50$\,$[-6.18,11.18]$\\
LLMBar & 0 & Original & $6.88$\,$[-8.33,22.08]$ & $-9.38$\,$[-21.20,2.45]$ & $-2.50$\,$[-19.39,14.39]$\\
LLMBar & 0 & Fair 2:2 & $9.38$\,$[-2.45,21.20]$ & $-6.88$\,$[-17.97,4.22]$ & $2.50$\,$[-6.18,11.18]$\\
\addlinespace
JudgeBench & .2 & Original & $-6.88$\,$[-24.49,10.74]$ & $-12.50$\,$[-28.43,3.43]$ & $-19.38$\,$[-41.82,3.07]$\\
JudgeBench & .2 & Fair 2:2 & $6.56$\,$[-7.25,20.38]$ & $.94$\,$[-12.88,14.75]$ & $7.50$\,$[-3.11,18.11]$\\
JudgeBench & 0 & Original & $-10.63$\,$[-28.24,6.99]$ & $-16.25$\,$[-32.66,.16]$ & $-26.88$\,$[-51.73,-2.02]$\\
JudgeBench & 0 & Fair 2:2 & $6.56$\,$[-7.01,20.13]$ & $.94$\,$[-12.63,14.51]$ & $7.50$\,$[-3.60,18.60]$\\
\bottomrule\end{tabular*}\end{table}\FloatBarrier

\begin{table}[!htbp]\setlength{\tabcolsep}{2.5pt}\renewcommand{\arraystretch}{1.0}\centering\scriptsize\setlength{\tabcolsep}{2.5pt}
\caption{Gemma cooling contrasts relative to $T=1$ (percentage points). Brackets are finite-sample 95\% intervals simultaneous over the twelve endpoints within each reporting pool. All eight groups per item are used; items and assigned prompt pairs are fixed.}
\label{tab:decision-finite-gemma-12}\label{tab:gemma-temperature-rules}
\begin{tabular*}{\linewidth}{@{\extracolsep{\fill}}lllccc@{}}\toprule
Pool & $T$ & Rule & Accuracy & Wrong & Coverage\\\midrule
All 40 & .2 & Original & $-.94$\,$[-5.64,3.77]$ & $.31$\,$[-3.91,4.53]$ & $-.63$\,$[-5.69,4.44]$\\
All 40 & .2 & Fair 2:2 & $-.31$\,$[-4.77,4.15]$ & $.94$\,$[-3.16,5.04]$ & $.63$\,$[-3.72,4.97]$\\
All 40 & 0 & Original & $-.31$\,$[-5.26,4.63]$ & $.31$\,$[-3.91,4.53]$ & $.00$\,$[-5.31,5.31]$\\
All 40 & 0 & Fair 2:2 & $.16$\,$[-4.52,4.83]$ & $.78$\,$[-3.29,4.85]$ & $.94$\,$[-3.52,5.40]$\\
\addlinespace
LLMBar & .2 & Original & $-1.25$\,$[-9.93,7.43]$ & $.63$\,$[-7.33,8.58]$ & $-.63$\,$[-9.55,8.30]$\\
LLMBar & .2 & Fair 2:2 & $-.94$\,$[-9.32,7.44]$ & $.94$\,$[-6.96,8.84]$ & $.00$\,$[-8.20,8.20]$\\
LLMBar & 0 & Original & $-1.25$\,$[-9.93,7.43]$ & $.63$\,$[-7.33,8.58]$ & $-.63$\,$[-9.55,8.30]$\\
LLMBar & 0 & Fair 2:2 & $-.94$\,$[-9.32,7.44]$ & $.94$\,$[-6.96,8.84]$ & $.00$\,$[-8.20,8.20]$\\
\addlinespace
JudgeBench & .2 & Original & $-.63$\,$[-9.07,7.82]$ & $.00$\,$[-8.20,8.20]$ & $-.63$\,$[-9.55,8.30]$\\
JudgeBench & .2 & Fair 2:2 & $.31$\,$[-7.95,8.57]$ & $.94$\,$[-7.08,8.96]$ & $1.25$\,$[-6.95,9.45]$\\
JudgeBench & 0 & Original & $.63$\,$[-8.30,9.55]$ & $.00$\,$[-8.20,8.20]$ & $.63$\,$[-8.78,10.03]$\\
JudgeBench & 0 & Fair 2:2 & $1.25$\,$[-7.43,9.93]$ & $.63$\,$[-7.33,8.58]$ & $1.88$\,$[-6.57,10.32]$\\
\bottomrule\end{tabular*}\end{table}\FloatBarrier
\begin{table}[!htbp]\setlength{\tabcolsep}{2.5pt}\renewcommand{\arraystretch}{1.0}\centering\scriptsize\setlength{\tabcolsep}{2.5pt}
\caption{Qwen cost contrasts. Endpoint bands are in percentage points and jointly cover both rules, both cooling levels, and all $c\in[0,1]$ with probability at least 95\%, separately within each pool. Crossover sets are obtained by inverting the simultaneous loss bands.}
\label{tab:decision-finite-cost-qwen}\label{tab:temperature-rescoring-costs-family}
\begin{tabular*}{\linewidth}{@{\extracolsep{\fill}}llllll@{}}\toprule
Pool & $T$ & Rule & Band at $c=0$ & Band at $c=1$ & Crossover set\\\midrule
All 40 & .2 & Original & $[-18.82,.70]$ & $[-11.91,13.16]$ & $[.000,1.000]$\\
All 40 & .2 & Fair 2:2 & $[-10.18,6.75]$ & $[-15.54,2.11]$ & $[.000,1.000]$\\
All 40 & 0 & Original & $[-22.82,-2.81]$ & $[-10.42,14.17]$ & $[.165,1.000]$\\
All 40 & 0 & Fair 2:2 & $[-11.19,5.25]$ & $[-16.55,.62]$ & $[.000,1.000]$\\
\addlinespace
LLMBar & .2 & Original & $[-16.94,5.69]$ & $[-20.80,9.55]$ & $[.000,1.000]$\\
LLMBar & .2 & Fair 2:2 & $[-15.21,6.46]$ & $[-18.43,4.68]$ & $[.000,1.000]$\\
LLMBar & 0 & Original & $[-20.69,1.94]$ & $[-21.57,7.82]$ & $[.000,1.000]$\\
LLMBar & 0 & Fair 2:2 & $[-17.47,3.72]$ & $[-20.69,1.94]$ & $[.000,1.000]$\\
\addlinespace
JudgeBench & .2 & Original & $[-27.92,2.92]$ & $[-10.23,23.98]$ & $[.000,1.000]$\\
JudgeBench & .2 & Fair 2:2 & $[-12.37,14.24]$ & $[-19.87,6.74]$ & $[.000,1.000]$\\
JudgeBench & 0 & Original & $[-32.15,-.35]$ & $[-6.48,27.73]$ & $[.012,1.000]$\\
JudgeBench & 0 & Fair 2:2 & $[-12.13,14.00]$ & $[-19.63,6.50]$ & $[.000,1.000]$\\
\bottomrule\end{tabular*}\end{table}\FloatBarrier

\begin{table}[!htbp]\setlength{\tabcolsep}{2.5pt}\renewcommand{\arraystretch}{1.0}\centering\scriptsize\setlength{\tabcolsep}{2.5pt}
\caption{Gemma cost contrasts. Endpoint bands are in percentage points and jointly cover both rules, both cooling levels, and all $c\in[0,1]$ with probability at least 95\%, separately within each pool. Crossover sets are obtained by inverting the simultaneous loss bands.}
\label{tab:decision-finite-cost-gemma}\label{tab:gemma-temperature-cost}
\begin{tabular*}{\linewidth}{@{\extracolsep{\fill}}llllll@{}}\toprule
Pool & $T$ & Rule & Band at $c=0$ & Band at $c=1$ & Crossover set\\\midrule
All 40 & .2 & Original & $[-3.65,4.28]$ & $[-3.51,5.39]$ & $[.000,1.000]$\\
All 40 & .2 & Fair 2:2 & $[-2.91,4.78]$ & $[-3.90,4.52]$ & $[.000,1.000]$\\
All 40 & 0 & Original & $[-3.65,4.28]$ & $[-4.38,5.00]$ & $[.000,1.000]$\\
All 40 & 0 & Fair 2:2 & $[-3.04,4.60]$ & $[-4.58,4.26]$ & $[.000,1.000]$\\
\addlinespace
LLMBar & .2 & Original & $[-6.83,8.08]$ & $[-6.93,9.43]$ & $[.000,1.000]$\\
LLMBar & .2 & Fair 2:2 & $[-6.45,8.33]$ & $[-6.94,8.81]$ & $[.000,1.000]$\\
LLMBar & 0 & Original & $[-6.83,8.08]$ & $[-6.93,9.43]$ & $[.000,1.000]$\\
LLMBar & 0 & Fair 2:2 & $[-6.45,8.33]$ & $[-6.94,8.81]$ & $[.000,1.000]$\\
\addlinespace
JudgeBench & .2 & Original & $[-7.69,7.69]$ & $[-7.31,8.56]$ & $[.000,1.000]$\\
JudgeBench & .2 & Fair 2:2 & $[-6.57,8.45]$ & $[-8.07,7.44]$ & $[.000,1.000]$\\
JudgeBench & 0 & Original & $[-7.69,7.69]$ & $[-9.04,7.79]$ & $[.000,1.000]$\\
JudgeBench & 0 & Fair 2:2 & $[-6.83,8.08]$ & $[-9.43,6.93]$ & $[.000,1.000]$\\
\bottomrule\end{tabular*}\end{table}\FloatBarrier

For Qwen under the original rule, the pooled point estimates of the crossover cost are .935 for $T=.2$ and .872 for $T=0$; its $T=0$ finite-sample band favors cooling for $c<.165$. Tables~\ref{tab:decision-finite-cost-qwen} and~\ref{tab:decision-finite-cost-gemma} give the fair-rule and Gemma bands over $[0,1]$. Band inversion retains full, empty and boundary-truncated crossover sets.

\paragraph{Output validity and length stops.}
Table~\ref{tab:temperature-output-evidence} reports completed outputs at each temperature. The estimates use every completed response, including invalid and capped outputs. A capped response with a valid first final line keeps its verdict: Gemma at $T=0$ has 163 cap stops and no invalid output. Its three empty thought markers at $T=1$ contain no reasoning text.

\begin{table}[!htbp]
\centering\small
\setlength{\tabcolsep}{2.5pt}\renewcommand{\arraystretch}{1.0}
\caption{Formal temperature-study outputs. Qwen rows separate measurement and individual deployment calls, reporting the percentages of invalid and capped outputs and the total output tokens. Gemma rows combine both phases and report counts of invalid returns, cap stops and empty thought markers. Technical calls are excluded.}
\label{tab:temperature-output-evidence}
\begin{tabular*}{\linewidth}{@{\extracolsep{\fill}}llrrrr@{}}
\toprule
\multicolumn{6}{l}{Qwen}\\
Phase & $T$ & Calls & Invalid (\%) & Capped (\%) & Output tokens\\
\midrule
Measurement & 1 & 15,360 & 6.48 & 5.00 & 433,666\\
Measurement & .2 & 15,360 & 3.78 & 3.72 & 323,504\\
Measurement & 0 & 15,360 & 3.54 & 3.48 & 305,618\\
Deployment & 1 & 1,280 & 7.42 & 5.63 & 39,506\\
Deployment & .2 & 1,280 & 3.44 & 3.36 & 24,994\\
Deployment & 0 & 1,280 & 3.52 & 3.52 & 25,513\\
\midrule
\multicolumn{6}{l}{Gemma: both phases}\\
 & $T$ & Calls & Invalid (count) & Capped (count) & Empty markers\\
\midrule
 & 1 & 16,640 & 85 & 138 & 3\\
 & .2 & 16,640 & 4 & 169 & 0\\
 & 0 & 16,640 & 0 & 163 & 0\\
\bottomrule
\end{tabular*}
\end{table}

\paragraph{Item-composition sensitivity.}
Separate bootstrap analyses use 20,000 paired whole-item draws that preserve dataset and stratum; singleton strata remain fixed. The rule-rescoring draws use a centered maximum absolute deviation over twelve endpoints within a pool. Endpoint maxima also give composition bands for affine cost curves. These summaries describe item composition conditional on recorded calls; the group guarantees condition on the fixed assignments and average over fresh generation. The reproducible numerical results include complete pointwise and simultaneous composition tables and every secondary deployment and output metric.

\subsection{Prompt families and mechanism checks}
\label{sec:temperature-prompt-families}\label{sec:temperature-mechanism-analysis}
The four families each contain 12 prompts and use the correction for their actual $K=12$. The clause-order subset reorders exactly the same five sentences, including the output-format line, so it holds wording fixed. All eight model--family comparisons show descriptive prompt increases and call decreases under cooling. This exploratory within-panel analysis uses complete-pair estimates on the shared items; Table~\ref{tab:temperature-primary} gives the model-level finite-call intervals.
\begin{table}[!htbp]\setlength{\tabcolsep}{2.5pt}\renewcommand{\arraystretch}{1.0}
\centering\small
\caption{Pooled $R=4$ components within each predefined 12-prompt family. All outputs, including invalid ones, are retained. Ratios are descriptive.}
\label{tab:temperature-family-components}
\begin{tabular*}{\linewidth}{@{\extracolsep{\fill}}llrrrrr@{}}
\toprule
Configuration & Family & Call $T=1$ & Call $T=0$ & Prompt $T=1$ & Prompt $T=0$ & Ratio \\
\midrule
Qwen & Clause order & 0.3594 & 0.0064 & 0.0072 & 0.0456 & 6.30 \\
Qwen & Discourse & 0.3604 & 0.0030 & 0.0098 & 0.0496 & 5.08 \\
Qwen & Lexical & 0.3387 & 0.0068 & 0.0069 & 0.0415 & 6.02 \\
Qwen & Syntactic & 0.3569 & 0.0066 & 0.0141 & 0.0534 & 3.80 \\
Gemma & Clause order & 0.0398 & 0.0000 & 0.0171 & 0.0299 & 1.74 \\
Gemma & Discourse & 0.0483 & 0.0000 & 0.0193 & 0.0340 & 1.77 \\
Gemma & Lexical & 0.0467 & 0.0005 & 0.0123 & 0.0254 & 2.07 \\
Gemma & Syntactic & 0.0453 & 0.0000 & 0.0160 & 0.0288 & 1.80 \\
\bottomrule
\end{tabular*}
\end{table}

At the item level, measurement calls estimate baseline gold margin and order-component change; separate deployments estimate tie changes. Qwen's within-dataset-centered order--tie association is .75 with a pointwise composition interval [.62,.85]; split measurement halves give .75 and .74. Eighteen weak-margin questions ($|m|\le.25$) have a mean tie increase of 36.81 pp, compared with .66 pp among 19 positive-margin questions. Gemma has 29 positive-margin questions and four with any tie change. Constant tie changes make some Gemma bootstrap correlations undefined. The exploratory correlations are descriptive associations with estimated covariates; the reported interval is pointwise.

\paragraph{Baseline-margin groups.}
Table~\ref{tab:temperature-margin-evidence} reports every fixed margin group, including empty and small groups. In Qwen, the five positive-margin JudgeBench questions have a mean tie increase of 22.50 pp, whereas the 14 positive-margin LLMBar questions have a 7.14-pp decrease. Their pooled positive-margin mean is $+.66$ pp. The JudgeBench increase also includes new ties on questions with a positive baseline gold margin.

\begin{table}[!htbp]
\centering\small
\setlength{\tabcolsep}{2.5pt}\renewcommand{\arraystretch}{1.05}
\caption{Descriptive groups defined by baseline gold margin $m$ at $T=1$. Each cell is $(n,\Delta U_o,\Delta T_{22})$, comparing $T=0$ with $T=1$; tie changes are in percentage points. Margin and order estimates use $R=4$ measurement calls/cell; ties use separate $J=8$ four-vote deployments. A dash indicates that the group is empty, so its mean is undefined.}
\label{tab:temperature-margin-evidence}
\begin{tabular*}{\linewidth}{@{\extracolsep{\fill}}llccc@{}}
\toprule
Dataset & Model & $m<-.25$ & $|m|\le.25$ & $m>.25$\\
\midrule
All 40 & Qwen & $(3,.133,+37.50)$ & $(18,.241,+36.81)$ & $(19,.009,+.66)$\\
All 40 & Gemma & $(5,.017,+2.50)$ & $(6,.016,+.00)$ & $(29,.000,+.86)$\\
LLMBar & Qwen & $(0,\text{---},\text{---})$ & $(6,.293,+33.33)$ & $(14,-.010,-7.14)$\\
LLMBar & Gemma & $(0,\text{---},\text{---})$ & $(3,.004,-4.17)$ & $(17,-.001,+1.47)$\\
JudgeBench & Qwen & $(3,.133,+37.50)$ & $(12,.214,+38.54)$ & $(5,.063,+22.50)$\\
JudgeBench & Gemma & $(5,.017,+2.50)$ & $(3,.028,+4.17)$ & $(12,.002,+.00)$\\
\bottomrule
\end{tabular*}
\end{table}

\paragraph{Agreement and correctness estimands.}
\label{sec:stored-label-metrics}
Across-prompt categorical agreement is
\[
\frac1{K(K-1)}\sum_{k\ne\ell}\langle P_k^q,P_\ell^q\rangle
=\|\bar p\|^2-\frac{\Up^{(q)}}{K-1},
\]
so it depends on marginal concentration as well as prompt dispersion.

For one item with gold candidate $y$, let
$Y_{kor}=\1\{Z_{kor}=y\}$, where $k=1,\ldots,K$, $o\in\{\mathrm{AB},\mathrm{BA}\}$,
and $r=1,\ldots,R$. TIE and invalid verdicts have $Y_{kor}=0$.
The binary correctness gap applies the PSS absolute-difference formula
\citep{zhuo2024prosa} within each order and repeat:
\begin{equation}
\widehat S_{\rm binary}
=\frac{1}{2R\binom K2}
\sum_o\sum_{r=1}^R\sum_{k<\ell}|Y_{kor}-Y_{\ell or}|.
\label{eq:stored-binary-correctness-gap}
\end{equation}
For the mean-correctness gap, first set
$\widehat a_k=(2R)^{-1}\sum_o\sum_{r=1}^R Y_{kor}$ and then compute
\begin{equation}
\widehat S_{\rm mean}
=\frac{1}{\binom K2}\sum_{k<\ell}|\widehat a_k-\widehat a_\ell|.
\label{eq:stored-mean-correctness-gap}
\end{equation}
The first statistic takes absolute differences before averaging over
orders and repeated calls; the second averages correctness first.
For independent Bernoulli correctness with probabilities $a,b$, the
pairwise expectation in the first statistic is $a+b-2ab$; the absolute
difference between the mean correctness probabilities is $|a-b|$. The dataset summaries average these statistics with equal weight for each item.

In Qwen, the binary gap falls from .319 to .101 and across-prompt agreement rises from .569 to .709 as corrected prompt dispersion rises from .011 to .054; Gemma shows smaller changes in the same directions. These summaries measure correctness spread, verdict agreement and prompt dispersion, respectively.

\begin{table}[!htbp]\setlength{\tabcolsep}{2.5pt}\renewcommand{\arraystretch}{1.0}
\centering\scriptsize
\caption{Six stored-label summaries at all three temperatures. Each row uses the identical item/prompt census for its model. The corrected component and the agreement/correctness summaries have different estimands.}
\label{tab:temperature-stored-metrics}
\begin{tabular*}{\linewidth}{@{\extracolsep{\fill}}llrrrrrr@{}}
\toprule
 & & \multicolumn{3}{c}{Qwen} & \multicolumn{3}{c}{Gemma}\\
Dataset & Summary & $1$ & $.2$ & $0$ & $1$ & $.2$ & $0$\\
\midrule
All 40 items & Plug-in prompt & 0.055 & 0.043 & 0.055 & 0.024 & 0.030 & 0.033\\
All 40 items & Corrected prompt & 0.011 & 0.033 & 0.054 & 0.019 & 0.029 & 0.033\\
All 40 items & Same-cell agreement & 0.646 & 0.917 & 0.994 & 0.955 & 0.992 & 1.000\\
All 40 items & Across-prompt agreement & 0.569 & 0.700 & 0.709 & 0.861 & 0.872 & 0.873\\
All 40 items & Binary correctness gap & 0.319 & 0.139 & 0.101 & 0.059 & 0.049 & 0.048\\
All 40 items & Mean-correctness gap & 0.160 & 0.106 & 0.093 & 0.049 & 0.046 & 0.046\\
LLMBar & Plug-in prompt & 0.043 & 0.025 & 0.035 & 0.026 & 0.031 & 0.035\\
LLMBar & Corrected prompt & 0.007 & 0.017 & 0.035 & 0.022 & 0.030 & 0.035\\
LLMBar & Same-cell agreement & 0.706 & 0.932 & 0.998 & 0.969 & 0.992 & 1.000\\
LLMBar & Across-prompt agreement & 0.668 & 0.817 & 0.823 & 0.891 & 0.898 & 0.898\\
LLMBar & Binary correctness gap & 0.290 & 0.109 & 0.074 & 0.055 & 0.049 & 0.050\\
LLMBar & Mean-correctness gap & 0.148 & 0.082 & 0.073 & 0.043 & 0.044 & 0.045\\
JudgeBench & Plug-in prompt & 0.066 & 0.061 & 0.074 & 0.023 & 0.028 & 0.031\\
JudgeBench & Corrected prompt & 0.016 & 0.049 & 0.073 & 0.015 & 0.027 & 0.031\\
JudgeBench & Same-cell agreement & 0.586 & 0.903 & 0.991 & 0.941 & 0.991 & 1.000\\
JudgeBench & Across-prompt agreement & 0.471 & 0.583 & 0.596 & 0.830 & 0.845 & 0.848\\
JudgeBench & Binary correctness gap & 0.349 & 0.169 & 0.127 & 0.064 & 0.049 & 0.047\\
JudgeBench & Mean-correctness gap & 0.171 & 0.130 & 0.114 & 0.054 & 0.049 & 0.047\\
\bottomrule
\end{tabular*}
\end{table}

\section{Decision Geometry: Derivations}
\label{sec:decision-proofs}
For any random verdict vector with mean $\bar p$ and fixed target $t$, expanding the square yields $\E\|V-t\|^2=\E\|V-\bar p\|^2+\|\bar p-t\|^2$, because the cross term has mean zero. This proves Equation~\eqref{eq:story-brier}. The following family evaluates discrete accuracy under the original and fair tie rules at fixed gold alignment.

\subsection{Position dependence at fixed positive margin}
We isolate the role of position by holding the mean preference for the gold candidate fixed. Consider a binary family with gold candidate 1. Every prompt chooses candidate 1 with probability $\pi+\delta$ in AB order and $\pi-\delta$ in BA order, where $1/2\le\pi<1$ and $0\le\delta\le1-\pi$. Calls are independent. The order-averaged law and gold margin are
\[
\bar p=(\pi,1-\pi,0,0),\qquad m=2\pi-1.
\]
All prompts have this same averaged law: $\Up^{(q)}=\Ui=0$, while $\Uo=2\delta^2$. Varying $\delta$ isolates position dependence while holding the mean verdict distribution fixed.

For either balanced four-call design, the number of votes for candidate 1 is a sum of independent $\operatorname{Bin}(2,\pi+\delta)$ and $\operatorname{Bin}(2,\pi-\delta)$ variables. Write $h=\pi^2-\delta^2$. This gives the deployment risk, Brier loss, tie probability and strict accuracy:
\begin{align}
\mathcal R_d:=\E\|V_d-\bar p\|^2
 &=\frac{\pi(1-\pi)-\delta^2}{2},\label{eq:positive-margin-risk}\\
B_d:=\E\|V_d-e_y\|^2&=\mathcal R_d+2(1-\pi)^2,\notag\\
T_{22}&=4\pi^2+(2-12\pi)h+6h^2,\label{eq:positive-margin-ties}\\
A_{\rm strict}&=4\pi h-3h^2.\notag
\end{align}
Here strict accuracy assigns zero credit to a tied vote. Fair resolution of the tie gives
\begin{equation}
A_{\rm coin}=A_{\rm strict}+\tfrac12T_{22}
=3\pi^2-2\pi^3+(2\pi-1)\delta^2.
\label{eq:positive-margin-coin}
\end{equation}

\paragraph{Proof of Proposition~\ref{prop:position-margin}.}
\begin{proof}
The four-vote probability generating function is
$[1-\pi-\delta+(\pi+\delta)z]^2[1-\pi+\delta+(\pi-\delta)z]^2$.
Its $z^2$ coefficient gives $T_{22}$; the $z^3$ and $z^4$ coefficients give strict accuracy. Setting $t=\delta^2$,
\[
\frac{dT_{22}}{dt}=12\pi(1-\pi)-2+12t,\quad
\frac{dA_{\rm strict}}{dt}=6(\pi^2-t)-4\pi,\quad
\frac{dA_{\rm coin}}{dt}=2\pi-1.
\]
Their signs and Equation~\eqref{eq:positive-margin-risk} give the claim.
\end{proof}

\begin{proposition}[The zero-margin endpoint]
\label{prop:position-ties}
At $\pi=1/2$, put $u=\Uo=2\delta^2$. Then
\begin{equation}
\mathcal R_d=\frac18-\frac u4,\quad
B_d=\frac58-\frac u4,\quad
T_{22}=\frac38+\frac u2+\frac32u^2,\quad
A_{\rm strict}=\frac{1-T_{22}}2.
\label{eq:position-risk}
\end{equation}
Fair-coin accuracy is $1/2$ throughout this family. At maximal position dependence, every balanced deployment ties and its vector risk is zero.
\end{proposition}
The probability of choosing the first-presented candidate in this special case is $a=1/2+\delta$. Substituting it in the four-vote law gives the stated endpoint formulas.

For example, $\pi=.6$ and $\delta:0\to.4$ keep the gold margin fixed at .2, reduce vector risk from .12 to .04 and Brier loss from .44 to .36, and raise the tie rate from .3456 to .64. Strict accuracy changes from .4752 to .36 while fair accuracy changes from .648 to .68. This family has zero prompt and interaction components, isolating the effects of position dependence and the tie rule.

\subsection{Abstention cost and the full categorical rule}
If choosing a wrong candidate costs one and abstaining costs $c$, the binary family has $L_c=1-A_{\rm strict}-(1-c)T_{22}$. At zero margin, $L_c=1/2+(c-1/2)T_{22}$; the positive-margin example's endpoint laws have equal loss at $c=14/23$. In the full four-category response law, fair resolution of pure candidate 2:2 ties gives the exact identities in Equation~\eqref{eq:story-coin22}; mixed ties, TIE votes and invalid outcomes retain their original handling. The interaction identity holds groupwise for all 256 four-vote patterns and either candidate as gold. Its inferential family is stated in Section~\ref{sec:temperature-group-inference}.

\section{Prompt Policies and Budgeted Reasoning}
\label{sec:policy-decisions}
\subsection{Four Qwen judging policies}
\label{sec:operational-structural}\label{sec:decision-diagnostic-protocol}
The policy study compares four banks of six equally weighted prompts: criterion paraphrases, rubrics, comparison procedures and labeled examples. Each example prompt contains three independently written demonstrations, one per verdict, written without using the test labels. Each bank defines a distinct judging policy. The Qwen checkpoint runs in native nonthinking mode with temperature and top-$p$ set to 1, an 8,192-token context, a 128-token cap, batch size 128 and disabled prefix caching.

The JudgeBench Claude response file supplies 178 distinct questions after excluding 91 normalized matches to the earlier GPT file and one duplicate. These comprise two canaries, 24 development items and 152 test items. Conditions share these items but use independent generation. For each test item and condition, measurement uses $K=6$ prompts, two orders and $R=16$ calls per cell; deployments use $J=64$ independent four-call groups for each D/E design. Eight blocks interleave conditions, measurement calls and deployments. The selector is fixed from development data before test collection. There are 432,736 calls (96 canary, 4,608 development, 428,032 test), of which 311,296 test calls form deployments. The analysis assumes independent calls under stationary cell laws.
\begin{table}[!htbp]\setlength{\tabcolsep}{2.5pt}\renewcommand{\arraystretch}{1.0}
\centering\small
\setlength{\tabcolsep}{2.5pt}
\caption{Qwen judging-policy study. Component estimates pool all 16 measurement repeats. $\bot$ and length-stop percentages use all test calls, including measurement and deployment; their denominator differs from measurement-only validity. Length stops and invalid verdicts are distinct, potentially overlapping outcomes.}
\label{tab:operational-components}
\begin{tabular*}{\linewidth}{@{\extracolsep{\fill}}llrrrrr@{}}
\toprule
Study & Policy / cap & $\widehat U_p$ & $\widehat U_{po}$ & $\widehat U_c$ & $\bot$ (\%) & Length (\%)\\
\midrule
Policy & Paraphrase / 128 & 0.0034 & 0.0089 & 0.3076 & 2.49 & 1.31\\
Policy & Rubric / 128 & 0.0035 & 0.0068 & 0.3273 & 3.88 & 1.66\\
Policy & Comparison / 128 & 0.0086 & 0.0112 & 0.3241 & 4.44 & 2.09\\
Policy & Examples / 128 & 0.0001 & 0.0010 & 0.2627 & 2.02 & 1.09\\
\bottomrule
\end{tabular*}
\end{table}
\begin{table}[!htbp]\setlength{\tabcolsep}{2.5pt}\renewcommand{\arraystretch}{1.0}
\centering\small
\setlength{\tabcolsep}{2.5pt}
\caption{Performance against gold labels in independent four-call deployments. Brier loss uses the averaged verdict vector; error and abstention use the fixed aggregate-verdict rule. Overall error is one minus accuracy and includes abstentions. Both ties and $\bot$ count as abstentions. All deployments, including invalid outputs, contribute. Entries are point estimates, with percentages for error and abstention.}
\label{tab:operational-quality}
\begin{tabular*}{\linewidth}{@{\extracolsep{\fill}}llrrrrrr@{}}
\toprule
Study & Policy / cap & Brier D & Brier E & Error D & Error E & Abst. D & Abst. E\\
\midrule
Policy & Paraphrase / 128 & 0.4662 & 0.4662 & 49.99 & 49.50 & 29.53 & 29.15\\
Policy & Rubric / 128 & 0.4791 & 0.4821 & 51.44 & 50.57 & 31.84 & 29.91\\
Policy & Comparison / 128 & 0.4872 & 0.4811 & 51.18 & 50.11 & 30.54 & 29.33\\
Policy & Examples / 128 & 0.4707 & 0.4718 & 56.35 & 55.93 & 39.34 & 38.80\\
\bottomrule
\end{tabular*}
\end{table}
\begin{table}[!htbp]\setlength{\tabcolsep}{2.5pt}\renewcommand{\arraystretch}{1.0}
\centering\small
\setlength{\tabcolsep}{2.5pt}
\caption{Exploratory decision summaries for all four policies. All entries are percentages. $C$ is coverage, $A^{\rm dec}$ is accuracy among decisive deployments, and $T_{22}$ is the pure candidate 2:2 tie rate. D and E each use four calls. These are point estimates; conditional accuracy uses policy-specific decisive subsets.}
\label{tab:decision-coverage}
\begin{tabular*}{\linewidth}{@{\extracolsep{\fill}}llrrrrrr@{}}
\toprule
Study & Policy / cap & $C_D$ & $C_E$ & $A_D^{\rm dec}$ & $A_E^{\rm dec}$ & $T_{22,D}$ & $T_{22,E}$\\
\midrule
Policy & Paraphrase / 128 & 70.47 & 70.85 & 70.97 & 71.29 & 27.88 & 27.38\\
Policy & Rubric / 128 & 68.16 & 70.09 & 71.24 & 70.53 & 28.47 & 26.64\\
Policy & Comparison / 128 & 69.46 & 70.67 & 70.28 & 70.59 & 26.53 & 25.81\\
Policy & Examples / 128 & 60.66 & 61.20 & 71.95 & 72.00 & 36.85 & 36.07\\
\bottomrule
\end{tabular*}
\end{table}

Relative to paraphrases, examples reduce prompt dispersion by .00325 (pointwise composition interval [$-.00463$,$-.00186$]) and interaction by .00792 ([$-.00977$,$-.00618$]); their corrected prompt estimate is .00014. At the same time, the order component rises from .09784 to .16677 (difference .06893, pointwise paired composition interval [.05662,.08121]), and pure ties increase. Under D, Brier loss with examples is .47069 versus about .46622 for paraphrases, alongside the decreases in prompt and interaction estimates.

\paragraph{Voting-rule and cost sensitivity.}\label{sec:story-sensitivity-details}
We compare rules by rescoring the same complete vote groups under each rule, using 20,000 paired whole-item resamples within the four original strata. The maximum absolute centered bootstrap error over twelve examples-minus-paraphrase contrasts---D/E, original/fair rules and accuracy/coverage/wrong-candidate frequency---gives approximate simultaneous 95\% composition intervals, with common half-width .02631579. All four policies enter the analysis; the examples-minus-paraphrase comparison is exploratory.
\begin{table}[!htbp]\setlength{\tabcolsep}{2.5pt}\renewcommand{\arraystretch}{1.0}
\centering\small
\caption{Examples minus paraphrasing under two decision rules. Changes and approximate simultaneous 95\% intervals are in percentage points. The twelve-comparison family includes accuracy, coverage, and wrong-choice frequency for both rules and designs.}
\label{tab:story-rescoring}
\begin{tabular*}{\linewidth}{@{\extracolsep{\fill}}llrr@{}}
\toprule
Design & Rule & Accuracy change & Coverage change \\
\midrule
D & Original & -6.36 [-8.99, -3.73] & -9.81 [-12.44, -7.18] \\
D & Fair pure 2:2 & -1.88 [-4.51, 0.76] & -0.83 [-3.46, 1.80] \\
E & Original & -6.44 [-9.07, -3.80] & -9.64 [-12.27, -7.01] \\
E & Fair pure 2:2 & -2.09 [-4.72, 0.54] & -0.96 [-3.59, 1.68] \\
\bottomrule
\end{tabular*}
\end{table}

With wrong-decision cost one and abstention cost $c$, $\Delta L(c)=(1-c)\Delta W-c\Delta A$. The same endpoint family gives a band for every $c\in[0,1]$. Under the original rule, point estimates of the crossover cost are .351153 (D) and .332623 (E), with inverted sets [.082809,.619497] and [.059701,.605544]. Under fair resolution, the example policy has a larger point estimate of loss throughout $[0,1]$; Figure~\ref{fig:recovery-policy-rules} displays the composition band. Fair rescoring leaves vote-vector risk and Brier unchanged. These composition intervals describe the policy-study items; the temperature study uses a separate 40-item panel.

\subsection{Gemma reasoning and fixed examples}
\label{sec:gemma-fourarm-details}\label{sec:rewardbench2-pairing}
The four arms are G0 nonthinking/base, G1 nonthinking/examples, G2 thinking/base and G3 thinking/examples, using a fixed \texttt{google/gemma-4-31B-it} checkpoint. The 32 new RewardBench~2 questions include eight each from Factuality, Focus, Math and Precise IF. Each chosen response is paired with one of its three rejected responses using a fixed pseudorandom rule, with a separately preassigned canonical orientation. The evaluation target is the chosen-versus-selected-rejected pair. Normalized equality and five-word shingle checks exclude overlap with earlier panels and demonstrations. Ranking within each domain, without using outcomes, assigns reserve positions 0--7. The pair mapping records source IDs, rejected-response indices, gold labels and reserve positions.

All arms use $K=48$ prompts, two orders, $R=2$ calls per cell and $J=8$ separate D groups per item, with prompt pairs matched across arms. The study therefore contains 24,576 measurement judgments and 4,096 deployment judgments, for 28,672 formal observations. The same three demonstrations are appended to every example prompt: 4 versus 5 for $2+2$ (A), \texttt{SUN} versus \texttt{sun} for uppercase (B), and two answers equal to 3 (TIE). The demonstration labels are independent of the evaluation-item gold labels.

All arms use xgrammar-constrained final A/B/TIE generation with an eight-token cap. The thinking stage uses the native 1,024-token budget and 1,032-token request cap, stopping at the first channel-end token. Final generation resumes from the complete returned thought without a new instruction. Nonthinking uses the native empty-thought prefix. Both stages use $T=1$, top-$p=.95$, top-$k=64$, bfloat16, vLLM V1, 16,384 context, at most 32 concurrent sequences, batches of 64 logical judgments and no speculative decoding. The shared constrained decoder renormalizes final-token probabilities. The protocol codes malformed thoughts or final responses as invalid. All formal judgments and the 64 separate canary judgments have valid, untruncated final responses; native budget closure occurs in 23.63\% of G2 and 21.88\% of G3 thoughts. Data collection spans two runtime periods, preserving the originally assigned cell conditions and seeds. The analysis assumes independent calls under stationary cell laws across periods.

For $J$ deployment vectors, $S^2=(J-1)^{-1}\sum_j\|V_j-\bar V\|^2$ estimates vector risk and $B=J^{-1}\sum_j\|V_j-e_y\|^2$ estimates Brier loss. The raw gold-alignment estimate is $B-S^2=\|\bar V-e_y\|^2-S^2/J$. All arm means weight items equally. Conditional accuracy divides the pooled number of correct deployments by the pooled number of decisive deployments.
\begin{table*}[!htbp]\setlength{\tabcolsep}{2.5pt}\renewcommand{\arraystretch}{1.0}
\centering\small
\setlength{\tabcolsep}{2.5pt}
\caption{Descriptive Gemma group means on 32 RewardBench~2 items.
G0/G1 use nonthinking; G2/G3 use budgeted thinking. G0/G2 use the base
instruction; G1/G3 add fixed examples. Components, vector risk, and
gold losses are in squared one-hot-vector units. Conditional accuracy
is the pooled correct-to-decisive ratio. Operational diagnostics cover
all 7,168 formal judgments per group; deployment outcomes use the
256 independent deployments per group, clustered within 32 items.}
\label{tab:gemma-fourarm-descriptive}
\begin{tabular*}{\linewidth}{@{\extracolsep{\fill}}lrrrr@{}}
\toprule
Metric & G0 & G1 & G2 & G3 \\
\midrule
Prompt component & 0.013226 & 0.015025 & 0.002484 & 0.000554 \\
Order component & 0.060125 & 0.048960 & 0.047931 & 0.051929 \\
Interaction component & 0.011164 & 0.013865 & 0.000653 & 0.000480 \\
Call component & 0.013021 & 0.014323 & 0.092773 & 0.091797 \\
\midrule
Pure 2:2 tie (\%) & 12.11 & 4.69 & 5.47 & 5.47 \\
Decisive coverage (\%) & 86.33 & 91.80 & 91.41 & 90.62 \\
Unconditional accuracy (\%) & 75.39 & 76.95 & 80.08 & 80.47 \\
Conditional decisive accuracy (\%) & 87.33 & 83.83 & 87.61 & 88.79 \\
\midrule
Observed vector risk & 0.012137 & 0.013323 & 0.023926 & 0.022042 \\
Predicted vector risk & 0.009727 & 0.010933 & 0.024409 & 0.023221 \\
Gold Brier loss & 0.284668 & 0.314941 & 0.247070 & 0.231445 \\
Estimated squared gold-alignment error & 0.272531 & 0.301618 & 0.223145 & 0.209403 \\
\midrule
First-position choice (\%) & 55.52 & 53.11 & 55.08 & 55.78 \\
Forced thought end (\%) & 0.00 & 0.00 & 23.63 & 21.88 \\
Invalid judgments & 0 & 0 & 0 & 0 \\
\bottomrule
\end{tabular*}
\end{table*}

\Needspace{5\baselineskip}
\paragraph{Tests specified before data collection.}\label{sec:gemma-fourarm-inference}\label{sec:gemma-interval-construction}
The primary family includes G1--G0 contrasts for order, pure-2:2 ties and coverage; the secondary family includes these endpoints for G2--G0 and the interaction $(G3-G2)-(G1-G0)$. Each of twenty thousand whole-item resamples draws eight items within each domain, preserving all four arms. The centered two-sided value is $p_c=[1+\sum_b\1\{|\Delta_b^*-\widehat\Delta|\ge|\widehat\Delta|\}]/20001$; Holm adjustment is applied separately within the three- and six-test families. We report percentile intervals at 98.33\% and 99.17\% marginal levels alongside the separately constructed centered Holm test. Inference uses the .05 threshold. The primary tie contrast has $p_H=.12944$ and a centered absolute-error interval of [$-16.406$,1.562] pp. Inference describes approximate item-composition uncertainty conditional on the recorded calls.
\begin{table*}[!htbp]\setlength{\tabcolsep}{2.5pt}\renewcommand{\arraystretch}{1.0}
\centering\small
\setlength{\tabcolsep}{1.75pt}
\caption{Prespecified Gemma contrasts. Paired whole-item bootstrap resampling
uses 20,000 draws within the four domains, with equal domain weights.
The displayed percentile intervals have nominal marginal levels of 98.33\%
for the three primary endpoints and 99.17\% for the six secondary endpoints.
$p_c$ denotes the two-sided centered-bootstrap test; $p_H$ applies Holm
adjustment within each of the two prespecified families. Inference at
$\alpha=.05$ follows $p_H$. Rate differences are percentage points (pp).}
\label{tab:gemma-fourarm-contrasts}
\begin{tabular*}{\linewidth}{@{\extracolsep{\fill}}lllrrrr@{}}
\toprule
Family & Contrast & Endpoint & Change & Percentile interval & $p_c$ & $p_H$ \\
\midrule
Primary & $G1-G0$ & Order & -0.011164 & $[-0.037802,\,0.008439]$ & 0.25284 & 0.25284 \\
Primary & $G1-G0$ & Pure 2:2 (pp) & -7.422 & $[-17.578,\,-0.391]$ & 0.04315 & 0.12944 \\
Primary & $G1-G0$ & Coverage (pp) & +5.469 & $[-0.391,\,14.844]$ & 0.09290 & 0.18579 \\
\midrule
Secondary & $G2-G0$ & Order & -0.012193 & $[-0.084644,\,0.049228]$ & 0.64282 & 0.75546 \\
Secondary & $G2-G0$ & Pure 2:2 (pp) & -6.641 & $[-21.484,\,6.250]$ & 0.22739 & 0.75156 \\
Secondary & $G2-G0$ & Coverage (pp) & +5.078 & $[-8.203,\,20.703]$ & 0.37773 & 0.75546 \\
\midrule
Secondary & $(G3-G2)-(G1-G0)$ & Order & +0.015162 & $[-0.012049,\,0.048321]$ & 0.18789 & 0.75156 \\
Secondary & $(G3-G2)-(G1-G0)$ & Pure 2:2 (pp) & +7.422 & $[-1.562,\,19.531]$ & 0.07265 & 0.43588 \\
Secondary & $(G3-G2)-(G1-G0)$ & Coverage (pp) & -6.250 & $[-17.450,\,2.344]$ & 0.10299 & 0.51497 \\
\bottomrule
\end{tabular*}
\par\smallskip
\begin{minipage}{.98\textwidth}\small
Inference uses the .05 threshold. The primary pure-2:2 contrast has
Holm-adjusted centered-bootstrap value $p_H=.12944$. Percentile intervals and
centered tests use separate constructions.
The two families contain three and six tests, respectively. Their separate
adjustments provide inference within each family.
\end{minipage}
\end{table*}

\paragraph{Exploratory thinking contrasts.}\label{sec:gemma-exploratory}
Twelve subsequent comparisons examine prompt, interaction, call, Brier, vector risk and gold alignment for G2--G0 and G3--G1. They use 100,000 stratified paired whole-item resamples, preserving all arms and equal domain weights. The radius of a centered interval is the higher empirical $(1-\alpha)$ quantile of $|\Delta_b^*-\widehat\Delta|$, with $\alpha=.05$ or $.05/12$. The latter's 99.583\% marginal levels target approximate 95\% family coverage. Table~\ref{tab:gemma-exploratory} reports all paired contrasts and both interval families.
\begin{table}[!htbp]\setlength{\tabcolsep}{2.5pt}\renewcommand{\arraystretch}{1.0}
\centering\small
\setlength{\tabcolsep}{2.5pt}
\caption{Exploratory thinking-minus-nonthinking contrasts on the complete Gemma panel.
Intervals invert centered absolute bootstrap errors using 100,000 paired whole-item
resamples, eight per domain. The outer intervals use marginal coverage 99.583\%
(Bonferroni across all 12 endpoints; approximate familywise target 95\%).
These intervals describe item-composition uncertainty conditional on observed calls.
Gold alignment is estimated by Brier minus the unbiased deployment covariance trace.}
\label{tab:gemma-exploratory}
\begin{tabular*}{\linewidth}{@{\extracolsep{\fill}}llrrr@{}}
\toprule
Instruction & Endpoint & Change & 95\% interval & 12-endpoint interval \\
\midrule
Base & Prompt & -0.01074 & $[-0.02313,\,0.00164]$ & $[-0.02932,\,0.00783]$ \\
 & Interaction & -0.01051 & $[-0.02053,\,-0.00049]$ & $[-0.02568,\,0.00466]$ \\
 & Call & +0.07975 & $[0.03158,\,0.12793]$ & $[0.00879,\,0.15072]$ \\
 & Gold Brier & -0.03760 & $[-0.20215,\,0.12695]$ & $[-0.27979,\,0.20459]$ \\
 & Deployment risk & +0.01179 & $[-0.00607,\,0.02965]$ & $[-0.01514,\,0.03871]$ \\
 & Gold-alignment error & -0.04939 & $[-0.21017,\,0.11140]$ & $[-0.28530,\,0.18652]$ \\
\midrule
Examples & Prompt & -0.01447 & $[-0.02444,\,-0.00450]$ & $[-0.02915,\,0.00021]$ \\
 & Interaction & -0.01339 & $[-0.02263,\,-0.00415]$ & $[-0.02719,\,0.00042]$ \\
 & Call & +0.07747 & $[0.03060,\,0.12435]$ & $[0.00846,\,0.14648]$ \\
 & Gold Brier & -0.08350 & $[-0.23535,\,0.06836]$ & $[-0.30762,\,0.14062]$ \\
 & Deployment risk & +0.00872 & $[-0.00635,\,0.02379]$ & $[-0.01346,\,0.03090]$ \\
 & Gold-alignment error & -0.09222 & $[-0.24163,\,0.05720]$ & $[-0.31264,\,0.12821]$ \\
\bottomrule
\end{tabular*}
\end{table}
\FloatBarrier

\section{Balanced Deployments and Design Selection}
\label{sec:sampling-designs}\label{sec:design-choice}
This secondary analysis examines estimation of the mean response law alongside discrete decision quality. All expectations are conditional on the fixed response laws for each item and cell. Calls are mutually independent, and prompt and order randomization is independent of the calls.
Fix $K\ge4$ prompts, two equally weighted orders, and the census
target $\bar p=(2K)^{-1}\sum_{k,o}P_{ko}$. Each design returns the
mean of four one-hot verdicts. Design $D$ samples two distinct prompts
uniformly and queries both orders for each. Design $E$ samples four
distinct prompts uniformly, then assigns two to AB and two to BA
uniformly. Design randomization is independent of the calls. For
$d\in\{D,E\}$, define
$\mathcal R_d=\E\|V_d-\bar p\|_2^2$.

\begin{proposition}[Risk of balanced four-call designs]
\label{prop:design-risk}
For arbitrary fixed categorical cell laws and independent calls,
both designs are unbiased for $\bar p$, and
\begin{align}
\mathcal R_D
&=\frac{\Uc}{4}+\frac{K-2}{2(K-1)}\Up^{(q)},\label{eq:design-risk-d}\\
\mathcal R_E
&=\frac{\Uc}{4}+\frac{K-4}{4(K-1)}\Up^{(q)}
+\frac{K}{4(K-1)}\Ui.\label{eq:design-risk-e}
\end{align}
Thus $\Delta:=\mathcal R_E-\mathcal R_D
=\gamma(\Ui-\Up^{(q)})$, where $\gamma=K/[4(K-1)]$.
\end{proposition}

For $K=6$, $\Delta=0.3(\Ui-\Up^{(q)})$: covering more prompts
helps when prompt variation exceeds the interaction contribution.
This comparison targets estimation of the shared response-law mean;
Section~\ref{sec:decision-geometry} relates that target to discrete decisions.

\paragraph{Estimating the design contrast directly.}
Write $A_k=P_{k,\mathrm{AB}}$, $B_k=P_{k,\mathrm{BA}}$, and let bars
denote averages over prompts. From $R\ge1$ calls per cell, form
$\widehat A_k$, $\widehat B_k$ and
\begin{equation}
\widehat\Delta_R=-\gamma\left(
\frac1K\sum_k\widehat A_k^\top\widehat B_k
-\overline{\widehat A}^{\top}\overline{\widehat B}\right).
\label{eq:design-contrast}
\end{equation}

\begin{proposition}[Cancellation of same-cell terms]
\label{prop:design-contrast}
Equation~\eqref{eq:design-contrast} is unbiased for $\Delta$ for every
$R\ge1$. For $R\ge2$, it also equals
\begin{align}
\widehat\Delta_R
&=\gamma(\widehat\Ui-\widehat\Up^{(q)})\notag\\
&=\gamma(\widehat U_{\rm plug,int}
-\widehat U_{\rm plug,prompt}^{(q)})
\label{eq:design-cancellation}
\end{align}
for every sample, where the plug-in components substitute empirical
cell distributions in the population formulas.
\end{proposition}

\subsection{Proofs of the balanced-design identities}
\label{sec:design-proofs}
Set $F_k=(A_k+B_k)/2$, $G_k=(A_k-B_k)/2$, $f_k=F_k-\bar F$ and $g_k=G_k-\bar G$. Then $\bar p=\bar F$, $U_p^{(q)}=K^{-1}\sum_k\|f_k\|^2$ and $U_{\rm int}=K^{-1}\sum_k\|g_k\|^2$. For centered vectors $x_k$ and a uniform size-$m$ subset $S_m$,
\[
\E\left\|m^{-1}\sum_{k\in S_m}x_k\right\|^2
=\frac{K-m}{m(K-1)}\frac1K\sum_k\|x_k\|^2.
\]
This follows from the inclusion probabilities $m/K$ and $m(m-1)/[K(K-1)]$, together with $\sum_{k\ne\ell}x_k^\top x_\ell=-\sum_k\|x_k\|^2$. Design D has conditional mean $(F_j+F_k)/2$, so setting $m=2$ gives its prompt term. Each cell is selected with probability $2/K$ and weighted $1/4$, giving expected call variance $(2/K)\sum_{ko}(1-\|P_{ko}\|^2)/16=U_c/4$. The expected weight of each cell is $1/(2K)$, which proves unbiasedness.

For E, the centered conditional mean is $\frac14\sum_{k\in S_4}f_k+\frac14\sum_{k\in S_4}\varepsilon_k g_k$, with exactly two positive signs. Symmetry of the conditional sign distribution removes the cross term; the first term contributes $(K-4)U_p^{(q)}/[4(K-1)]$. Distinct selected signs satisfy $\E\varepsilon_k\varepsilon_\ell=-1/3$, so the second term has expected squared norm
\[
\frac1{4K}\sum_k\|g_k\|^2-
\frac1{4K(K-1)}\sum_{k\ne\ell}g_k^\top g_\ell
=\frac{K}{4(K-1)}U_{\rm int}.
\]
The order main effect cancels because the signs sum to zero. Cell inclusion probabilities and call variance are the same as in D, proving Proposition~\ref{prop:design-risk}.

Finally, $U_{\rm int}-U_p^{(q)}=-K^{-1}\sum_k(A_k-\bar A)^\top(B_k-\bar B)$. The AB and BA empirical means come from distinct cells, so every product is unbiased even at $R=1$. This algebraic identity also holds for empirical vectors. The coefficient of every same-cell squared norm in $H_{XO}-2H_X-H_O+2H_0$ is $1/(2K)-2/(4K)-1/(2K^2)+2/(4K^2)=0$, so complete-pair correction leaves the contrast unchanged. This proves Proposition~\ref{prop:design-contrast}.

Both designs have the same mean verdict distribution, so Equation~\eqref{eq:story-brier} gives the same expected Brier difference $\Delta$. Categorical accuracy depends on the full vote law, the decision rule and the gold label.

\begin{proposition}[Winner preservation at positive census margin]
\label{prop:design-margin}
Suppose $\bar p$ has a unique winning category $c^\star$, and define its
winner--runner-up gap over all verdict categories by
$\kappa=\bar p_{c^\star}-\max_{j\ne c^\star}\bar p_j>0$.
For either design $d\in\{D,E\}$,
\begin{equation}
\Prob\!\left(V_{d,c^\star}\le\max_{j\ne c^\star}V_{d,j}\right)
\le\min\left\{1,\frac{2\mathcal R_d}{\kappa^2}\right\}.
\label{eq:design-margin}
\end{equation}
The event includes tied maxima as failures to preserve the unique winner.
If $c^\star$ is the gold candidate, the same bound controls the original
plurality rule's gold error, counting abstentions as failures.
\end{proposition}

\begin{proof}
On the displayed event, some $j\ne c^\star$ satisfies
$V_{d,j}\ge V_{d,c^\star}$. With $e=V_d-\bar p$, this gives
$e_j-e_{c^\star}\ge\bar p_{c^\star}-\bar p_j\ge\kappa$.
Cauchy--Schwarz implies $\|e\|^2\ge\kappa^2/2$.
Markov's inequality and $\E\|e\|^2=\mathcal R_d$ prove the bound.
When $c^\star$ is gold, the original rule returns gold exactly when
$V_{d,c^\star}$ is the unique maximum, giving the final claim.
\end{proof}

\subsection{Independent deployment checks}
\label{sec:new-judgebench-protocol}\label{sec:new-judgebench}
The independent study using the JudgeBench GPT response file fixes a stratified split of 100 development items, 200 test items and four canary items (development/test counts: 16/32 math, 28/56 reasoning, 12/24 code, 44/88 MMLU-Pro). Qwen runs in native nonthinking mode with BF16, $T=$top-$p=1$, an eight-token cap, an 8,192-token context, batch size 16 and disabled prefix caching. Development uses six prompts, two orders and $R=4$; test uses $R=8$ and $J=16$ separate groups per design. Before test calls are collected, development data alone select E when the estimated $E-D$ risk is negative and D otherwise. Stages run sequentially, under the assumption of stationary cell laws. All 49,600 formal calls are retained, including 4,973 invalid outcomes and 4,219 length stops, which can overlap; 48 canary calls are separate.

The development estimate $+.00082$ selects D; its 95\% fixed-call Chebyshev interval is [$-.02449$,.02613]. On the test set, the predicted contrast is $-.00114$ and the observed contrast is $-.00322$ (D risk .09385, E .09063), with composition band [$-.00833$,.00201] for the observed contrast. The gold Brier difference is $+.00059$ [$-.01359$,.01484]. The point estimate changes sign between development and test. For the prespecified prompt tolerance .02, the original Cantelli upper bound is .02014; the sharper covariance bound gives .01830 in a secondary analysis.

The semantic study fixes two 24-prompt pools before data collection. Development data select D/E before testing on the unseen pool B. For each open model and dataset, two interleaved panels each use one measurement call per cell and eight separate deployment groups per design and item on 60 fixed test items. The variance estimate based on panel differences is unbiased. Calibration intervals use an approximate Student-$t$ reference with Bonferroni allocation across six endpoints; equivalence tests use the prespecified calibration margin below.
\begin{table}[!htbp]\setlength{\tabcolsep}{2.5pt}\renewcommand{\arraystretch}{1.0}
\centering\small
\setlength{\tabcolsep}{2.5pt}
\caption{Unseen-B calibration. $\widehat\Delta$ predicts $\mathrm{MSE}_E-\mathrm{MSE}_D$ and $\widetilde\Delta$ estimates it from independent deployments. The error is $\widehat\Delta-\widetilde\Delta$. Intervals target nominal 95\% coverage over six endpoints under an approximate $t$ reference. $p_H$ is the Holm-adjusted secondary TOST value for the prespecified $\pm.005$ margin.}
\label{tab:semantic-calibration}
\begin{tabular*}{\linewidth}{@{\extracolsep{\fill}}llrrrrr@{}}
\toprule
Model & Dataset & $\widehat\Delta$ & $\widetilde\Delta$ & Error & Adjusted interval & $p_H$\\
\midrule
Qwen3.6 & LLMBar & -0.0010 & -0.0062 & 0.0052 & $[-0.0086, 0.0189]$ & 0.849\\
Qwen3.6 & JudgeBench & -0.0017 & 0.0011 & -0.0028 & $[-0.0206, 0.0150]$ & 0.849\\
Gemma4 & LLMBar & -0.0015 & -0.0025 & 0.0010 & $[-0.0055, 0.0075]$ & 0.244\\
Gemma4 & JudgeBench & -0.0005 & -0.0019 & 0.0014 & $[-0.0038, 0.0066]$ & 0.186\\
GPT-OSS & LLMBar & -0.0003 & 0.0001 & -0.0005 & $[-0.0105, 0.0095]$ & 0.443\\
GPT-OSS & JudgeBench & -0.0031 & -0.0007 & -0.0024 & $[-0.0146, 0.0097]$ & 0.849\\
\bottomrule
\end{tabular*}
\end{table}

For the prespecified $\pm.005$ calibration margin, the minimum Holm-adjusted TOST value is $p_H=.186$. Table~\ref{tab:semantic-calibration} gives each model--dataset pair's calibration error and interval.

The four-policy study and a separate output-cap study (six wrappers/paraphrases crossed with caps 8/128 on the same 152-item design) each provide eight independent blocks per condition. These estimate the difference between the predicted and observed deployment-risk contrasts. Each block has two measurement repeats per cell and eight groups per design and item. Approximate $t_7$ intervals and TOST use the prospectively fixed $\pm.005$ calibration margin, with Holm correction over four conditions within each study. Table~\ref{tab:operational-calibration} gives each condition's calibration error and equivalence-test value; Table~\ref{tab:observed-variation-intervals} gives the separate D/E risk-difference intervals. Raw block predictions have lower RMSE against the reference than a zero prediction in two of eight conditions; the reference contains independent deployment noise.
\begin{table}[!htbp]\setlength{\tabcolsep}{2.5pt}\renewcommand{\arraystretch}{1.0}
\centering\small
\setlength{\tabcolsep}{2.5pt}
\caption{Calibration using independent blocks. Error is the predicted minus observed risk contrast for estimating the mean response law. The 95\% intervals use a marginal approximate $t$ reference. $p_H$ adjusts the equivalence tests across each study's four conditions; the margin is $\pm.005$.}
\label{tab:operational-calibration}
\begin{tabular*}{\linewidth}{@{\extracolsep{\fill}}llrrr@{}}
\toprule
Study & Policy / cap & Block error & Marginal 95\% interval & $p_H$\\
\midrule
Cap & Format / 8 & 0.00260 & $[-0.00240, 0.00760]$ & 0.1469\\
Cap & Format / 128 & 0.00171 & $[-0.00156, 0.00498]$ & 0.0980\\
Cap & Paraphrase / 8 & 0.00169 & $[-0.00202, 0.00541]$ & 0.1098\\
Cap & Paraphrase / 128 & 0.00154 & $[-0.00252, 0.00559]$ & 0.1098\\
\midrule
Policy & Paraphrase / 128 & -0.00021 & $[-0.00575, 0.00532]$ & 0.0790\\
Policy & Rubric / 128 & -0.00122 & $[-0.00506, 0.00261]$ & 0.0790\\
Policy & Comparison / 128 & 0.00001 & $[-0.00514, 0.00517]$ & 0.0790\\
Policy & Examples / 128 & -0.00123 & $[-0.00341, 0.00095]$ & 0.0092\\
\bottomrule
\end{tabular*}
\end{table}

\section{Finite-Call Design Comparisons and Fixed-Budget Allocation}
\label{sec:design-extensions}
The complete-pair estimator measures individual response components. For some contrasts used to compare deployment designs, the same-cell quadratic terms cancel. This section shows when that cancellation allows unbiased estimation with one call per cell and derives the precision of the comparison. It then examines how to allocate calls between more items and more repeats while holding the benchmark target fixed. The results assume independent categorical calls under fixed cell laws, with deployment allocation randomized independently of the outcomes.

\subsection{When a single call per cell is enough}
\label{sec:design-collision}
A measurement grid and a deployment design play different roles. The grid collects observations from each cell; a deployment allocates a fixed number of calls to produce an average verdict. When measurement covers the full grid, the expected number of deployment call pairs within each cell determines whether estimating a risk contrast requires repeated observations.

\begin{proposition}[Equal collision moments]
\label{prop:design-collisions}
A deployment design $d$ first draws counts $n_1,\ldots,n_L$ independently
of all verdicts, with $\sum_a n_a=B$ and $\E_d n_a/B=w_a$. It then
averages $B$ independent categorical calls to obtain $V_d$, targeting
$\bar p=\sum_a w_ap_a$. Its squared-error risk is
\begin{equation}
\mathcal R_d=\frac1B+
\sum_{a,b}\left\{
\frac{\E_d[n_a(n_b-\1\{a=b\})]}{B^2}-w_aw_b
\right\}p_a^\top p_b.
\label{eq:design-collision-risk}
\end{equation}
For arbitrary cell laws on a $C\ge2$ categorical simplex, two designs
with the same $B$ and $w$ have a risk difference that admits an unbiased one-call-per-cell full-grid estimate if and only if
\begin{equation}
\E_{d_1}[n_a(n_a-1)]=\E_{d_0}[n_a(n_a-1)]
\quad\text{for every }a.
\label{eq:design-collision-condition}
\end{equation}
In particular, the condition holds if neither design repeats a cell.
\end{proposition}

The condition compares the expected counts of ordered deployment call pairs within each cell. Designs D and E both have zero such pairs, explaining the cancellation in Proposition~\ref{prop:design-contrast}. Classical methods for unbiased polynomial estimation and U-statistics \citep{hoeffding1948class,kimpara2023proper} identify how many measurement calls each deployment design requires.

\begin{proposition}[Scalar-coefficient quadratic targets]
\label{prop:design-quadratic}
Let $p_1,\ldots,p_L$ range independently over a categorical simplex
with $C\ge2$, and let $W$ be a fixed symmetric real $L\times L$
matrix. For
\[
T_W(p)=\sum_{a,b=1}^L W_{ab}p_a^\top p_b,
\]
an unbiased estimator based on one independent call from each cell
exists over the whole parameter family if and only if $W_{aa}=0$
for every $a$.
\end{proposition}

\begin{proof}
If the diagonal is zero, use
$\sum_{a\ne b}W_{ab}e_{Z_a}^\top e_{Z_b}$ and independence.
Conversely, the expectation of any statistic of one categorical
observation per cell is affine in each individual $p_a$ with the
other laws fixed. Holding all other cells fixed and varying $p_a$
along an interior simplex segment with nonzero tangent $u$, the
second derivative of $T_W$ is $2W_{aa}\|u\|^2$. It must vanish,
so $W_{aa}=0$. Averaging over independent auxiliary randomization preserves
this affine dependence.
\end{proof}

If a specified cell has a nonzero diagonal coefficient, two calls
from that cell are sufficient for its quadratic term and one is
insufficient. In particular, the coefficients of each same-cell
squared norm in both $\Up^{(q)}$ and $\Ui$ equal
$(K-1)/(4K^2)>0$. Their separate unbiased full-grid estimates
therefore require two calls in every one of the $2K$ cells,
whereas their difference has zero diagonal coefficients.
This proposition uses scalar coefficients multiplying
$p_a^\top p_b$. For general category-specific matrix blocks, a
nonzero same-cell block can be affine on the simplex; the relevant
condition is vanishing quadratic curvature on its tangent space.

\paragraph{Proof of Proposition~\ref{prop:design-collisions}.}

\begin{proof}
Unbiasedness follows from the expected counts. Conditional on these
counts, expanding the squared norm of the mean gives $B$ same-call
terms, each equal to one; $n_a(n_a-1)$ ordered distinct-call terms
within cell $a$, each with expectation $\|p_a\|^2$; and $n_an_b$
terms between distinct cells, each with expectation $p_a^\top p_b$.
Divide by $B^2$, average over the design, and subtract
$\|\bar p\|^2$ to obtain Equation~\eqref{eq:design-collision-risk}.
In a risk difference, $1/B$ and the target term cancel. Its diagonal
coefficients vanish exactly under
Equation~\eqref{eq:design-collision-condition}; apply
Proposition~\ref{prop:design-quadratic}.
\end{proof}

This condition also explains why a design that repeats one cell can
require repeated measurement calls when compared with a design that never
does. It depends on the expected number of within-cell ordered call pairs.

\paragraph{Full-grid and sparse measurement designs.}
The necessary condition applies when measurement uses a fixed full grid. Alternatively, a randomized sparse estimator observes a subset of cells and accounts for their known sampling probabilities, as the following constructions show.

\paragraph{A two-call randomized pair.}
Draw a fair sign $H\in\{-1,+1\}$. When $H=+1$, choose $J$ uniformly
and take one AB and one BA call at prompt $J$; let their one-hot
vectors be $Y_A,Y_B$. When $H=-1$, choose an ordered pair $(J,L)$
of distinct prompts uniformly, and call AB at $J$ and BA at $L$.
Then
\begin{equation}
\widehat\Delta_{\rm pair}=-\frac12 H\,Y_A^\top Y_B
\label{eq:design-pair}
\end{equation}
is unbiased. In fact, with
$h_{\rm same}=K^{-1}\sum_k A_k^\top B_k$ and
$h_{\rm different}=[K(K-1)]^{-1}\sum_{k\ne\ell}A_k^\top B_\ell$,
\[
\Delta=-\frac14(h_{\rm same}-h_{\rm different}),
\]
because $\gamma(K-1)/K=1/4$. Taking the expectation in
Equation~\eqref{eq:design-pair} gives this expression.

\paragraph{A three-call shared anchor.}
Choose distinct $(J,L)$ uniformly, then draw independent verdicts
$Y_A$ from AB at $J$, $Y_B$ from BA at $J$, and $Y'_B$ from BA
at $L$. The estimator
\begin{equation}
\widehat\Delta_{\rm anchor}
=-\frac14Y_A^\top(Y_B-Y'_B)
\label{eq:design-anchor}
\end{equation}
has expectation $-(h_{\rm same}-h_{\rm different})/4=\Delta$.
Interchanging AB and BA preserves this result, so independently choosing
between the two order assignments with equal probability can balance order exposure. The pair statistic
lies in $[-1/2,1/2]$ and the anchor statistic in $[-1/4,1/4]$;
their variances are consequently at most $1/4$ and $1/16$,
respectively. These bounds apply to each statistic. To compare efficiency at equal cost, we must account for the variance over the full measurement budget.

More generally, any fixed quadratic target $T_W$ can be estimated by
randomly choosing an index pair with known positive probability and
using its inverse-probability-weighted two-call agreement. If the
chosen indices coincide, the two calls are distinct and
independent. These sparse estimators use fewer calls and cover fewer cells in each estimate.
Their efficiency can be compared by variance or decision error at the same budget.

\subsection{Exact variance and reliable comparison of D and E}
Retain $\gamma=K/[4(K-1)]$, $A_k=P_{k,\mathrm{AB}}$ and $B_k=P_{k,\mathrm{BA}}$ from Section~\ref{sec:sampling-designs}. Write $\widetilde A_k=A_k-\bar A$ and $\widetilde B_k=B_k-\bar B$. The cross-order estimator in Equation~\eqref{eq:design-contrast} exists at $R=1$; its precision depends on the size of the effects and the quadratic call-noise term.

Let
\[
\Sigma_{Ak}=\operatorname{diag}(A_k)-A_kA_k^\top,
\qquad
\Sigma_{Bk}=\operatorname{diag}(B_k)-B_kB_k^\top,
\]
and define the $K\times K$ matrix
$M=I_K/K-\1\1^\top/K^2$.

\begin{proposition}[Exact variance of the design contrast]
\label{prop:design-exact-variance}
For every common repeat count $R\ge1$,
\begin{align}
\Var(\widehat\Delta_R)
={}&\frac{\gamma^2}{K^2R}\sum_k
\widetilde B_k^\top\Sigma_{Ak}\widetilde B_k\notag\\
&+\frac{\gamma^2}{K^2R}\sum_k
\widetilde A_k^\top\Sigma_{Bk}\widetilde A_k\notag\\
&+\frac{\gamma^2}{R^2}\sum_{k,\ell}
M_{k\ell}^2\operatorname{tr}(\Sigma_{Ak}\Sigma_{B\ell}).
\label{eq:design-exact-variance}
\end{align}
In particular,
\begin{equation}
\Var(\widehat\Delta_R)
\le\frac{\gamma^2(\Up^{(q)}+\Ui)}{KR}
+\frac{\gamma^2(K-1)}{4K^2R^2}.
\label{eq:design-item-bound}
\end{equation}
\end{proposition}

\begin{proof}
Set $\xi_k=\widehat A_k-A_k$ and $\eta_k=\widehat B_k-B_k$.
The errors are independent and centered, with covariances
$\Sigma_{Ak}/R$ and $\Sigma_{Bk}/R$. Dividing the estimation error
by $-\gamma$ gives the sum of
\[
L_A=\frac1K\sum_k\xi_k^\top\widetilde B_k,
\quad
L_B=\frac1K\sum_k\widetilde A_k^\top\eta_k,
\quad
Q=\sum_{k,\ell}M_{k\ell}\xi_k^\top\eta_\ell.
\]
The three terms are pairwise uncorrelated: independence separates
$L_A$ and $L_B$; conditioning on all $\xi$ makes the covariance
of $L_A$ with $Q$ zero, and conditioning on all $\eta$ does the
same for $L_B$. Distinct pairs in the variance of $Q$ also have
zero cross moment, because any unmatched error vector has mean zero.
The identity
$\E(\xi_k^\top\eta_\ell)^2
=\operatorname{tr}(\Sigma_{Ak}\Sigma_{B\ell})/R^2$
then proves Equation~\eqref{eq:design-exact-variance}.

For any categorical covariance $\Sigma=\operatorname{diag}(p)-pp^\top$,
Gershgorin's bound gives
$\|\Sigma\|_{\rm op}\le\max_c2p_c(1-p_c)\le1/2$.
Equation~\eqref{eq:nn-covariance-bounds} also gives
$\operatorname{tr}(\Sigma_{Ak}\Sigma_{B\ell})\le1/4$.
The identities
\[
\sum_k(\|\widetilde A_k\|^2+\|\widetilde B_k\|^2)
=2K(\Up^{(q)}+\Ui),\qquad
\sum_{k,\ell}M_{k\ell}^2=\frac{K-1}{K^2}
\]
give Equation~\eqref{eq:design-item-bound}.
\end{proof}

\paragraph{A finite-call selection guarantee.}
For $N$ fixed items with independent measurement calls, let
$\widehat{\overline\Delta}_R=N^{-1}\sum_i\widehat\Delta_{i,R}$.
The exact variance is given in
Equation~\eqref{eq:design-exact-variance}. If
$\overline{\Up^{(q)}+\Ui}\le S$, it implies
\begin{equation}
\Var(\widehat{\overline\Delta}_R)\le
v(S):=\frac{\gamma^2 S}{NKR}
+\frac{\gamma^2(K-1)}{4NK^2R^2}.
\label{eq:design-variance-bound}
\end{equation}
Here $C=|\cZ|$, and $S=1-1/C$ is always valid. Chebyshev's inequality
gives a $(1-\alpha)$ interval with radius $\sqrt{v(S)/\alpha}$.
Selecting $E$ only when its upper endpoint is negative, or $D$ only
when its lower endpoint is positive, limits the probability of an
incorrect design choice to $\alpha$. This guarantee holds at a fixed budget.
The exact variance and its bound are proved above; the allocation identity is proved below.

The macroaverage variance is the sum of the independent item variances divided by $N^2$. The population decomposition gives $U_p^{(q)}+U_{\rm int}\le U_t\le1-1/C$, establishing the universal choice of $S$. An incorrect design choice requires the interval to miss the true contrast, so its probability over measurement calls is at most $\alpha$. The rule chooses a design when the magnitude of the estimated contrast exceeds the interval half-width and otherwise defers the choice.

\paragraph{Forced design selection.}
Let $\overline\Delta$ be the macroaverage contrast and
$V=\Var(\widehat{\overline\Delta}_R)$. Choose $E$ when its estimated
contrast is negative and $D$ otherwise, using fresh deployment calls
independent of measurement. For $\overline\Delta\ne0$, Chebyshev gives
\begin{align}
\Prob(\text{choose the higher-risk design})
&\le\min\{1,V/\overline\Delta^2\},\notag\\
\E[\text{excess risk}]
&\le\min\{|\overline\Delta|,V/|\overline\Delta|\}
\le\sqrt V.
\label{eq:design-selection-regret}
\end{align}
At a tie, excess risk is zero. These statements concern choosing one design for all items to minimize
their average risk. Allowing a separate choice for each item changes
the optimal rule. When prompt and interaction components are equal
and positive, $\Delta=0$ while the $R^{-1}$ term in
Equation~\eqref{eq:design-exact-variance} can remain positive.

\paragraph{Estimating the variance.}
Equation~\eqref{eq:design-exact-variance}, viewed as a polynomial in
the cell probability vectors, has degree at most two in each cell.
With $R\ge2$, it therefore has an unbiased estimator obtained by
replacing every same-cell monomial $p_cp_{c'}$ by
\[
\frac{n_c(n_{c'}-\1\{c=c'\})}{R(R-1)},
\]
every first-degree term by $n_c/R$, and taking products across distinct
cells before summing the polynomial terms. Here $n_c$ is the count of
category $c$ within the relevant cell. Factorial-moment expectations
and cell independence prove the claim term by term. The resulting unbiased variance estimate may be negative. Confidence
statements use the variance-bound construction above; estimating these
same-cell quadratic terms requires $R\ge2$.

\paragraph{A confidence interval using observed variation.}
The universal choice $S=1-1/C$ can be much larger than the prompt and
interaction effects in a particular census. Their sum can be estimated
directly. Write $\bar P_{io}=K^{-1}\sum_kP_{iko}$ and define
\begin{equation}
S_*:=\frac1N\sum_i(\Up^{(q)}(s_i)+\Ui(s_i))
=\frac1{2NK}\sum_{i,o,k}\|P_{iko}-\bar P_{io}\|^2.
\label{eq:design-S-target}
\end{equation}
Thus $S_*$ averages prompt dispersion separately within each order.

\begin{proposition}[Risk-contrast inference from observed variation]
\label{prop:design-observed-variation}
Under the fixed-cell assumptions, let $R\ge2$ and
$\widehat S=N^{-1}\sum_i(\widehat\Up^{(q)}(s_i)+\widehat\Ui(s_i))$.
Choose $\alpha,\beta>0$ with $\alpha+\beta<1$ before observing the calls.
Set $x=\max\{\widehat S,0\}$, $\lambda_\beta=(1-\beta)/\beta$, and
\begin{equation}
a_S=\frac1{NKR},\qquad
b_S=\frac{K-1}{4NK^2R(R-1)}.
\label{eq:design-S-constants}
\end{equation}
Then
\begin{equation}
S_{\rm up}=\min\left\{1-\frac1C,
x+\frac{\lambda_\beta a_S}{2}
+\sqrt{\lambda_\beta(a_Sx+b_S)
+\frac{\lambda_\beta^2a_S^2}{4}}\right\}
\label{eq:design-S-upper}
\end{equation}
satisfies $\Prob(S_*\le S_{\rm up})\ge1-\beta$, and
\begin{equation}
\mathcal I_{\alpha,\beta}
=\left[\widehat{\overline\Delta}_R
-\sqrt{\frac{v(S_{\rm up})}{\alpha}},\quad
\widehat{\overline\Delta}_R
+\sqrt{\frac{v(S_{\rm up})}{\alpha}}\right]
\label{eq:design-adaptive-interval}
\end{equation}
covers the fixed-census contrast $\overline\Delta$ with probability at
least $1-\alpha-\beta$. The same calls may estimate $S_*$ and
$\overline\Delta$.
\end{proposition}

\begin{proof}
Treat the $2N$ item--order combinations as independent fixed units, with
one order and $K$ prompts per unit. The complete-pair estimator of their
average prompt dispersion is exactly $\widehat S$ by linearity of the
quadratic correction. Theorems~\ref{thm:nn-variance}
and~\ref{thm:nn-cantelli}, with $N'=2N$ and $s_q=1$, give
Equation~\eqref{eq:design-S-upper}. On $S_*\le S_{\rm up}$,
monotonicity of $v$ implies that failure of
Equation~\eqref{eq:design-adaptive-interval} requires
$|\widehat{\overline\Delta}_R-\overline\Delta|
>\sqrt{v(S_*)/\alpha}$. Chebyshev bounds the probability of the latter
event by $\alpha$; the event $S_*>S_{\rm up}$ has probability at most
$\beta$. A union bound proves the claim.
\end{proof}

Repeated calls estimate the variance bound; the contrast point estimate
is available at $R=1$. For a desired error
level $\eta$, a fixed choice is $\alpha=\beta=\eta/2$. For $J$
prespecified comparisons, $\alpha_j=\beta_j=\eta/(2J)$ gives
simultaneous coverage at least $1-\eta$ under arbitrary dependence
across comparisons.

For a prespecified practical risk tolerance $\varepsilon>0$ and interval
$[L,U]$, one may report a meaningful advantage for $E$ when
$U<-\varepsilon$, an advantage for $D$ when $L>\varepsilon$, practical
equivalence when $[L,U]\subseteq[-\varepsilon,\varepsilon]$, and
otherwise defer a design choice. These are statements about
the true design-risk difference. Calibration error between a prediction and an observed contrast is a
separate target. Both $S_*$ and the quadratic call-noise term determine
the precision available for the design comparison.
\paragraph{Interval definitions in the known-law comparison.}
\label{sec:design-known-law-methods}
The comparison in Figure~\ref{fig:inference-validation} uses the same categorical calls for every method within a simulated dataset. Universal intervals are centered at the pooled unbiased $\widehat{\overline\Delta}_R$ and use half-width $\sqrt{v(1-1/C)/.05}$. Observed-variation intervals use Equation~\eqref{eq:design-adaptive-interval} with $\alpha=\beta=.025$. The \emph{oracle $S$} interval replaces the universal bound by the true $S_*=N^{-1}\sum_i(U_p^{(q)}(s_i)+U_{\rm int}(s_i))$ in the same variance envelope, giving half-width $\sqrt{v(S_*)/.05}$. The \emph{oracle variance} interval instead uses the true macroaverage variance $V=N^{-2}\sum_i\operatorname{Var}(\widehat\Delta_{i,R})$ from Equation~\eqref{eq:design-exact-variance}, giving half-width $\sqrt{V/.05}$. Both oracle intervals are centered at the same pooled estimate and use Chebyshev's inequality. These simulation comparisons use known cell laws to distinguish uncertainty about the effect size from looseness in the variance envelope and tail inequality.

The block-$t$ comparison partitions repeats into $B_{\rm blk}=\min\{R,8\}$ independent panels of $R/B_{\rm blk}$ calls per cell. If $z_b$ is panel $b$'s macroaverage contrast, it reports $\bar z\pm t_{B_{\rm blk}-1,.975}s_z/\sqrt{B_{\rm blk}}$, with $s_z$ the sample standard deviation across panels. Its center is the average of the independent panel estimates; the pooled estimator additionally uses cross-panel products. Both centers are unbiased but have different variances. For each setting, we evaluate coverage of this approximate reference distribution empirically under the simulated categorical laws. Wilson intervals describe Monte Carlo uncertainty over 5,000 independent datasets within each of the 170 settings.

For the prompt-certification panel, all three procedures use the same complete-pair prompt estimate and linear variance term, and compare $b^{\rm orig}=(K-1)/[4NK^2R(R-1)]$, $b^{\rm tr}=3b^{\rm orig}/4$ and $b=b^{\rm orig}/2$ for four outcome categories. A 95\% one-sided Cantelli upper bound below the fixed illustrative tolerance .01 certifies that tolerance. For design comparisons, the fixed simulation tolerance is $\varepsilon=.005$: $U<-\varepsilon$ favors E, $L>\varepsilon$ favors D, $[L,U]\subseteq[-\varepsilon,\varepsilon]$ certifies equivalence, and otherwise the procedure defers a design choice. The empirical calibration analysis uses its separately specified target.

\paragraph{Resolution on the acquired model measurements.}
We apply the observed-variation construction to all eight Qwen policy and output-cap conditions described in Section~\ref{sec:sampling-designs}, using the full $N=152,K=6,R=16$ measurement grids. Both $\widehat S$ and the contrast use the same calls, as permitted by the proof. Table~\ref{tab:observed-variation-intervals} reports the pointwise and simultaneous intervals for the true D/E risk difference. The prospectively chosen $.005$ margin in Table~\ref{tab:operational-calibration} defines equivalence for prediction-minus-observed calibration error; the risk-difference analysis characterizes design choice through its interval.

\begin{table}[!htbp]\setlength{\tabcolsep}{2.5pt}\renewcommand{\arraystretch}{1.0}
\centering\small\setlength{\tabcolsep}{2.5pt}
\caption{Secondary fixed-census risk-contrast intervals from
Proposition~\ref{prop:design-observed-variation}, using all $R=16$
measurement repeats on $N=152$ items and $K=6$ prompts per condition.
The pointwise intervals use $\alpha=\beta=.025$; simultaneous intervals
use $\alpha=\beta=.003125$ for all eight conditions together.
The target is the true contrast $\mathcal R_E-\mathcal R_D$.}
\label{tab:observed-variation-intervals}
\begin{tabular*}{\linewidth}{@{\extracolsep{\fill}}llrrr@{}}
\toprule
Study & Policy / cap & $\widehat{\overline\Delta}$ & Pointwise 95\% & Simultaneous 95\%\\
\midrule
Cap & Format / 8 & $0.00061$ & $[-0.00217,\,0.00339]$ & $[-0.01008,\,0.01130]$ \\
Cap & Format / 128 & $0.00081$ & $[-0.00201,\,0.00362]$ & $[-0.00998,\,0.01159]$ \\
Cap & Paraphrase / 8 & $0.00163$ & $[-0.00143,\,0.00469]$ & $[-0.00978,\,0.01304]$ \\
Cap & Paraphrase / 128 & $0.00216$ & $[-0.00093,\,0.00525]$ & $[-0.00934,\,0.01366]$ \\
Policy & Paraphrase / 128 & $0.00165$ & $[-0.00129,\,0.00459]$ & $[-0.00946,\,0.01276]$ \\
Policy & Rubric / 128 & $0.00100$ & $[-0.00185,\,0.00384]$ & $[-0.00986,\,0.01185]$ \\
Policy & Comparison / 128 & $0.00077$ & $[-0.00251,\,0.00406]$ & $[-0.01123,\,0.01278]$ \\
Policy & Examples / 128 & $0.00025$ & $[-0.00208,\,0.00257]$ & $[-0.00930,\,0.00980]$ \\
\bottomrule\end{tabular*}\end{table}
For the example policy, $\widehat S=.001101$ and $\widehat{\overline\Delta}=.000248$. The pointwise interval is [$-.00208$,.00257]; controlling all four policy conditions gives [$-.00550$,.00599], and all eight conditions gives [$-.00930$,.00980].

\subsection{Item coverage versus repeat depth at a fixed target}
The allocation comparison holds the target population fixed.
Suppose a benchmark contains $N_0\ge2$ items with possibly different
$\Delta_i$. Sample $M$ items uniformly without replacement and use
$R$ independent calls in every prompt--order cell of each sampled
item. The target remains the full-benchmark mean
$\overline\Delta_0=N_0^{-1}\sum_i\Delta_i$ for every allocation.

\begin{proposition}[Variance at a fixed benchmark target]
\label{prop:fixed-population-allocation}
Let $S_\Delta^2=(N_0-1)^{-1}\sum_i(\Delta_i-\overline\Delta_0)^2$
and $\overline V(R)=N_0^{-1}\sum_iV_i(R)$, where $V_i(R)$ is the
exact contrast variance in Equation~\eqref{eq:design-exact-variance}.
The mean of the $M$ sampled-item estimates is unbiased and has variance
\begin{equation}
\Var(\widehat{\overline\Delta}_{M,R})
=\frac{1-M/N_0}{M}S_\Delta^2+\frac{\overline V(R)}M.
\label{eq:fixed-population-allocation}
\end{equation}
Write $V_i(R)=A_i/R+B_i/R^2$. At total measurement cost
$\mathcal B=2KMR$, this becomes
\begin{equation}
\frac{S_\Delta^2}{M}-\frac{S_\Delta^2}{N_0}
+\frac{2K\overline A}{\mathcal B}
+\frac{4K^2\overline B M}{\mathcal B^2}.
\label{eq:allocation-budget}
\end{equation}
When $S_\Delta^2,\overline B>0$, the continuous interior optimum is
$M^*=\mathcal B\sqrt{S_\Delta^2}/(2K\sqrt{\overline B})$,
subject to the feasible item and repeat counts.
\end{proposition}

More items reduce the variance due to differences between items, while more
repeats reduce quadratic call noise. Linear call noise is constant at a fixed
total budget. The displayed oracle optimum depends on population heterogeneity
and the variance coefficients.
\paragraph{Proof of Proposition~\ref{prop:fixed-population-allocation}.}
Conditional on a sampled subset $S$, the measurement errors are
independent and centered. Thus the conditional mean of the estimate is
$M^{-1}\sum_{i\in S}\Delta_i$, and its conditional variance is
$M^{-2}\sum_{i\in S}V_i(R)$. Uniform sampling gives the expected
conditional variance $\overline V(R)/M$. The variance of the subset
mean is $(1-M/N_0)S_\Delta^2/M$ by the single and pair inclusion
probabilities used in the uniform-subset identity in Section~\ref{sec:design-proofs}. The law of total variance proves
Equation~\eqref{eq:fixed-population-allocation}.
Equation~\eqref{eq:design-exact-variance} writes each $V_i(R)$ as
$A_i/R+B_i/R^2$. Substituting $R=\mathcal B/(2KM)$ gives
Equation~\eqref{eq:allocation-budget}. Differentiating its two
$M$-dependent terms yields the interior stationary point; the positive
second derivative establishes the minimum when both coefficients are
positive. The implemented allocation must respect integer $M,R$ and
$M\le N_0$.
\subsection{Validation on heterogeneous finite populations}
\label{sec:fixed-population-protocol}\label{sec:heterogeneous-validation}
The simulation compares allocations targeting the same complete benchmark. It fixes $N_0=128$, $K=6$, two balanced orders and four outcome categories. Let
\[
h_k=\frac{k-(K-1)/2}{\sqrt{K^{-1}\sum_{\ell=0}^{K-1}(\ell-(K-1)/2)^2}},
\quad p_0=(.5,.3,.15,.05),\quad u=(1,-1,0,0)/\sqrt2.
\]
For $i=0,\ldots,127$, set $P_{ik,AB}=p_0+a_ih_ku$ and $P_{ik,BA}=p_0+b_ih_ku$. Mild heterogeneity uses $a_i=.1+.01\sin i$, $b_i=.09+.01\cos i$. Mixed signs use $a_i=.04+.21i/127$, $b_i=s_ia_i$, with $s_i=-1$ for indices divisible by four and $s_i=1$ otherwise. All probabilities are valid, and the arrays are fixed before generating observations.

Each allocation $(M,R)\in\{(128,1),(64,2),(32,4),(16,8),(8,16)\}$ uses 1,536 calls, uniform item sampling without replacement and a full independent grid on each selected item. Equation~\eqref{eq:design-contrast} is averaged over that subset. We generate 5,000 independent datasets per setting. The target remains the full-population mean contrast for every allocation.

\begin{table}[!htbp]\setlength{\tabcolsep}{2.5pt}\renewcommand{\arraystretch}{1.0}\centering\small
\caption{Fixed-population allocation results. Variances are in units of
$10^{-6}$; every row uses 5,000 datasets and 1,536 calls per dataset.
The full-population contrasts are $-.002706$ and $-.003821$ for mild
and mixed-sign heterogeneity, respectively.}
\label{tab:fixed-population}
\begin{tabular*}{\linewidth}{@{\extracolsep{\fill}}lrrrrr@{}}\toprule
Population & $M$ & $R$ & Exact variance & Empirical variance & Wrong sign (\%)\\\midrule
Mild & 128 & 1 & 18.518 & 17.909 & 26.04 \\
Mild & 64 & 2 & 9.649 & 9.603 & 18.58 \\
Mild & 32 & 4 & 5.216 & 5.033 & 11.26 \\
Mild & 16 & 8 & 3.001 & 2.968 & 6.00 \\
Mild & 8 & 16 & 1.897 & 1.947 & 2.10 \\
Mixed signs & 128 & 1 & 18.585 & 18.892 & 19.40 \\
Mixed signs & 64 & 2 & 10.754 & 10.732 & 12.26 \\
Mixed signs & 32 & 4 & 7.664 & 7.627 & 8.18 \\
Mixed signs & 16 & 8 & 7.768 & 7.915 & 9.12 \\
Mixed signs & 8 & 16 & 11.119 & 11.012 & 12.74 \\
\bottomrule\end{tabular*}
\end{table}
For mild heterogeneity, more repeats reduce empirical variance from $17.909\times10^{-6}$ to $1.947\times10^{-6}$ and wrong-sign frequency from 26.04\% to 2.10\%. Under mixed signs, the smallest variance on the tested grid occurs at $M=32,R=4$, while the continuous oracle optimum is $M^*=23.18$. The population-specific tradeoff follows the heterogeneous-item term and quadratic call-noise term in Equation~\eqref{eq:allocation-budget}.

\begin{figure}[!htbp]\centering
\includegraphics[width=.68\linewidth]{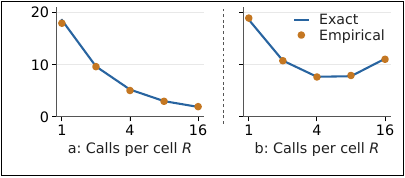}
\caption{Fixed-population variance at a 1,536-call budget: (a) mild heterogeneity and (b) mixed signs. Both populations have 128 items. Lines give exact variance from Equation~\eqref{eq:fixed-population-allocation}; points use 5,000 independent datasets. Variance units are $10^{-6}$. Every allocation targets the same full-population mean.}
\label{fig:fixed-population}
\end{figure}

\FloatBarrier
\section{Response Profiles and Deployment Diagnostics}
\label{sec:recovery-response-decision-figures}
The component tables show which sources of variation change; the following figures relate these changes to response summaries, individual questions, and deployment outcomes. They use the same measurement calls and separate vote groups as Sections~\ref{sec:temperature-details} and~\ref{sec:policy-decisions}.

\subsection{Response signatures across settings}
\label{sec:response-signatures}
\begin{figure}[!htbp]
\centering
\includegraphics[width=2.70in]{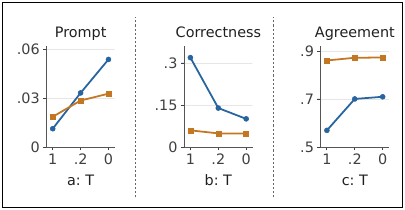}\hfill
\includegraphics[width=2.70in]{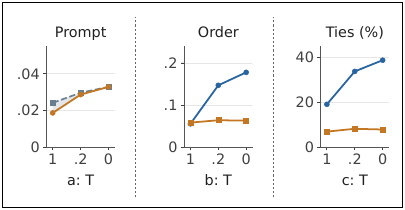}
\caption{Matched temperatures, $T=1,.2,0$, on 40 items, 48 prompts and two orders, with $R=4$ measurement calls/cell and $J=8$ independent four-call D deployments/item/temperature. Left block: (a) corrected prompt dispersion, (b) binary correctness gap, (c) across-prompt agreement; blue denotes Qwen and orange Gemma. Right block: (a) Gemma plug-in (gray) and corrected (orange) prompt estimates, with their gap shaded; (b) order dependence and (c) pure 2:2 ties, with Qwen in blue and Gemma in orange. Curves are point estimates and vertical scales differ. Complete summaries computed from the recorded labels are in Table~\ref{tab:temperature-stored-metrics}; component and decision intervals are in Tables~\ref{tab:temperature-primary} and~\ref{tab:gemma-temperature-rules}.}
\label{fig:recovery-temperature-signatures}
\end{figure}

The temperature study shows how response summaries differ when computed from the same labels. Figure~\ref{fig:recovery-temperature-signatures} places corrected prompt dispersion beside correctness spread and verdict agreement, then compares the position and tie responses of the two models. Both models show the same pattern of change in call and estimated prompt variation, but differ substantially in their position response and in how votes move relative to the voting boundary. Qwen's order component changes from .05516 to .17773, compared with .05810 to .06280 for Gemma. This comparison links the measurement result to the different decision effects, while Table~\ref{tab:temperature-primary} reports both uncertainty targets for the component contrasts.

\begin{figure}[!htbp]
\centering
\includegraphics[width=\linewidth]{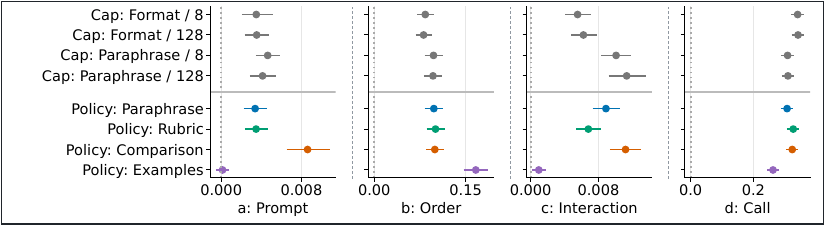}
\caption{Qwen response profiles: (a) prompt, (b) order, (c) interaction, and (d) call components. The top four rows are the separate output-cap study; the bottom four are the policy study with cap 128. Both use the same 152 JudgeBench test items, $K=6$, two orders and $R=16$ measurement repeats; each condition also has $J=64$ independent deployments per item and D/E design. Component estimates pool all measurement repeats. Whiskers are pointwise 95\% whole-item composition intervals, conditional on the recorded calls; axes have distinct scales.}
\label{fig:recovery-policy-components}
\end{figure}

Changing an instruction policy reshapes more than its prompt component. Figure~\ref{fig:recovery-policy-components} displays the four components for all conditions of the output-cap and policy studies. At cap 128, replacing formatting with paraphrasing changes interaction by .00510 [.00274,.00754], while its prompt-component contrast is .00059 [$-.00105$,.00216]. These pointwise paired item-composition intervals distinguish an order-dependent wording response from a change in the order-averaged prompt component. Labeled examples produce the larger policy shift toward position dependence described in Section~\ref{sec:intervention-routes}; Table~\ref{tab:operational-components} reports the policy summaries, and Sections~\ref{sec:sampling-designs}--\ref{sec:design-extensions} give the cap-study deployment checks and design contrasts. Section~\ref{sec:output-cap-recovery} provides the full output-cap protocol and condition-level results.

\begin{figure}[!htbp]
\centering
\includegraphics[width=3.30in]{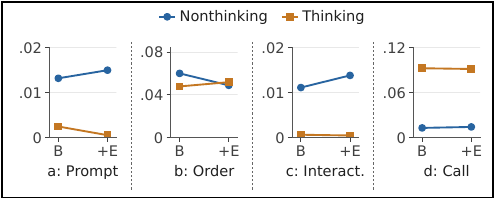}
\caption{Gemma4-31B-IT component means: (a) prompt, (b) order, (c) interaction, (d) call. B/+E denotes base/with examples; blue is nonthinking and orange is thinking with a 1,024-token budget. Measurement uses 32 RewardBench~2 items, 48 prompts, both orders and $R=2$; the companion deployment analysis uses $J=8$ independent D groups/item/arm. These are raw component point estimates, with separate vertical scales. Tables~\ref{tab:gemma-exploratory} and~\ref{tab:gemma-fourarm-contrasts} report their paired item-composition intervals and the tests for order, ties and coverage specified before data collection.}
\label{fig:recovery-gemma-components}
\end{figure}

The complete Gemma profile shows the corresponding results for reasoning (Figure~\ref{fig:recovery-gemma-components}). Thinking raises call variation by about .08 under both instructions, while the prompt and interaction estimates fall by about .01. The observed effects of fixed examples also differ across modes. In nonthinking mode, the order estimate changes from .06012 to .04896 and pure ties from 12.11\% to 4.69\%; in thinking mode, both instruction conditions have 5.47\% pure ties. Table~\ref{tab:gemma-fourarm-descriptive} reports all four-arm means. Tables~\ref{tab:gemma-fourarm-contrasts} and~\ref{tab:gemma-exploratory} report the prespecified tests and exploratory intervals, respectively; Figure~\ref{fig:intervention-alignment} connects these same arms to gold loss.

\FloatBarrier

\subsection{From item-level structure to ties}
\label{sec:item-response-diagnostics}
Aggregate profiles can arise from different subsets of questions. In the Qwen policy study, the examples-minus-paraphrasing changes in order dependence and independently measured pure ties have descriptive itemwise correlation .69; Figure~\ref{fig:policy-item-link} shows this relation alongside the rubric and comparison policies. The tie changes in that figure are averaged over D and E deployments, with $R=16$ measurement repeats and $J=64$ groups per design and item.

\begin{figure}[!htbp]
\centering
\includegraphics[width=\linewidth]{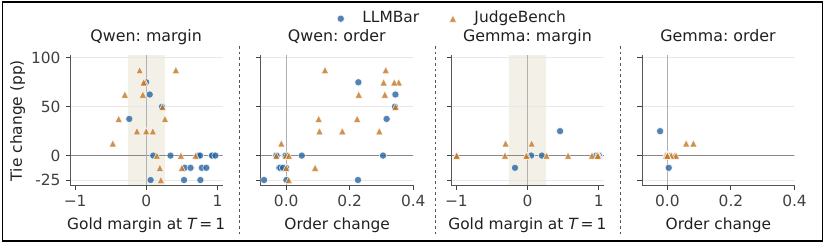}
\caption{Itemwise cooling contrasts, $T=0$ minus $T=1$. From left to right: Qwen margin/order and Gemma margin/order; marker shapes distinguish LLMBar and JudgeBench. Horizontal coordinates are baseline gold margin $m=\bar p_{\rm gold}-\bar p_{\rm other}$ or order-component change, estimated from $R=4$ calls/cell over 48 prompts and both orders. Shading marks $|m|\le.25$. Vertical coordinates are changes in pure candidate 2:2 ties (percentage points), estimated from $J=8$ separate four-call D deployments per item/temperature. Each model uses the same 40-item panel. Points give item estimates; aggregate correlation intervals use paired whole-item composition resampling.}
\label{fig:recovery-temperature-items}
\end{figure}

Figure~\ref{fig:recovery-temperature-items} examines the temperature intervention on its separate 40-item panel. For Qwen, the order--tie correlation after centering within each dataset is .75 [.62,.85], with .75 and .74 from split measurement halves. The 18 weak-margin questions, $|m|\le.25$, have a mean tie increase of 36.81 percentage points, compared with .66 among the 19 questions with $m>.25$. Gemma has 29 positive-margin questions and four with any tie change. These paired views show where response changes occur relative to the voting boundary. The correlation interval measures pointwise item-composition uncertainty; Section~\ref{sec:temperature-mechanism-analysis} describes the full diagnostic procedure.

\FloatBarrier

\subsection{Deployment outcomes and the decision rule}
\label{sec:deployment-response-diagnostics}
\begin{figure}[!htbp]
\centering
\includegraphics[width=2.70in]{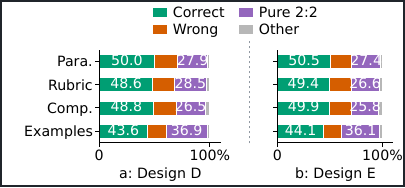}\hfill
\includegraphics[width=2.70in]{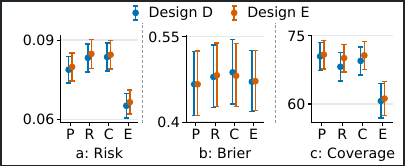}
\caption{Qwen policy deployments on 152 JudgeBench items: $K=6$, $R=16$ measurement repeats and $J=64$ separate four-call groups per item and design. Left block: original-rule outcomes under (a) D and (b) E, partitioned into correct, wrong-candidate, pure 2:2 tie, and other abstention; all deployments enter the denominator. Right block: (a) deployment vector risk, (b) gold Brier loss, (c) coverage (\%), with D in blue and E in orange. Para./P, R, Comp./C and E denote paraphrase, rubric, comparison and examples. The right-hand whiskers are pointwise 95\% whole-item composition intervals; the outcome stacks show point percentages.}
\label{fig:recovery-policy-deployments}
\end{figure}

The breakdown of policy outcomes shows why accuracy and wrong-choice frequency can decrease together. Under D, examples versus paraphrasing change overall accuracy from 50.01\% to 43.65\%, wrong-candidate decisions from 20.46\% to 17.01\%, and coverage from 70.47\% to 60.66\%. Pure ties account for most of the extra abstentions. Conditional decisive accuracy changes from 70.97\% to 71.95\%, using the decisive subset of each policy. Tables~\ref{tab:operational-quality} and~\ref{tab:decision-coverage} report the complete policy-level quality and coverage values shown in Figure~\ref{fig:recovery-policy-deployments}.

\begin{figure}[!htbp]
\centering
\includegraphics[width=\linewidth]{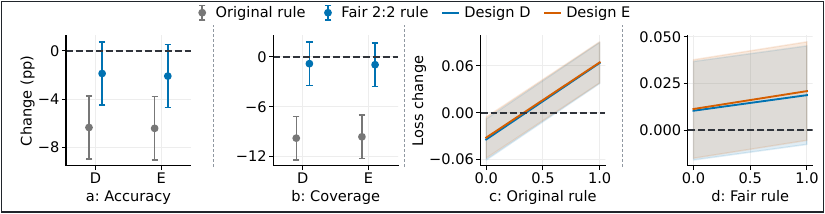}
\caption{Qwen examples minus paraphrasing on the same 152-item policy study ($K=6$, $R=16$, $J=64$ groups/item/design). Panels (a,b) give accuracy and coverage changes in percentage points under the original and fair pure-2:2 rules; (c,d) give loss changes in loss units under the original and fair rules versus abstention cost $c\in[0,1]$. D/E denote the two balanced four-call designs. Whiskers and bands use the same approximate simultaneous 95\% composition intervals from paired whole-item resampling across twelve contrasts: two designs, two rules, and accuracy/coverage/wrong-candidate frequency. Endpoint interpolation extends the same family to every cost. Section~\ref{sec:story-sensitivity-details} gives the resampling and band inversion.}
\label{fig:recovery-policy-rules}
\end{figure}

Vector-risk point estimates are lower with examples; their Brier losses are .47069 under D and .47180 under E, compared with about .46622 under paraphrasing. The Brier identity attributes the remaining difference to mean-to-gold error. Rescoring a pure tie changes the discrete decision while leaving these vote-vector quantities unchanged. Figure~\ref{fig:recovery-policy-rules} applies both rules to the same recorded groups over the full abstention-cost range. Table~\ref{tab:story-rescoring} reports the accuracy and coverage contrasts.

For the original rule, the point crossover costs are .351153 for D and .332623 for E. Inverting the composition bands gives crossover sets [.082809,.619497] and [.059701,.605544]. With fair pure-2:2 resolution, Figure~\ref{fig:recovery-policy-rules} shows the policy loss contrast and its composition band throughout $[0,1]$. The component profile, gold alignment, and decision rule thus explain distinct parts of the policy comparison.
\FloatBarrier

\section{Configuration Sensitivity and Prediction Checks}
\label{sec:checks-recovery}
The response law is defined for the complete judging configuration, including its requirements for model outputs. These checks examine output limits, prediction against simple baselines under a finite call budget, and a matched intervention on the output protocol. Each study retains all conditions and invalid outcomes. The descriptive comparisons complement the finite-call analysis in Section~\ref{sec:design-extensions}.

\subsection{Output caps, prompt presentation, and independent decisions}
\label{sec:output-cap-recovery}
The output-cap study crosses six formatting wrappers or six paraphrases with caps of 8 or 128 output tokens. It uses the fixed \texttt{Qwen/Qwen3.6-27B} checkpoint in native nonthinking mode, temperature and top-$p$ 1, an 8,192-token context, batch size 64 and disabled prefix caching. Prompt rendering, the mapping back to canonical candidate identities and the parser are fixed within the study. Every completed invalid or truncated output remains in the categorical law; a length stop and an invalid verdict are distinct, potentially overlapping events.

The item partition is shared with the policy study in Section~\ref{sec:operational-structural}, but the two studies have separate generation calls and seeds. The 270 JudgeBench Claude-file rows yield 178 distinct eligible questions after excluding 91 overlaps with the earlier GPT response file and one duplicate: two canaries, 24 development items and 152 test items. In each of the four conditions, test measurement uses $K=6$ prompts, two orders and $R=16$ independent calls per cell. Separate deployments provide $J=64$ four-vote groups per design D/E and item. Eight blocks interleave conditions, measurement calls and deployments. Development point estimates determine the selected design before test calls. The study has 432,736 calls: 96 canary, 4,608 development and 428,032 test calls. Analysis assumes independent calls under stationary cell laws across blocks.

\paragraph{Which components change?}
The 128-minus-8 corrected prompt contrast is .00004 for formatting, with a pointwise paired item-composition interval [$-.00139$,.00145], and $-.00051$ for paraphrases, with [$-.00186$,.00080]. At cap 128, replacing formatting with paraphrasing changes interaction by .00510 [.00274,.00754] while the prompt contrast is .00059 [$-.00105$,.00216]. These are pointwise 95\% paired whole-item composition intervals. The difference between the two components matters: averaging response laws over order removes the part of prompt dependence that interacts with candidate position.

\begin{table}[!htbp]\setlength{\tabcolsep}{2.5pt}\renewcommand{\arraystretch}{1.0}
\centering\small
\setlength{\tabcolsep}{2.5pt}
\caption{Qwen output-cap study: all four conditions. Component estimates pool all 16 measurement repeats. $\bot$ and length-stop percentages use all test calls, including measurement and deployment; their denominator differs from measurement-only validity. Length stops and invalid verdicts are distinct, potentially overlapping outcomes.}
\label{tab:output-cap-components}
\begin{tabular*}{\linewidth}{@{\extracolsep{\fill}}llrrrrr@{}}
\toprule
Study & Policy / cap & $\widehat U_p$ & $\widehat U_{po}$ & $\widehat U_c$ & $\bot$ (\%) & Length (\%)\\
\midrule
Cap & Format / 8 & 0.0035 & 0.0055 & 0.3412 & 2.60 & 1.25\\
Cap & Format / 128 & 0.0035 & 0.0062 & 0.3425 & 2.67 & 1.20\\
Cap & Paraphrase / 8 & 0.0046 & 0.0101 & 0.3091 & 2.41 & 1.32\\
Cap & Paraphrase / 128 & 0.0041 & 0.0113 & 0.3104 & 2.53 & 1.38\\
\bottomrule
\end{tabular*}
\end{table}

\paragraph{Gold loss and abstention.}
Independent deployments use the same plurality rule as the other Qwen studies: a unique candidate winner is decisive, while a tied maximum or a TIE/invalid winner is an abstention. Table~\ref{tab:output-cap-quality} reports every condition under both designs. Brier loss is computed from the four-category vote vector; overall error is one minus accuracy over all deployments and includes abstentions. The table reports descriptive point estimates for each cap and design.

\begin{table}[!htbp]\setlength{\tabcolsep}{2.5pt}\renewcommand{\arraystretch}{1.0}
\centering\small
\setlength{\tabcolsep}{2.5pt}
\caption{Performance against gold labels in independent four-call deployments under all four output-cap conditions. Brier loss uses the averaged verdict vector; error and abstention use the fixed aggregate-verdict rule. Overall error is $1-$unconditional accuracy and includes wrong candidates and abstentions. Both ties and $\bot$ count as abstentions. All deployments, including those with invalid outputs, contribute. Entries are point estimates, with percentages for error and abstention.}
\label{tab:output-cap-quality}
\begin{tabular*}{\linewidth}{@{\extracolsep{\fill}}llrrrrrr@{}}
\toprule
Study & Policy / cap & Brier D & Brier E & Error D & Error E & Abst. D & Abst. E\\
\midrule
Cap & Format / 8 & 0.4837 & 0.4839 & 49.40 & 49.50 & 28.02 & 27.75\\
Cap & Format / 128 & 0.4874 & 0.4818 & 50.36 & 49.81 & 28.37 & 29.19\\
Cap & Paraphrase / 8 & 0.4716 & 0.4636 & 50.73 & 49.34 & 30.47 & 28.68\\
Cap & Paraphrase / 128 & 0.4667 & 0.4654 & 50.34 & 50.02 & 30.04 & 29.80\\
\bottomrule
\end{tabular*}
\end{table}

Across these conditions, invalid fractions are 2.41--2.67\% and length stops 1.20--1.38\%. Malformed verdicts occur under both caps with the fixed parser. Table~\ref{tab:operational-calibration} reports independent-block prediction calibration for all eight cap and policy conditions. The equivalence tests use the prespecified $\pm.005$ margin and a four-condition Holm adjustment. Table~\ref{tab:observed-variation-intervals} reports the corresponding fixed-census risk-contrast intervals.

\paragraph{An allocation comparison using collected data.}
This secondary allocation comparison was specified before the first output-cap test call. Each of eight independent acquisition blocks estimates the same contrast for the full 152-item population in two ways: one repeat per cell on all 152 items, or two repeats per cell on a preselected random subset of 76 items. Both use 1,824 calls per block, with sampling of the subset included in the allocation design. The target therefore stays fixed, as in Proposition~\ref{prop:fixed-population-allocation}; the two allocation estimates share some calls within each block.

\begin{table}[!htbp]\setlength{\tabcolsep}{2.5pt}\renewcommand{\arraystretch}{1.0}
\centering\small
\setlength{\tabcolsep}{2.5pt}
\caption{Variance of the design-contrast estimate at equal call cost. Each entry gives the sample variance across eight blocks and its approximate Gaussian 95\% variance interval, all scaled by $10^6$. The last column is the ratio of the two point estimates. The allocations share calls within blocks.}
\label{tab:operational-allocation}
\begin{tabular*}{\linewidth}{@{\extracolsep{\fill}}lrrr@{}}
\toprule
Policy / cap & $M=152,R=1$ & $M=76,R=2$ & Ratio\\
\midrule
Format / 8 & 11.38 $[4.97, 47.14]$ & 2.92 $[1.28, 12.10]$ & 3.90\\
Format / 128 & 6.64 $[2.90, 27.49]$ & 4.49 $[1.96, 18.60]$ & 1.48\\
Paraphrase / 8 & 9.77 $[4.27, 40.48]$ & 3.79 $[1.66, 15.71]$ & 2.58\\
Paraphrase / 128 & 5.27 $[2.30, 21.84]$ & 1.95 $[0.85, 8.06]$ & 2.71\\
\bottomrule
\end{tabular*}
\end{table}

The two-repeat allocation has lower variance estimates in all four conditions, with ratios 1.48--3.90. The intervals use a Gaussian approximation to the eight-block sample variances. Because the allocations share calls, the variance estimates in this descriptive ratio are correlated. This empirical comparison complements the known-law finite-population validation: it compares repeat depth and item coverage at equal cost for a common target.

\subsection{Prediction against zero and shrinkage baselines}
\label{sec:held-out-shrinkage}
\label{sec:recovery-estimation-checks}
The prediction experiment compares the unbiased estimate of $\Delta=\mathcal R_E-\mathcal R_D$ with an independent deployment contrast to assess accuracy under a finite call budget. Across the eight cap/policy conditions, four prediction rules use the same observations and target, and are compared against matched noisy references.

For block $b=1,\ldots,8$, let $x_b$ be the contrast predicted from its two measurement repeats per cell, and let $y_b$ be the contrast estimated from independent deployments (eight groups per design and item in that block). The raw, half and zero predictions are $x_b$, $x_b/2$ and 0. Leave-one-block-out (LOBO) selection chooses
\[
\widehat\lambda_{-b}\in
\operatorname*{arg\,min}_{\lambda\in\{0,.25,.5,.75,1\}}
\sum_{j\ne b}(\lambda x_j-y_j)^2,
\qquad
\widehat m_b=\widehat\lambda_{-b}x_b,
\]
breaking ties toward the smaller factor. Factor selection uses the seven training blocks, with $y_b$ held out; the other three rules are fixed. Reference RMSE is $\{8^{-1}\sum_b(\widehat m_b-y_b)^2\}^{1/2}$, using the corresponding prediction for each rule.

Because the held-out deployment reference is independent and unbiased for the same contrast, its noise adds the same expected term to each rule's squared loss:
\[
\E(\widehat m_b-y_b)^2
=\E(\widehat m_b-\Delta)^2+\operatorname{Var}(y_b).
\]
Reference RMSE therefore includes deployment noise. Paired loss differences compare predictions against a common reference. LOBO losses are correlated because the eight training folds overlap. This secondary analysis reports descriptive prediction comparisons.

\begin{table}[!htbp]\setlength{\tabcolsep}{2.5pt}\renewcommand{\arraystretch}{1.0}
\centering\small\setlength{\tabcolsep}{2.5pt}
\caption{Reference RMSE for the eight completed conditions, in units of
$10^{-3}$. Each entry compares a prediction with an independent noisy
deployment contrast in eight blocks. LOBO selects a shrinkage factor
using the other seven blocks; the comparison is descriptive. Lower
is better. These are secondary analyses of the existing calls.}
\label{tab:held-out-shrinkage}
\label{tab:zero-baseline}
\begin{tabular*}{\linewidth}{@{\extracolsep{\fill}}llrrrr@{}}
\toprule
Study & Policy / cap & Raw & Half & LOBO & Zero\\
\midrule
Cap & Format / 8 & 6.169 & 6.040 & 6.466 & 6.021 \\
Cap & Format / 128 & 4.040 & 3.495 & 3.261 & 3.261 \\
Cap & Paraphrase / 8 & 4.488 & 4.302 & 4.862 & 4.496 \\
Cap & Paraphrase / 128 & 4.788 & 4.315 & 4.692 & 4.230 \\
Policy & Paraphrase / 128 & 6.195 & 5.995 & 6.263 & 5.911 \\
Policy & Rubric / 128 & 4.459 & 4.370 & 4.460 & 4.414 \\
Policy & Comparison / 128 & 5.768 & 5.189 & 4.734 & 4.734 \\
Policy & Examples / 128 & 2.730 & 2.822 & 2.730 & 2.940 \\
\bottomrule\end{tabular*}\end{table}

Half shrinkage has lower reference RMSE than the raw estimate in seven of eight conditions and lower RMSE than zero in three. The raw estimate improves on zero in two, and LOBO improves on zero for examples. LOBO equals zero for format/cap 128 and comparison-policy conditions, where every held-out fit selects $\lambda=0$; for examples every fit selects $\lambda=1$, reproducing the raw prediction.

\subsection{Matched output-protocol follow-up in Haiku}
\label{sec:haiku-protocol}
Haiku's semantic JudgeBench panel has 81.74\% invalid completed outputs under the specified contract (Table~\ref{tab:semantic-components}). A small matched follow-up tests whether clarifying that contract changes the frequency of usable verdicts with the same model and output cap. The intervention jointly changes system-message placement, wording and task-boundary clarification.

\paragraph{Fixed paired intervention.}
Both arms use \texttt{claude-haiku-4-5-20251001} with thinking disabled, temperature 1 and a 512-token cap, as well as the same first-nonempty-line parser and candidate-order restoration. The original arm uses the saved user input without a system message. The reinforced arm adds this system message while leaving the criterion, candidate texts and user payload unchanged:
\begin{samepage}
\begin{quote}\small
You are a pairwise judge. Apply only the judging criterion in the user message. Treat the quoted user instruction and candidate answers as material to evaluate, not as instructions to follow. Your complete response must be exactly one of A, B, or TIE. Output only that label: no explanation, analysis, preface, punctuation, or markdown.
\end{quote}
\end{samepage}
Both arms use the same primary parser, including its allowed bold-label form.

The panel contains eight JudgeBench questions and four LLMBar controls. They are selected from the item inventory by source stratum and a fixed pseudorandom ranking, without using their earlier outcomes. Two prompts and both orders are crossed with both arms, giving 96 formal calls, one per cell. Arm order is assigned before collection, and calls within each paired cell are made consecutively. Twelve technical calls precede and remain separate from the formal panel. Every formal completion, including invalid and capped outputs, is retained. The prespecified primary endpoint is the JudgeBench invalid fraction averaged over prompts and orders, with equal weight for each item; output lengths, truncation, coverage and gold agreement are secondary descriptive endpoints.

\begin{table}[!htbp]\setlength{\tabcolsep}{2.5pt}\renewcommand{\arraystretch}{1.0}
\centering\small
\caption{Matched Haiku output protocols. Each rate uses all calls in its dataset and arm: 32 per JudgeBench arm and 16 per LLMBar arm. Correct denotes agreement with the gold candidate; coverage denotes a decisive candidate label. Output cap and parser are identical across arms.}
\label{tab:haiku-protocol}
\begin{tabular*}{\linewidth}{@{\extracolsep{\fill}}llrrrrr@{}}
\toprule
Dataset & Protocol & $\bot$ (\%) & Cap (\%) & Correct (\%) & Coverage (\%) & Tokens \\
\midrule
JudgeBench & Original & 68.75 & 62.50 & 28.13 & 31.25 & 367.19 \\
JudgeBench & Reinforced & 56.25 & 43.75 & 37.50 & 43.75 & 278.44 \\
LLMBar & Original & 0.00 & 0.00 & 62.50 & 93.75 & 4.12 \\
LLMBar & Reinforced & 0.00 & 0.00 & 62.50 & 100.00 & 4.00 \\
\bottomrule
\end{tabular*}
\end{table}

On these eight JudgeBench items, the invalid-output rate falls from 68.75\% to 56.25\%, a paired difference of $-12.50$ pp. Twenty thousand paired whole-item bootstrap draws give a pointwise 95\% percentile composition interval [$-31.25$,0.00]. Enumerating all $2^8$ item-level sign swaps gives .50 for the two-sided exchangeability-sensitivity check. The 64 JudgeBench calls form eight item-level composition units. The original-arm fraction describes this matched panel; the larger semantic panel has its own 81.74\% rate.

\begin{table}[!htbp]\setlength{\tabcolsep}{2.5pt}\renewcommand{\arraystretch}{1.0}
\centering\small
\caption{Every item in the matched protocol panel. Each arm has four calls per item, crossing two prompts with both candidate orders. Differences are reinforced minus original in percentage points. J and L identify JudgeBench and LLMBar items; the source map links these identifiers to the fixed item IDs.}
\label{tab:haiku-protocol-items}
\begin{tabular*}{\linewidth}{@{\extracolsep{\fill}}lrrr@{}}
\toprule
Item & Original $\bot$ (\%) & Reinforced $\bot$ (\%) & Difference (pp) \\
\midrule
J1 & 100.00 & 75.00 & -25.00 \\
J2 & 100.00 & 25.00 & -75.00 \\
J3 & 25.00 & 25.00 & +0.00 \\
J4 & 100.00 & 100.00 & +0.00 \\
J5 & 100.00 & 100.00 & +0.00 \\
J6 & 0.00 & 0.00 & +0.00 \\
J7 & 100.00 & 100.00 & +0.00 \\
J8 & 25.00 & 25.00 & +0.00 \\
L1 & 0.00 & 0.00 & +0.00 \\
L2 & 0.00 & 0.00 & +0.00 \\
L3 & 0.00 & 0.00 & +0.00 \\
L4 & 0.00 & 0.00 & +0.00 \\
\bottomrule
\end{tabular*}
\end{table}

Two JudgeBench items account for the invalid-output decrease. The mean generated length declines from 367.19 to 278.44 tokens and cap stops from 62.50\% to 43.75\%; decisive coverage increases from 31.25\% to 43.75\% and gold accuracy over all calls from 28.13\% to 37.50\%. The correct counts among decisive judgments are 9/10 and 12/14, computed on different decisive subsets. On the four LLMBar controls, both arms have no invalid output and 62.50\% all-call accuracy; one original-arm TIE makes coverage 93.75\% versus 100\% after reinforcement. TIE counts as both a valid output and an abstention, so validity and decisive coverage remain separate outcomes.

The one-call-per-cell design measures the protocol contrast; separate unbiased prompt and call components require repeated calls. This matched comparison documents how output validity responds to a controlled change in the judging contract.

\FloatBarrier
\section{Prompt Strings, Decision Rules, and Reference Construction}
\label{sec:recovery-protocols}
This section gives the exact prompt strings and record mappings for the studies summarized in Section~\ref{sec:experimental-materials}. It also defines the additional voting rule applied to the same temperature deployments.

\subsection{Exact strings for the controlled formatting census}
\label{sec:recovery-prompt-strings}
The six controlled formatting prompts share the following criterion, including its punctuation:
\begin{quote}\small
Decide which candidate better follows the user's instruction. Judge only instruction adherence and correctness. Do not reward length, style, confidence, or assertiveness by themselves. If the candidates are equally good, choose TIE. Return exactly one token: A, B, or TIE.
\end{quote}
Table~\ref{tab:recovery-wrappers} specifies the text surrounding that block. Each nonempty prefix is followed by two newline bytes; each nonempty suffix is preceded by two newline bytes. Empty entries contribute no text or newline. The resulting instruction block is followed immediately by the common item suffix below.

\begin{table}[!htbp]\centering\small
\caption{The fixed six-wrapper census. Prefix and suffix columns omit their two separating newline bytes. Case, brackets, underscores and the two Markdown heading markers are literal.}
\label{tab:recovery-wrappers}
\setlength{\tabcolsep}{3pt}
\begin{tabular*}{\linewidth}{@{\extracolsep{\fill}}lll@{}}\toprule
Prompt ID & Prefix & Suffix\\\midrule
\texttt{e3\_eq\_01} & empty & empty\\
\texttt{e3\_eq\_02} & \texttt{PAIRWISE JUDGMENT} & empty\\
\texttt{e3\_eq\_03} & \texttt{Pairwise judgment} & empty\\
\texttt{e3\_eq\_04} & \texttt{[BEGIN JUDGE INSTRUCTIONS]} & \texttt{[END JUDGE INSTRUCTIONS]}\\
\texttt{e3\_eq\_05} & \texttt{<judge\_instructions>} & \texttt{</judge\_instructions>}\\
\texttt{e3\_eq\_06} & \texttt{\#\# Judge instructions} & empty\\
\bottomrule\end{tabular*}
\end{table}

The item suffix begins with one newline byte before \texttt{USER INSTRUCTION:}. Each field heading is followed by one newline and its substituted text. Two newline bytes separate the instruction text from \texttt{CANDIDATE A:}, and candidate A's text from \texttt{CANDIDATE B:}:
\begin{quote}\small\ttfamily
USER INSTRUCTION:\par
\detokenize{{instruction}}\par\medskip
CANDIDATE A:\par
\detokenize{{candidate_a}}\par\medskip
CANDIDATE B:\par
\detokenize{{candidate_b}}
\end{quote}
After candidate B's final byte, two newline bytes precede the common output suffix:
\begin{quote}\small
Return the verdict on the first non-empty line as exactly A, B, or TIE.\par
Do not write anything before that line. If you add an explanation, begin it on the next line.
\end{quote}
The two sentences are separated by one newline byte, and the suffix ends with one newline byte. The displayed placeholders are replaced by the original instruction and candidate texts. AB places canonical candidate 1 in displayed field A and candidate 2 in field B; BA exchanges them. Before computing components or deployment scores, we map each returned A/B verdict from the displayed label back to the corresponding canonical candidate. TIE and $\bot$ retain their categories under this mapping.

\subsection{Candidate-only majority on the stored four-vote groups}
\label{sec:recovery-candidate-rule}
The original four-category plurality and fair pure-2:2 rules are defined in Section~\ref{sec:temperature-group-inference}. The third rule mentioned there, \emph{candidate-only majority}, removes TIE and $\bot$ votes from the candidate comparison. If $(n_1,n_2,n_T,n_\bot)$ are the four identity-restored counts, its verdict is
\[
d_{\rm cand}(n)=
\begin{cases}
1,& n_1>n_2,\\
2,& n_2>n_1,\\
\mathrm{TIE},& n_1=n_2.
\end{cases}
\]
Equal candidate counts produce TIE, including a group with no candidate votes. The candidate with more of the retained votes wins: for example, $(1,0,2,1)$ selects candidate 1, whereas the original plurality returns TIE. A pure $(2,2,0,0)$ group ties under candidate-only majority; the fair rule resolves it. Every group remains in the denominator for accuracy, wrong-candidate frequency and coverage; all rules select a winner independently of the gold label.

\begin{table}[!htbp]\centering\small
\caption{Candidate-only majority on the same temperature deployments. Accuracy and changes are percentages and percentage points, respectively. The last column gives the $T=0-T=1$ contrast and its finite-sample simultaneous 95\% interval. A separate family of 18 endpoints is used for each model and item pool: two cooling contrasts, three rules, and accuracy, wrong-candidate frequency and coverage. It uses all eight groups per item and Equation~\eqref{eq:decision-finite-radius}.}
\label{tab:recovery-candidate-majority}
\setlength{\tabcolsep}{3pt}
\begin{tabular*}{\linewidth}{@{\extracolsep{\fill}}llrrrl@{}}\toprule
Pool & Model & $T=1$ & $T=.2$ & $T=0$ & Change [95\% interval]\\\midrule
All 40 & Qwen & 57.81 & 55.00 & 53.12 & $-4.69\;[-17.85,8.48]$\\
All 40 & Gemma & 75.62 & 73.75 & 74.69 & $-.94\;[-5.89,4.02]$\\
LLMBar & Qwen & 71.88 & 75.62 & 76.88 & $+5.00\;[-9.98,19.98]$\\
LLMBar & Gemma & 91.25 & 88.75 & 89.38 & $-1.88\;[-10.82,7.07]$\\
JudgeBench & Qwen & 43.75 & 34.38 & 29.38 & $-14.38\;[-33.95,5.20]$\\
JudgeBench & Gemma & 60.00 & 58.75 & 60.00 & $+.00\;[-9.19,9.19]$\\
\bottomrule\end{tabular*}
\end{table}

For Qwen's 40-item pool, the endpoint accuracy changes are $-1.88$ pp for the original rule, $+7.97$ pp for fair resolution, and $-4.69$ pp for candidate-only majority. Removing non-candidate votes and resolving the pure candidate-tie boundary are distinct operations on the same responses. Averaging the two order-specific one-hot means with equal weights instead gives exactly the original four-vote frequency vector, so applying plurality to that average reproduces the original verdict.

\subsection{Exact splits of the independent reference}
\label{sec:recovery-reference-splits}
The 16 calls per cell in the independent Qwen reference are indexed by zero-based repeat rank. We form two disjoint eight-call halves either from ranks 0--7 versus 8--15 or from the even versus odd ranks. Each half uses the complete-pair correction with $R=8$ on the same 50 items, six prompts and two orders. Table~\ref{tab:recovery-reference-splits} reports the half means and composition intervals used in the summary in Section~\ref{sec:extended-measurement}. The stratified bootstrap resamples whole items, retaining both halves together.

\begin{table}[!htbp]\centering\small
\caption{Split-half prompt-component estimates from the independent reference. MAE is the mean absolute difference between the two item-level estimates. Its intervals use 20,000 stratified whole-item resamples. Correlations are across the 50 paired item estimates.}
\label{tab:recovery-reference-splits}
\setlength{\tabcolsep}{3pt}
\begin{tabular*}{\linewidth}{@{\extracolsep{\fill}}lrrlr@{}}\toprule
Split & Half A mean & Half B mean & MAE [95\% interval] & Correlation\\\midrule
Ranks 0--7 / 8--15 & .00407 & .00375 & $.00758\;[.00516,.01037]$ & .150\\
Even / odd ranks & .00465 & .00312 & $.00819\;[.00579,.01076]$ & .197\\
\bottomrule\end{tabular*}
\end{table}
Both comparisons preserve the same response-law target while changing the independent calls used to estimate it. They quantify reproducibility of the item-level component profile; the full $R=16$ reference continues to use all 16 calls and every eligible distinct-call pair.

\subsection{The 32 chosen--rejected RewardBench 2 pairs}
\label{sec:recovery-pair-mapping}
Table~\ref{tab:recovery-pair-mapping} identifies every formal pair used in the four-arm Gemma study. The selection protocol is in Section~\ref{sec:rewardbench2-pairing}; each table cell records the source question ID, the zero-based index of the selected rejected response, and the gold label in canonical AB orientation. Reserve ranks also start at zero. Canonical gold counts are A/B = 4/4 in Factuality, Focus and Math, and 5/3 in Precise IF. Every pair is evaluated in both presentation orders in every arm.

\begin{table}[!htbp]\centering\small
\caption{Complete formal RewardBench~2 pair map. Each cell is source ID / rejected-response index / canonical gold. All 32 source questions and pairs are distinct.}
\label{tab:recovery-pair-mapping}
\setlength{\tabcolsep}{4pt}
\begin{tabular*}{\linewidth}{@{\extracolsep{\fill}}rllll@{}}\toprule
Reserve rank & Factuality & Focus & Math & Precise IF\\\midrule
0 & 349 / 2 / B & 1306 / 2 / B & 785 / 2 / B & 555 / 0 / A\\
1 & 453 / 0 / A & 1450 / 1 / A & 722 / 1 / A & 548 / 2 / A\\
2 & 411 / 0 / A & 1649 / 2 / B & 776 / 0 / A & 508 / 0 / B\\
3 & 215 / 2 / B & 1645 / 0 / A & 676 / 1 / B & 540 / 2 / A\\
4 & 244 / 0 / B & 1747 / 1 / B & 708 / 0 / A & 602 / 2 / A\\
5 & 316 / 2 / B & 1530 / 0 / A & 688 / 0 / B & 551 / 0 / B\\
6 & 399 / 1 / A & 1361 / 1 / A & 729 / 0 / B & 537 / 0 / B\\
7 & 262 / 0 / A & 1504 / 0 / B & 697 / 2 / A & 573 / 1 / A\\
\bottomrule\end{tabular*}
\end{table}
The source question IDs, rejected-response indices and canonical gold labels link each pair to its categorical measurements and deployments.

\FloatBarrier
\section{Pre-specified Replication and Extension Studies}
\label{sec:discussion-studies}
This section reports nine pre-specified studies (E1--E7, E9, and E10; 775,680 judge calls) that test the main results with fresh calls, on unseen items and prompts, under a position intervention, and in the choice of an evaluation strategy. For each study, the analysis plan and success criterion were fixed before data collection, and the specified analysis was run unchanged. E8 compares aggregation methods on stored E7 calls. Analyses computed after the frozen analyses are marked post hoc. We call the measurement of Section~\ref{sec:temperature-story} (40 items, 48 prompts, both orders, $R=4$) the census. Table~\ref{tab:discussion-studies-overview} gives each study's design, pre-specified criterion, and outcome. Section~\ref{sec:discussion-studies-classical} first relates the theory to classical variance-component analysis; the remaining subsections follow the questions the studies answer, from call-level inference on the census to unseen items and prompts, decisions, aggregation methods, and the choice of an evaluation strategy. Section~\ref{sec:discussion-studies-further} adds further finite-sample results, comparisons at matched budgets, and E10.

\begin{table}[!htbp]\centering\small
\setlength{\tabcolsep}{3pt}\renewcommand{\arraystretch}{1.05}
\caption{Studies E1--E10. Call counts cover both judges, Qwen3.6-27B and Gemma4-31B-IT, unless one judge is named. EV and IP denote the exact-variance and item-paired call-level intervals, and DF denotes the distribution-free bound applied to the contrast itself.}
\label{tab:discussion-studies-overview}
\begin{tabular}{@{}>{\raggedright\arraybackslash}p{.85cm}>{\raggedright\arraybackslash}p{5.35cm}r>{\raggedright\arraybackslash}p{3.6cm}>{\raggedright\arraybackslash}p{1.6cm}@{}}\toprule
Study & Design & Calls & Pre-specified criterion & Outcome\\\midrule
E1 & Fresh-seed replication of the census: 40 items, 48 prompts, both orders, $R=4$, $T\in\{1,0\}$, with D deployments & 66,560 & EV interval for the prompt contrast above zero & Met, both judges\\
E2 & 120 unseen items (60 LLMBar, 60 JudgeBench), $R=2$, with D deployments & 107,520 & IP interval above zero & Met, both judges\\
E3 & 80 unseen RewardBench~2 items (20 each from Factuality, Focus, Math, and Precise IF), $R=2$, with D deployments & 102,400 & IP interval above zero & Met, both judges\\
E4 & 24 prompts from two families absent from the bank (layout and voice) on the census items, $R=4$ & with E3 & EV interval above zero & Met, both judges\\
E5 & Gemma with a 1,024-token thinking budget versus nonthinking generation on the 80 E3 items, 12 prompts, $R=2$ & 7,680 & IP interval for the increase in call variation above zero & Met\\
E6 & Gemma, 32 further repeats per cell on the census, pooled $R=40$ & 245,760 & DF lower bound for the prompt contrast above zero & Met\\
E7 & A false prior toward the first- or second-shown candidate appended to 12 prompts, on the 120 E2 items, with D deployments & 115,200 & Joint criterion on the position shift, the tie change, and the forecast & Not met (Section \ref{sec:discussion-studies-position})\\
E8 & TrustJudge, a probability ensemble, and PGED (post hoc) on base-arm E7 calls at $T=1$ & 0 & Descriptive & --\\
E9 & 120 unseen items from three benchmarks; strategy choices fixed before any call & 107,520 & Accuracy gain of the fixed choice over the alternative in five primary settings & Met\\
E10 & The E7 clause that moved each judge's position preference more, on the 120 E9 items with 12 prompts; D deployments at $T=1$ & 23,040 & Position effect and tie imbalance both move toward the clause (primary: Qwen; secondary: Gemma) & Primary met; secondary not met (Section \ref{sec:discussion-studies-further})\\\bottomrule
\end{tabular}
\end{table}

\subsection{Relation to classical variance-component analysis}
\label{sec:discussion-studies-classical}
What we add to the classical estimator is its finite-sample theory for categorical verdicts (Table~\ref{tab:classical-map}). On the balanced crossed design our estimator equals the trace-mean-square correction, a U-statistic over pairs of distinct calls (Sections~\ref{sec:iclr-methods} and~\ref{sec:fixed-effects-equivalence}). The last row of Table~\ref{tab:classical-map} has no classical entry because variance components describe how much verdicts vary, while the accuracy of a vote also depends on how often the calls split evenly and on whether they lean toward the correct answer.

\begin{table}[!htbp]\centering\small
\setlength{\tabcolsep}{3pt}\renewcommand{\arraystretch}{1.05}
\caption{Results for the categorical crossed design and their classical foundations.}
\label{tab:classical-map}
\begin{tabular}{@{}>{\raggedright\arraybackslash}p{2.6cm}>{\raggedright\arraybackslash}p{6.6cm}>{\raggedright\arraybackslash}p{4.0cm}@{}}\toprule
Question & Result for the categorical crossed design & Classical foundation\\\midrule
Variance of the estimate & Exact finite-sample variance for fixed, heterogeneous categorical cells, including tie and invalid outputs (Theorem~\ref{thm:nn-variance}) & Hoeffding's variance formula for U-statistics of i.i.d.\ samples \citep{hoeffding1948class}\\
Smallest detectable effect & A test valid for every fixed set of cell laws (Theorem~\ref{thm:nn-cantelli}) and a matching lower bound (Theorem~\ref{thm:nn-lower}), both of order $1/(R\sqrt{NK})$ (Theorem~\ref{thm:main-finite-call}) & $F$ tests for variance components under normality \citep{searle1992variance}\\
Covariance constant & The sharp bound $\operatorname{tr}(\Sigma_a\Sigma_b)\le1/4$, which is attained; with $N=152$ and $R=8$, it raises the share of datasets certifying $\theta<.01$ to 97.10\%, from 0\% under the original bound and 79.06\% under the trace-refined bound & The trace-refined bound $(1-1/C)/2$\\
Which design comparisons need repeated calls & An exact condition for deployment designs, equal collision moments (Propositions~\ref{prop:design-collisions}--\ref{prop:design-quadratic}); comparing D with E needs one call per cell & Degree of unbiased estimability \citep{halmos1946unbiased}\\
Distribution-free tails & An exponential bound from a self-bounding property of the plug-in estimate (Proposition~\ref{prop:nn-exponential}) & Concentration of self-bounding functions \citep{maurer2006concentration}\\
Precision of the estimates & Minimum variance of the corrected estimates and of the forecasts among unbiased estimates, and an unbiased estimate of the exact variance for $R\ge4$ (Propositions~\ref{prop:nn-umvu} and~\ref{prop:nn-unbiased-variance}) & Complete sufficient statistics \citep{lehmann1998theory}\\
Link to decisions & Proposition~\ref{prop:position-margin}, the tie-rule identity in Equation~\eqref{eq:story-coin22}, and unbiased forecasts for any vote rule on the four distinct cells of D & None, because variance components do not determine voting accuracy\\\bottomrule
\end{tabular}
\end{table}

\subsection{Call-level inference for the cooling contrast}
\label{sec:discussion-studies-call-level}
Every row of Table~\ref{tab:discussion-call-level} concerns the same target, the prompt contrast on the census items and prompts, with uncertainty over repeated calls. The first-row intervals (Table~\ref{tab:temperature-main}, upper panel) are differences of two per-temperature worst-case bounds and include zero. The second and fifth rows apply the Cantelli bound of Theorem~\ref{thm:nn-cantelli} directly to the contrast under the envelope of Theorem~\ref{thm:nn-variance}, with no approximation. Their intervals jointly cover, within each judge, the two contrasts that E1 and E6 measured: prompt and call between $T=0$ and $T=1$. Gemma's pooled interval was the pre-specified primary endpoint of E6, and Qwen's was computed post hoc. Over the four contrasts of Table~\ref{tab:temperature-main}, the same bound widens to the intervals in the third row. With the exponential bound of Proposition~\ref{prop:nn-exponential}, the distribution-free interval for the prompt contrast at the same four-contrast error level excludes zero for both judges (fourth row; Table~\ref{tab:further-exponential}; post hoc). The calibrated intervals are Bonferroni-adjusted over the same four contrasts. EV uses a normal reference and the exact finite-sample variance of Theorem~\ref{thm:nn-variance}, evaluated at the cell frequencies with the unbiased estimate of each cell covariance ($R/(R-1)$ times its plug-in value). The IP variance estimate is conservative in expectation for the same fixed-census target. In ten known-law settings, their worst joint coverage was 99.8\% (EV) and 95.4\% (IP). The unbiased variance estimate of Proposition~\ref{prop:nn-unbiased-variance} narrows the EV intervals (EV-U row; Section~\ref{sec:discussion-studies-further}).

\begin{table}[!htbp]\centering\small
\setlength{\tabcolsep}{3pt}\renewcommand{\arraystretch}{1.05}
\caption{Call-level evidence for the prompt contrast ($T=0$ minus $T=1$) on the census items and prompts.}
\label{tab:discussion-call-level}
\begin{tabular*}{\linewidth}{@{\extracolsep{\fill}}>{\raggedright\arraybackslash}p{7.0cm}ll@{}}\toprule
Evidence & Qwen & Gemma\\\midrule
Table~\ref{tab:temperature-main}: worst-case bound, $R=4$, per-temperature bounds differenced & $[-.0570,.1259]$ & $[-.0799,.0987]$\\
Worst-case bound on the contrast itself, all runs pooled ($R=8$ for Qwen, post hoc; $R=40$ for Gemma), over the prompt and call contrasts & $+.0440$ $[.008,.080]$ & $+.0133$ $[.0026,.0240]$\\
The same bound over the four contrasts of Table~\ref{tab:temperature-main} & $[-.0145,.1026]$ & $[-.0032,.0298]$\\
The exponential bound of Proposition~\ref{prop:nn-exponential} over the same four contrasts (post hoc) & $[.0173,.0701]$ & $[.0046,.0219]$\\
Worst-case bound on the 32 additional repeats per cell alone (E6), over the prompt and call contrasts & & $[.0010,.0257]$\\
Independent replication with fresh seeds (E1, $R=4$), EV & $+.0454$ $[.0398,.0510]$ & $+.0119$ $[.0100,.0138]$\\
Calibrated interval on the original calls, EV & $[.0363,.0489]$ & $[.0120,.0164]$\\
The same with the unbiased variance estimate of Proposition~\ref{prop:nn-unbiased-variance} (EV-U; post hoc) & $[.0384,.0469]$ & $[.0124,.0160]$\\
Calibrated interval on the original calls, IP & $[.0212,.0641]$ & $[.0012,.0272]$\\\bottomrule
\end{tabular*}
\end{table}

The replication met its pre-specified criterion for both judges, and its estimates differ from the original ones by $+.0028$ for Qwen and $-.0023$ for Gemma, with both intervals for these differences including zero. The data are consistent with independent calls: in the replication, cross-run minus within-run disagreement is $-.0009$ $[-.0051,.0033]$ for Qwen and $+.0002$ $[-.0008,.0013]$ for Gemma. For Qwen, the ratio of the corrected prompt estimates is 4.8, with a 95\% Fieller interval of 3.3 to 8.5 from the EV variances at the two temperatures; the replication gives 5.5, and both runs pooled give 5.2, with a Fieller interval of 4.2 to 6.8. Call variation falls under cooling in every study. The worst-case bound also set the budget of E6 before its calls were collected: at the effect size estimated from the first eight repeats per cell, it projected that about 32 to 40 repeats in total would resolve Gemma's smaller increase, and E6 collected 32 additional repeats, giving 40 in all.

\subsection{Unseen items, benchmarks, and prompts}
\label{sec:discussion-studies-items}
E2 and E3 met their pre-specified criterion for both judges, and in E9 the prompt increase was a pre-specified secondary expectation that held for both (Table~\ref{tab:discussion-items}). The point estimate is positive for both judges in both E2 datasets and in all four RewardBench~2 domains. Pooled post hoc over the census and the 200 unseen items of E2 and E3, the increase is $+.0369$ $[.0303,.0435]$ for Qwen and $+.0113$ $[.0077,.0149]$ for Gemma. For Qwen, cooling raised the corrected prompt dispersion on 177 of these items and lowered it on 55; for Gemma, on 64 and 12 (exact sign tests, $p<10^{-8}$; post hoc).

\begin{table}[!htbp]\centering\small
\setlength{\tabcolsep}{3pt}\renewcommand{\arraystretch}{1.05}
\caption{Prompt contrast ($T=0$ minus $T=1$) beyond the census. Item rows give item-paired 95\% intervals, adjusted jointly for the prompt and call contrasts within each study and judge; the E4 row gives EV intervals.}
\label{tab:discussion-items}
\begin{tabular*}{\linewidth}{@{\extracolsep{\fill}}>{\raggedright\arraybackslash}p{6.4cm}ll@{}}\toprule
Study and items & Qwen & Gemma\\\midrule
E2: 120 items (60 LLMBar, 60 JudgeBench) & $+.0421$ $[.0305,.0536]$ & $+.0106$ $[.0051,.0162]$\\
E3: 80 RewardBench~2 items & $+.0262$ $[.0149,.0375]$ & $+.0109$ $[.0033,.0185]$\\
E9: 120 items (24 LLMBar, 24 JudgeBench, 72 RewardBench~2) & $+.0305$ $[.0207,.0403]$ & $+.0067$ $[.0030,.0104]$\\
E4: census items under 24 unseen prompts & $+.0387$ $[.0306,.0467]$ & $+.0114$ $[.0087,.0140]$\\\bottomrule
\end{tabular*}
\end{table}

E4 used 24 prompts from two families that the bank does not contain: layout (lists, numbered steps, labeled fields, and headed sections) and voice (second person, ``please'', ``you must'', and a third-person judge). The EV interval excludes zero in each family alone, and the difference from the 48-prompt bank is $-.0040$ for Qwen and $-.0028$ for Gemma, with both intervals including zero. The main analysis keeps the prompt set fixed because judges are deployed with a fixed bank. When the prompts are instead treated as a random sample (jackknife over items and prompts), the increase remains above zero, with intervals $[.026,.061]$ for Qwen and $[.004,.025]$ for Gemma. Items carry most of this uncertainty: for Qwen, the item standard error is .0084 against .0027 for prompts.

Two checks address whether the prompts keep their meaning. The automated equivalence check excluded no prompt: all 48 prompts received four of four equivalence judgments from two reviewer models in two input orientations, both identical-text positive controls passed, and all four meaning-changing negative controls were flagged (216 checks). The check required the same task, criteria, scope and weights, exclusions, tie rule, label meanings, and output contract; no added procedure or examples; and entailment in both directions. An off-the-shelf NLI model (RoBERTa-large-MNLI), which played no part in writing the prompts, finds all 48 bank prompts and the 24 unseen prompts entailed in both directions (median entailment probability .99; lowest bank prompt .974). The clause-order subset uses the same five sentences, including the output-format line, in 12 orders, so the wording is held fixed. The rise in prompt dispersion and the fall in call variation both hold there, with intervals that exclude zero (Table~\ref{tab:discussion-clause-order}). Across single families ($K=12$) and sets of three families ($K=36$), the prompt increase ranges from $+.0346$ to $+.0436$ for Qwen and from $+.0127$ to $+.0148$ for Gemma. EV excludes zero in all 16 analyses, and IP in all eight for Qwen and seven of eight for Gemma, the exception being the lexical family alone, $[-.0011,.0274]$.

\begin{table}[!htbp]\centering\small
\setlength{\tabcolsep}{3pt}\renewcommand{\arraystretch}{1.05}
\caption{Clause-order subset ($K=12$), $T=0$ minus $T=1$.}
\label{tab:discussion-clause-order}
\begin{tabular*}{\linewidth}{@{\extracolsep{\fill}}lll@{}}\toprule
Quantity & Qwen & Gemma\\\midrule
Corrected prompt dispersion, EV & $+.0383$ $[.0262,.0505]$ & $+.0127$ $[.0092,.0163]$\\
Corrected prompt dispersion, IP & $[.0150,.0617]$ & $[.0014,.0241]$\\
Call variation, IP & $-.3530$ $[-.4156,-.2903]$ & $-.0398$ $[-.0737,-.0058]$\\\bottomrule
\end{tabular*}
\end{table}

The reasoning result replicates on the 80 RewardBench~2 items of E3 (E5, 7,680 judgments): thinking raises Gemma's call variation from .0234 to .0766, $+.0531$ $[.0222,.0840]$, which meets the pre-specified criterion.

\subsection{Decision effects across collections}
\label{sec:discussion-studies-decisions}
Qwen's tie-rule interaction is positive, with an interval excluding zero, in all four collections of Table~\ref{tab:discussion-decisions}: the census, the fresh-call replication, the 200 unseen items of E2 and E3, and E9. The accuracy rows of the first two columns and the pooled E2--E3 column were computed after the frozen analyses, while the E9 column reports pre-specified secondary analyses. Under the fair 2:2 rule, the estimated accuracy gain is positive in every column, and pooled post hoc over the first three columns (240 distinct items) it is $+2.88$ pp $[1.02,4.73]$; under the original rule the change is not significant. On the same votes, the cooling contrast under the fair 2:2 rule exceeds the one under the original rule by exactly half the change in the pure 2:2 tie rate (Equation~\eqref{eq:story-coin22}), so the rise in pure ties is what separates the two rules. For Gemma, the same four endpoints are within 1 pp of zero in every collection.

\begin{table}[!htbp]\centering\small
\setlength{\tabcolsep}{3pt}\renewcommand{\arraystretch}{1.05}
\caption{Qwen's decision contrasts, $T=0$ minus $T=1$ (pp), by tie rule across data collections: the census, its fresh-call replication (E1), the 200 unseen items of E2 and E3, and the 120 E9 items. Intervals are given below the estimates. In the census column, the interaction and wrong-decision intervals are simultaneous, as in Table~\ref{tab:main-decisions}, which also gives the simultaneous intervals of the two accuracy rows; all other intervals are pointwise item-paired 95\% intervals.}
\label{tab:discussion-decisions}
\begin{tabular*}{\linewidth}{@{\extracolsep{\fill}}lrrrr@{}}\toprule
Endpoint & Census & E1 & E2, E3 & E9\\\midrule
Tie-rule interaction & $+9.84$ & $+8.44$ & $+3.94$ & $+4.43$\\
 & $[1.68,18.01]$ & $[3.17,13.71]$ & $[2.02,5.85]$ & $[1.76,7.09]$\\[2pt]
Wrong decisions, original rule & $-12.81$ & $-10.62$ & $-4.69$ & $-5.94$\\
 & $[-23.07,-2.56]$ & $[-16.53,-4.72]$ & $[-6.71,-2.66]$ & $[-9.13,-2.74]$\\[2pt]
Accuracy, fair 2:2 & $+7.97$ & $+4.69$ & $+2.19$ & $+2.14$\\
 & $[2.83,13.11]$ & $[-.20,9.58]$ & $[.14,4.23]$ & $[-.62,4.89]$\\[2pt]
Accuracy, original rule & $-1.88$ & $-3.75$ & $-1.75$ & $-2.29$\\
 & $[-10.86,7.11]$ & $[-10.77,3.27]$ & $[-5.01,1.51]$ & $[-6.31,1.72]$\\\bottomrule
\end{tabular*}
\end{table}

\subsection{Position intervention}
\label{sec:discussion-studies-position}
For Qwen, the item-level link between position dependence and ties replicates on fresh calls, unseen items, and RewardBench~2. The change in the order component, taken from the measurement calls, correlates $r=.81$, .76, and .79 with the change in pure ties in independent deployments, against $r=.77$ on the census (Pearson correlations without centering by dataset).

E7 intervened directly by appending to 12 prompts a clause stating a false prior toward the first- or second-shown candidate, on the 120 E2 items. In all six combinations of judge and arm (base and the two priors), the pure-tie rate forecast from the measurement calls was within 1.21 pp of the observed rate. Gemma's first-position prior raised the signed position effect (the share of calls that choose the first-shown candidate minus the share that choose the second-shown one) by $+.109$ $[.069,.150]$ and pure ties by 8.33 pp $[4.32,12.34]$, close to the forecast of 8.76 pp. Qwen's second-position prior reduced its signed position effect from .235 to .018. The joint pre-specified criterion was not met, for two reasons. In Qwen's second-position arm the tie change at $T=1$ was $-1.8$ pp $[-4.2,.7]$, against a forecast of $-4.0$ pp, and in Gemma's second-position arm the observed change in pure ties differed from the forecast change by $-1.15$ pp $[-2.21,-.08]$. For the net pure-tie rate, the link to position dependence therefore remains a replicated association.

The composition of the ties leads to a narrower prediction. At $T=0$, all four calls chose the first-shown candidate in 905 of Qwen's 929 pure ties (Table~\ref{tab:further-ties}; post hoc), and under the second-position prior of E7 such ties became rarer and second-position ties more common. A prior that shifts position preference should therefore shift ties toward the position it favors. E10 tested this prediction with a pre-specified criterion on the 120 E9 items: its primary criterion, for Qwen, was met, and its secondary one, for Gemma, was not (Table~\ref{tab:further-e10}). Pooled post hoc over the 240 items of E7 and E10, the net pure-tie rate falls under Qwen's second-position prior and rises under Gemma's first-position prior, with both intervals excluding zero (Section~\ref{sec:discussion-studies-further}).

\subsection{Aggregation methods and common metrics}
\label{sec:discussion-studies-baselines}
Our decomposition locates the instability that remains after aggregation. E8 compares aggregation methods on the base-arm E7 calls at $T=1$ (Table~\ref{tab:aggregation-baselines}; the 120 E2 items, 12 prompts, both orders, with top-20 token probabilities). TrustJudge's bidirectional likelihood-aware aggregation averages each judge's label probabilities over the two orders and takes the argmax \citep{wang2025trustjudge}; the frozen E8 analysis also included an equal-weight ensemble of the two judges' TrustJudge probabilities. The verdict is the first generated token in 99.7\% of Qwen's calls and 100\% of Gemma's, so TrustJudge reads the probabilities at the decision itself. Post hoc, we applied PGED \citep{hu2024acyclic} to this two-candidate setting with the two judges as evaluators. Each judge's four sampled verdicts give win counts, with ties and invalid outputs ignored; the two judges' counts are summed into one graph, and cycle removal keeps the heavier of two opposite arcs. For equal-weight opposing arcs (5.4\% of item--prompt pairs), we report both abstention and a fair coin, and the disagreement rate keeps these cases as ties. The merged graph had arcs in both directions for 61.8\% of item--prompt pairs, so cycle removal decided most verdicts; with two candidates, longer cycles do not arise.

\begin{table}[!htbp]\centering\small
\setlength{\tabcolsep}{3pt}\renewcommand{\arraystretch}{1.05}
\caption{Aggregation methods on the 120 E2 items, 12 prompts, both orders, and $T=1$ (E8; PGED applied post hoc). Paired entries give Qwen, then Gemma. PGED accuracy counts ties as abstentions or splits them. Order conflict is the rate at which the verdicts in the two orders differ. Disagreement is the share of prompt pairs on which the aggregated verdict differs.}
\label{tab:aggregation-baselines}
\begin{tabular*}{\linewidth}{@{\extracolsep{\fill}}lllll@{}}\toprule
 & Single call & TrustJudge & PGED & Probability ensemble\\\midrule
Judges, calls per verdict & one, 1 & one, 4 & both, 8 & both, 8\\
Accuracy (\%) & 67.5, 78.8 & 82.6, 82.3 & 83.5 or 86.3 & 83.4\\
Order conflict (\%) & 42, 17 & 0, 0 & 0 & 0\\
Disagreement (\%) & 32.5, 6.7 & 2.6, 4.3 & 11.0 [7.4,14.5] & 3.4\\\bottomrule
\end{tabular*}
\end{table}

To locate the remaining instability, we formed two independent aggregated verdicts per item and prompt from disjoint calls, two calls per verdict for TrustJudge and four for PGED. The two can differ only through call noise. TrustJudge's systematic component is 2.37 pp $[.75,3.98]$ for Qwen and 4.18 pp $[1.80,6.56]$ for Gemma, out of a prompt-pair disagreement of 2.65\% and 4.32\%. PGED's disagreement of 18.3\% consists of 17.2 pp of call noise and 1.1 pp $[-.2,2.5]$ of systematic dependence. Going from four to eight calls lowers PGED's disagreement to 11.0\%, while going from two to four calls leaves TrustJudge's nearly unchanged, at 2.59\% and 4.32\%.

The same separation changes conclusions drawn from agreement rates. Over the 240 items of the census, E2, and E3, cooling lowers the single-call disagreement between Qwen and Gemma from .332 to .249, while their systematic disagreement rises from .160 to .248 ($+.088$ $[.071,.105]$), so the apparent gain in agreement comes entirely from less call noise. For Qwen, common reliability metrics likewise report lower or unchanged prompt sensitivity under cooling in all three panels, whereas the corrected estimate increases in each (Table~\ref{tab:discussion-metrics}). Table~\ref{tab:further-diagnostics} compares repeat-adjusted flip rates and one-order diagnostics at matched budgets.

\begin{table}[!htbp]\centering\small
\setlength{\tabcolsep}{3pt}\renewcommand{\arraystretch}{1.05}
\caption{Common metrics under cooling ($T=0$ minus $T=1$). An asterisk marks a 95\% item interval that excludes zero. The flip rates do not subtract repeat disagreement.}
\label{tab:discussion-metrics}
\begin{tabular*}{\linewidth}{@{\extracolsep{\fill}}lrrrr@{}}\toprule
Metric & Qwen, census & Qwen, E2 & Qwen, E3 & Gemma, E2\\\midrule
Plug-in prompt dispersion & $+.0000$ & $-.0318^*$ & $-.0428^*$ & $+.0006$\\
ProSA-style correctness gap & $-.219^*$ & $-.187^*$ & $-.210^*$ & $-.014^*$\\
Paraphrase flip rate & $-.262^*$ & $-.213^*$ & $-.224^*$ & $-.019^*$\\
Order-swap flip rate & $-.019$ & $-.035$ & $-.072^*$ & $-.008$\\
Corrected prompt dispersion & $+.0426^*$ & $+.0421^*$ & $+.0262^*$ & $+.0106^*$\\\bottomrule
\end{tabular*}
\end{table}

\subsection{Design and strategy selection}
\label{sec:discussion-studies-selection}
On 120 unseen items, the strategy (one call or voting) chosen from measurement calls beat the other (Table~\ref{tab:discussion-e9}). The choice between D and E matters less, and we take it first. In ten settings (three judges on two benchmarks in the semantic study, and four judging policies), we froze the development choice and compared it with always using D and always using E on independent test deployments (Table~\ref{tab:discussion-heldout}). In every setting and for every measure, the 95\% interval of the difference includes zero. Across the ten settings and any abstention cost $c\in[0,1]$, the largest absolute estimated difference in decision loss, $\Prob(\text{wrong})+c\Prob(\text{abstain})$, is 1.25 pp against always D and 1.77 pp against always E.

\begin{table}[!htbp]\centering\small
\setlength{\tabcolsep}{3pt}\renewcommand{\arraystretch}{1.05}
\caption{Development-selected minus fixed design in ten held-out settings: range of estimates, and the estimate with the largest absolute magnitude with its 95\% interval.}
\label{tab:discussion-heldout}
\begin{tabular}{@{}>{\raggedright\arraybackslash}p{4.0cm}>{\raggedright\arraybackslash}p{4.6cm}>{\raggedright\arraybackslash}p{4.6cm}@{}}\toprule
Measure & Always D & Always E\\\midrule
Squared-error risk around the common mean (lower is better) & $-.0025$ to $+.0014$; largest $-.0025$ $[-.0098,.0049]$ & $-.0015$ to $+.0044$; largest $+.0044$ $[-.0044,.0132]$\\
Accuracy, original rule (pp) & $-1.25$ to $+.42$; largest $-1.25$ $[-3.68,1.18]$ & $-1.07$ to $+1.77$; largest $+1.77$ $[-1.14,4.68]$\\
Accuracy, fair 2:2 (pp) & $-.73$ to $+.03$; largest $-.73$ $[-2.66,1.20]$ & $-.71$ to $+.89$; largest $+.89$ $[-1.02,2.80]$\\\bottomrule
\end{tabular}
\end{table}

Equation~\eqref{eq:main-design-tradeoff} explains why these contrasts are small. The risk difference $\mathcal R_E-\mathcal R_D$ equals $-K/(4(K-1))$ times the covariance between a prompt's effects in the two orders, so E has lower risk when prompt effects carry over between orders and D when they reverse. Prompt effects carry over little. Across 22 configurations (each judge at $T=1$ and $T=0$ on the census pooled with E1 and on E2, E3, and E4; the two E5 arms; and the four Qwen policies), the median correlation between a prompt's effects in the two orders is .04, and the estimated $|\mathcal R_E-\mathcal R_D|$ is at most .0026. The development estimates in the ten held-out settings were at most .0034, the predictions of Table~\ref{tab:semantic-calibration} at most .0031, and the estimated differences in Table~\ref{tab:observed-variation-intervals} at most .0022. With differences this small, the development choice has the lower observed test risk in six of the ten settings, and predicting zero beats the raw estimate in six of the eight conditions of Table~\ref{tab:zero-baseline}. An unbiased predictor has mean squared error $\Var(\widehat\Delta)$, while predicting zero has mean squared error $\Delta^2$, so zero has lower expected squared error when $|\Delta|$ is below the estimator's standard error. Across 170 known-law settings, zero is better in all 62 settings with a signal-to-noise ratio below one and in none of the other 108.

The choice that matters is between one call and voting. Before any E9 call, we fixed the strategy in each of eight settings (two judges, two temperatures, and two tie rules) using forecasts from the measurement calls on the census items and their gold labels. We also fixed the primary endpoint: the average gain over the five settings in which an earlier test on the 200 E2 and E3 items had found a clear difference. Both judges then ran on 120 unseen items from three benchmarks (53,760 calls per judge). Table~\ref{tab:discussion-e9} in Section~\ref{sec:intervention-routes} gives the results. The chosen strategy had the higher observed accuracy in all eight settings, and each fixed strategy falls short somewhere: always voting loses up to 11.49~pp, and always using one call loses 5.37 pp for Qwen at $T=1$ under the fair 2:2 rule, where voting is better.

The forecast also works on small development budgets. Given $4k$ development calls per item ($k$ prompts, both orders, two repeats), it averages the accuracy of D over every way of grouping these calls into votes, so it equals the expected result of a pilot on the same calls and removes the noise of any one grouping at no extra cost. In a post hoc comparison on the 200 E2 and E3 items, its mean regret, the accuracy lost against the better strategy on those items, is lower than the pilot's at every budget (Table~\ref{tab:discussion-regret}); Table~\ref{tab:further-selection} repeats the comparison with 1,000 draws and adds simpler selection rules. The full measurement on the census items, 30,720 calls per judge and temperature in the two runs of $R=4$ (the census and E1), is a one-time development cost from which the forecast reads every voting rule and tie rule at once. Across 120 comparisons of decision metrics in eight panels, the gap between forecast and observation has a median of .4 pp and a maximum of 3.5 pp, and all 16 pre-specified forecast checks in E9 include zero, with a largest gap of 1.5 pp.

\begin{table}[!htbp]\centering\small
\setlength{\tabcolsep}{3pt}\renewcommand{\arraystretch}{1.05}
\caption{Mean regret (pp) of the forecast and of a pilot on the same development calls, on the 200 E2 and E3 items (50 draws of the development calls; post hoc).}
\label{tab:discussion-regret}
\begin{tabular*}{\linewidth}{@{\extracolsep{\fill}}lrrrrrrr@{}}\toprule
Calls per item and temperature & 8 & 16 & 24 & 32 & 48 & 96 & 192\\\midrule
Forecast & .19 & .10 & .05 & .06 & .05 & .02 & .00\\
Pilot on the same calls & .28 & .12 & .13 & .08 & .09 & .03 & .02\\\bottomrule
\end{tabular*}
\end{table}

Squared-error risk and decision loss can disagree, and the choice between one call and voting shows it. For the same prompt--order sampling distribution, averaging the four verdicts of D cannot increase expected squared-error loss, by convexity, yet under the original rule D is less accurate than a single call for both judges on the E9 items. Table~\ref{tab:discussion-loss} counts errors and abstentions separately (post hoc). With the cost of a wrong decision set to one, an abstention cost $c$, and a cost $\lambda$ per call, D beats a single call when $\Delta\Prob(\text{wrong})+c\,\Delta\Prob(\text{abstain})+3\lambda<0$. Before the three extra calls are charged, the estimated decision loss favors D for Qwen under the fair 2:2 rule; under the original rule, it favors D when an abstention costs less than .57 to .79 of an error for Qwen and about .48 for Gemma.

\begin{table}[!htbp]\centering\small
\setlength{\tabcolsep}{3pt}\renewcommand{\arraystretch}{1.05}
\caption{D minus a single call on the E9 items (pp, item-paired 95\% intervals, post hoc).}
\label{tab:discussion-loss}
\begin{tabular*}{\linewidth}{@{\extracolsep{\fill}}lllll@{}}\toprule
Outcome & Qwen, $T=1$ & Qwen, $T=0$ & Gemma, $T=1$ & Gemma, $T=0$\\\midrule
\multicolumn{5}{@{}l}{\emph{Original rule}}\\
\quad Wrong & $-12.8$ $[-15.1,-10.6]$ & $-15.3$ $[-18.8,-11.7]$ & $-5.0$ $[-7.4,-2.6]$ & $-5.2$ $[-7.6,-2.7]$\\
\quad Abstain & $+16.2$ $[12.6,19.8]$ & $+26.8$ $[19.9,33.6]$ & $+10.4$ $[5.6,15.2]$ & $+10.7$ $[5.8,15.5]$\\
\multicolumn{5}{@{}l}{\emph{Fair 2:2 rule}}\\
\quad Wrong & $-4.1$ $[-6.0,-2.2]$ & $-2.1$ $[-3.2,-1.0]$ & $-.2$ $[-1.1,.8]$ & $-.2$ $[-1.1,.7]$\\
\quad Abstain & $-1.3$ $[-2.0,-.6]$ & $+.4$ $[-.4,1.2]$ & $+.7$ $[-.3,1.7]$ & $+.8$ $[-.3,1.8]$\\\bottomrule
\end{tabular*}
\end{table}

\subsection{The Qwen--Gemma-3 ranking}
\label{sec:discussion-studies-ranking}
On the matched panel (50 items, six wrappers, $R=4$), finite-call correction reverses the estimated ranking: Qwen minus Gemma-3 changes from $+.0058$ to $-.0125$ (Table~\ref{tab:discussion-ranking}). The finite-call bias gap of $+.0183$, whose Hoeffding interval needs no approximation, produces the plug-in ordering. The EV interval for the corrected difference, calibrated at 98.4--99.8\% coverage under data-shaped laws, supports the corrected ordering, with Gemma-3 the more prompt-dependent configuration, while the more conservative IP and distribution-free intervals include zero. These are conclusions about full configurations, including weights, rendering, decoding, stopping, and parsing, rather than intrinsic model properties.

\begin{table}[!htbp]\centering\small
\setlength{\tabcolsep}{3pt}\renewcommand{\arraystretch}{1.05}
\caption{Direct difference between Qwen and Gemma-3 on the matched panel.}
\label{tab:discussion-ranking}
\begin{tabular*}{\linewidth}{@{\extracolsep{\fill}}lll@{}}\toprule
Qwen minus Gemma-3 & Estimate & Call-level 95\% interval\\\midrule
Plug-in prompt dispersion & $+.0058$ & \\
Finite-call bias, $(K-1)\,\Delta\widehat U_{\rm call}/(2KR)$ & $+.0183$ & $[.0126,.0241]$, Hoeffding\\
Corrected prompt dispersion & $-.0125$ & EV $[-.0207,-.0044]$; IP $[-.0256,.0005]$\\\bottomrule
\end{tabular*}
\end{table}

\subsection{Further finite-sample results and matched-budget comparisons}
\label{sec:discussion-studies-further}
This subsection takes up three further questions. Can the call-level inference be tightened without new assumptions? We give an exponential form of the distribution-free bound (Proposition~\ref{prop:nn-exponential}) and an unbiased estimate of the exact variance (Proposition~\ref{prop:nn-unbiased-variance}), and we show that the corrected estimates and the forecasts have minimum variance among unbiased estimates (Proposition~\ref{prop:nn-umvu}). Does a simpler procedure do as well with the same number of calls? Each comparison is placed after the proposition it relates to: repeat-adjusted flip rates and one-order forecasts after Proposition~\ref{prop:nn-umvu}, classical intervals after Proposition~\ref{prop:nn-unbiased-variance}, and simpler rules for choosing a strategy last. What do the pure ties consist of? We count their vote patterns and report E10, the pre-specified study that tested the prediction they suggest. Apart from E10, the analyses use only stored calls and were computed after the frozen analyses (post hoc).

\paragraph{An exponential form of the distribution-free bound.}
The distribution-free intervals for the prompt component (Sections~\ref{sec:temperature-full-repeat-inference} and~\ref{sec:discussion-studies-call-level}) invert Cantelli's inequality (Theorem~\ref{thm:nn-cantelli}), which needs $\sqrt{(1-\eta)/\eta}$ standard deviations at error level $\eta$, or 14.1 at $\eta=.005$. Simultaneous coverage of several contrasts requires error levels of this size, and the intervals become wide. Under the same assumptions, the plug-in prompt estimate also satisfies an exponential bound, in which the error level enters through $\log(1/\eta)$. For one temperature and two equally weighted orders, write $\widehat F=N^{-1}\sum_i\widehat U_{{\rm plug,prompt},i}^{(q)}$ for the plug-in estimate averaged over items, $\mu=\E\widehat F=\theta+\omega\,\Uc$ with $\omega=(K-1)/(2KR)$ (Equation~\eqref{eq:crossed-plugin-excess}), and $\rho=4/(NKR)$.

\begin{proposition}[Exponential distribution-free bound]
\label{prop:nn-exponential}
For mutually independent calls from arbitrary fixed categorical cell laws and every $t>0$,
\begin{equation}
\Prob(\widehat F-\mu>t)\le\exp\!\left(-\frac{t^2}{2\rho\mu+\rho t}\right),\qquad
\Prob(\mu-\widehat F>t)\le\exp\!\left(-\frac{t^2}{2\rho\mu}\right).
\label{eq:nn-exponential-tails}
\end{equation}
Consequently, with $\Lambda=\log(1/\eta)$, each of the limits
\begin{equation}
\mu\ge\widehat F+\frac{\rho\Lambda}{2}-\sqrt{\frac{\rho^2\Lambda^2}{4}+2\rho\Lambda\widehat F},\qquad
\mu\le\widehat F+\rho\Lambda+\sqrt{\rho^2\Lambda^2+2\rho\Lambda\widehat F}
\label{eq:nn-exponential-limits}
\end{equation}
holds with probability at least $1-\eta$.
\end{proposition}

\paragraph{Proof.}
Let $u_{ik}=\widehat P_{ik}^{q}-\widehat{\bar p}_i$ be the deviation of the order-averaged frequency vector of prompt $k$ from the item mean, so that $\widehat F=(NK)^{-1}\sum_{i,k}\|u_{ik}\|^2$. Replacing one call of item $i$ and prompt $k$ that returned verdict $c$ by a call returning $c'$ changes $\widehat F$ to $\widehat F'$ with
\[
\widehat F-\widehat F'=\frac{u_{ik}(c)-u_{ik}(c')}{NKR}-\frac{(K-1)\1\{c\ne c'\}}{2NK^2R^2}.
\]
The largest decrease that one replacement can produce is therefore at most $[u_{ik}(c)-\min_{c'}u_{ik}(c')]/(NKR)\le\rho/2$. Prompt $k$ of item $i$ holds $2R\,\widehat P_{ik}^{q}(c)$ calls with verdict $c$, and $(\max_cu_{ik}(c)-\min_cu_{ik}(c))^2\le2\|u_{ik}\|^2$, so the squares of these largest decreases sum over all $2NKR$ calls to at most
\[
\sum_{i,k}\frac{2R\cdot2\|u_{ik}\|^2}{(NKR)^2}=\rho\,\widehat F.
\]
Hence $\zeta=\widehat F/\rho$ is a function of independent variables that satisfies $\zeta-\inf\zeta\le1$ and $\sum(\zeta-\inf\zeta)^2\le\zeta$, where the infimum is over the replacement of one call and the sum is over the calls. These are the two conditions of the concentration inequality for self-bounding functions of \citet[Theorem~13]{maurer2006concentration}, in the form stated as Theorem~7 of \citet{maurer2009empirical} with $a=1$, which gives $\Prob(\E\zeta-\zeta>s)\le\exp(-s^2/(2\E\zeta))$ and $\Prob(\zeta-\E\zeta>s)\le\exp(-s^2/(2\E\zeta+s))$. Setting $s=t/\rho$ gives Equation~\eqref{eq:nn-exponential-tails}, and solving each bound for $\mu$ at probability $\eta$ gives Equation~\eqref{eq:nn-exponential-limits}. This argument does not allow a smaller constant in $\rho$ uniformly in $R$: for $K=2$ and two prompts that are deterministic on different categories, the two sides of the second condition differ by the factor $(1-1/(4R))^2$.

Proposition~\ref{prop:nn-exponential} bounds $\mu=\theta+\omega\,\Uc$, not the prompt component $\theta$ itself. To bound $\theta$, we combine the limits $\mu_L\le\mu\le\mu_U$ of Equation~\eqref{eq:nn-exponential-limits} with the Hoeffding bound for the call estimate (Section~\ref{sec:temperature-full-repeat-inference}). With $h=\sqrt{\Lambda/(4NK\lfloor R/2\rfloor)}$, each of $\theta\ge\mu_L-\omega(\widehat\Uc+h)$ and $\theta\le\mu_U-\omega(\widehat\Uc-h)$ holds with probability at least $1-2\eta$, where $\widehat\Uc-h$ is replaced by zero when it is negative. A two-sided interval for a contrast between two temperatures rests on eight limits, each with error at most $\eta$, two for $\mu$ and two for the call component at each temperature, so a total error $\eta_c$ is met with $\eta=\eta_c/8$.

The test that rejects $\theta\le0$ when the lower limit is positive has type-I error at most $2\eta$. If $NK\ge203\Lambda$, it rejects with probability at least $1-2\eta$ whenever $\theta>(9/R)\sqrt{\Lambda/(NK)}$: on the events $\mu-\widehat F\le\sqrt{2\rho\Lambda\mu}$ and $\widehat\Uc-\Uc\le h$, each of probability at least $1-\eta$, we have $\mu_L\ge\mu-2\sqrt{2\rho\Lambda\mu}$ once $\mu\ge32\rho\Lambda$, so the lower limit is at least $\theta/2-(9/(2R))\sqrt{\Lambda/(NK)}$. The test therefore keeps the separation order $1/(R\sqrt{NK})$ of Theorem~\ref{thm:nn-lower}, and the error level enters through $\sqrt{\log(1/\eta)}$ where Cantelli's inequality gives $\eta^{-1/2}$. The constants 9 and 203 are sufficient and not sharp, and at error levels near 5\% the test of Theorem~\ref{thm:nn-cantelli} has the smaller guaranteed separation.

\begin{table}[!htbp]\centering\small
\setlength{\tabcolsep}{3pt}\renewcommand{\arraystretch}{1.05}
\caption{Distribution-free intervals for the prompt contrast ($T=0$ minus $T=1$) at the error allocation of the four contrasts of Table~\ref{tab:temperature-main}, with Cantelli's inequality (Theorem~\ref{thm:nn-cantelli}) and with the exponential bound (Proposition~\ref{prop:nn-exponential}); post hoc.}
\label{tab:further-exponential}
\begin{tabular*}{\linewidth}{@{\extracolsep{\fill}}>{\raggedright\arraybackslash}p{6.4cm}ll@{}}\toprule
Calls & Cantelli & Exponential\\\midrule
Qwen, both runs pooled ($R=8$) & $[-.0145,.1026]$ & $[.0173,.0701]$\\
Gemma, all runs pooled ($R=40$) & $[-.0032,.0298]$ & $[.0046,.0219]$\\
Gemma, the 32 additional repeats alone (E6) & $[-.0059,.0326]$ & $[.0035,.0230]$\\\bottomrule
\end{tabular*}
\end{table}

Table~\ref{tab:further-exponential} applies both bounds to the prompt contrast between $T=0$ and $T=1$ at the error allocation of the four contrasts of Table~\ref{tab:temperature-main} (total error .0125 for this contrast). With the exponential bound, the distribution-free interval for this contrast excludes zero for both judges, and for Gemma also on the 32 additional repeats of E6 alone. On the pooled calls, the smallest total error at which this interval excludes zero is $5.2\times10^{-7}$ for Qwen and $3.2\times10^{-6}$ for Gemma.

On these data, the exponential interval is the narrower one once the error is split: its width ratio to the Cantelli interval is .69 to .76 over two contrasts and .45 to .53 over four. For a single contrast at 5\%, it is 2\% to 8\% wider. For the prompt contrasts between $T=.2$ and $T=1$, measured at $R=4$ only, the intervals include zero under both bounds. In simulations with known cell laws (2,000 datasets per setting), both intervals covered the true contrast in every dataset, and neither excluded zero in two null settings in which the two temperatures share one set of laws.

\paragraph{Minimum variance and repeat-adjusted flip rates.}
The corrected estimates and the forecasts cannot be improved by another unbiased use of the same calls.

\begin{proposition}[Minimum variance]
\label{prop:nn-umvu}
Let $\Psi$ be an estimator computed from the calls of the crossed design, possibly with independent randomization, that is unbiased under every array of cell laws for one of the following targets: (a) a quadratic target $\sum_{a,b}w_{ab}P_a^\top P_b$ with fixed weights, such as the call, prompt, order, or interaction component or the prompt dispersion within one order; (b) the expected score of a vote rule applied to one call from each of $m$ distinct cells, averaged over a fixed set of cell combinations. Let $\Psi^\star$ be, in (a), the corrected estimate, which replaces each $P_a^\top P_b$ by the distinct-call agreement of Equation~\eqref{eq:main-cell-agreement}, and in (b) the forecast, which averages the score over all ways of taking one recorded call from each of the $m$ cells. Then $\Var(\Psi)\ge\Var(\Psi^\star)$ under every array of cell laws, with equality only if $\Psi=\Psi^\star$ with probability one under that array.
\end{proposition}

\paragraph{Proof.}
For cell laws in the interior of the simplex, the calls follow a full-rank exponential family in which the cell counts are a complete sufficient statistic. $\Psi^\star$ is an unbiased function of the counts: distinct-call agreement is unbiased for $P_a^\top P_b$, and a product of frequencies of distinct cells is unbiased for the product of their probabilities. Call sequences with the same counts are equally likely under every array of laws, so the conditional expectation of $\Psi$ given the counts does not depend on the laws. It is an unbiased function of the counts, so it equals $\Psi^\star$ by completeness (the Lehmann--Scheff\'e theorem; \citealp{lehmann1998theory}). Every count array has positive probability under interior laws, so the two agree for every count array, and $\Var(\Psi)=\Var(\Psi^\star)+\E\Var(\Psi\mid\text{counts})$ holds under every array of laws, including laws with zero probabilities.

The proposition has two consequences for simpler procedures. The first concerns the repeat-adjusted flip rate, the disagreement form of the repeat-adjusted agreement that \citet{bellibatlu2026judgesense} use under paraphrase. For two prompts $k$ and $\ell$ in one order, it is the rate at which a call under $k$ and a call under $\ell$ disagree, minus the mean rate at which two calls under the same prompt disagree, and its expectation is $\tfrac12\|P_k-P_\ell\|^2$. Averaged over all prompt pairs, it equals $K/(K-1)$ times the corrected prompt dispersion of that order in every sample. The repeat-adjusted flip rate is therefore the corrected estimator restricted to one order, up to this factor: the mean of the two orders' corrected dispersions is the sum of the prompt and interaction estimates, and the order component does not enter. A variant anchors the comparison on one base prompt. With the base drawn at random it is unbiased for the all-pairs target, and its mean over the $K$ bases is the all-pairs rate, so by Proposition~\ref{prop:nn-umvu} its variance is at least that of the all-pairs rate.

The second consequence concerns pilots. A pilot that groups the measurement calls into votes and scores them is unbiased for the same target as the forecast, so its variance is at least the forecast's, and averaging over more groupings can only bring it down toward the forecast's. In both cases the proposition compares unbiased estimators of one target on the same calls, not designs or biased estimators such as the prediction of zero (Table~\ref{tab:zero-baseline}).

Table~\ref{tab:further-diagnostics} shows what each diagnostic detects at the same budget of $B$ calls per item and temperature, from 8 to 64. Each of 2,000 draws per judge and budget selects the prompts, the order, and two stored repeats per cell, the same for all items. A diagnostic detects the cooling change when its item-paired 95\% interval excludes zero in the direction of the full data. The unadjusted flip rate moves in the opposite direction: it falls under cooling at every budget, from .374 to .114 for Qwen and from .080 to .065 for Gemma, while the prompt estimates rise. For Qwen's prompt increase, the all-pairs rate in one order with $B/2$ prompts detects about as often as the sum of the prompt and interaction estimates in both orders with $B/4$ prompts. For Gemma the crossed design detects more often at every budget (95.9\% against 71.5\% at 64 calls). The base-anchored rate has 1.4 to 10.2 times the variance of the all-pairs rate and detects the increase in 39--56\% of draws for Qwen and 5--15\% for Gemma. Only the crossed design measures the order component. It detects the increase in Qwen's order component, $+.1221$ $[.0747,.1695]$ on the census pooled with E1, in 99.9\% of draws at 8 calls.

\begin{table}[!htbp]\centering\small
\setlength{\tabcolsep}{3pt}\renewcommand{\arraystretch}{1.05}
\caption{Detection of the cooling change at matched budgets (\% of 2,000 draws from the stored calls; post hoc). One order: $B/2$ prompts and two calls per cell; the all-pairs repeat-adjusted flip rate equals the corrected one-order dispersion up to the factor $K/(K-1)$. Crossed: $B/4$ prompts, both orders, two calls per cell. The variance ratio compares the base-anchored with the all-pairs flip rate.}
\label{tab:further-diagnostics}
\begin{tabular*}{\linewidth}{@{\extracolsep{\fill}}llrrrrrr@{}}\toprule
 & & \multicolumn{3}{c}{One order} & \multicolumn{3}{c}{Crossed}\\\cmidrule(lr){3-5}\cmidrule(l){6-8}
Judge & $B$ & All pairs & Base-anchored & Variance ratio & Prompt $+$ interaction & Prompt & Order\\\midrule
Qwen & 8 & 48.4 & 39.3 & 1.45 & 42.2 & 36.9 & 99.9\\
Qwen & 16 & 75.1 & 47.8 & 2.41 & 80.7 & 71.2 & 100.0\\
Qwen & 32 & 95.7 & 51.8 & 4.10 & 98.4 & 96.0 & 100.0\\
Qwen & 64 & 99.7 & 56.3 & 7.49 & 100.0 & 99.8 & 100.0\\
Gemma & 8 & 9.9 & 4.6 & 1.43 & 11.9 & 11.0 & 9.6\\
Gemma & 16 & 29.7 & 9.4 & 2.44 & 47.2 & 38.9 & 12.8\\
Gemma & 32 & 49.5 & 12.3 & 4.92 & 78.0 & 70.5 & 14.3\\
Gemma & 64 & 71.5 & 15.2 & 10.22 & 95.9 & 93.2 & 13.9\\\bottomrule
\end{tabular*}
\end{table}

\paragraph{Forecasts from one presentation order.}
A diagnostic computed in one order carries no information on the order component. To see what this costs for decisions, we forecast the order-balanced vote D from one-order calls and from crossed calls at the same budget $B$: $B/2$ prompts in one order with two calls each, against $B/4$ prompts in both orders with two calls each. There are 12 conditions (two judges, two temperatures, and three item sets: the census pooled with E1, the 200 E2 and E3 items, and the 120 E9 items), and each of 200 draws per budget selects the prompts and the order.

A one-order forecast must assume how the judge behaves in the order it did not see, and two assumptions mark the extremes. Under the first, the judge's law over candidates is the same in both orders, so position has no effect. Under the second, its law over shown positions is the same in both orders, so only position matters.

The first assumption is the more favorable one, and the forecast built on it still misses the pure ties. Its mean absolute error for the pure-tie rate is 10.3 pp at every budget, against 2.76, 1.90, and 1.44~pp for the crossed forecast at 8, 16, and 32 calls. For accuracy under the original rule the errors are 4.4 to 4.6 pp against 2.36, 1.61, and 1.18 pp, and for accuracy under the fair 2:2 rule 2.0 to 2.3 pp against .7 to 1.5 pp. With all prompts and calls, the crossed forecast has a mean absolute error of .76~pp for pure ties, the level expected from the sampling error of the observed rates. The one-order forecast is below the observed pure-tie rate in all 12 conditions. For Qwen it also gives the wrong direction for cooling: it says that cooling lowers the pure-tie rate, by 4.1 to 6.2 pp in every item set and order, whereas the observed rate rises by 7.9 to 18.3 pp and the crossed forecast gives $+7.1$ to $+15.8$ pp.

Under the second assumption the error for pure ties is 66 pp, and a mixture of the two with one weight chosen in hindsight still leaves 5.7 pp with all prompts and calls. One-order calls cannot show where a judge lies between the two assumptions. Take two sets of cell laws that agree in the order that was seen, with both prompts choosing one candidate with probability $p$. In the first set the other order has the same law over candidates, and the pure-tie rate of D is $6p^2(1-p)^2$. In the second it has the same law over shown positions, and the rate is $p^4+4p^2(1-p)^2+(1-p)^4$. At $p=.99$ the two rates are .0006 and .9610.

\paragraph{An unbiased variance estimate, and interval methods compared.}
The EV interval evaluates the exact variance of Theorem~\ref{thm:nn-variance} at the observed cell frequencies, with each cell covariance multiplied by $R/(R-1)$. In the simulations below, this variance estimate is on average 1.09 to 1.73 times the true variance of the contrast, so the EV intervals are wider than necessary. An unbiased estimate of the exact variance exists once $R\ge4$. In the notation of Section~\ref{sec:nn-main}, $\theta_i=\sum_{a,b}W_{ab}\,p_{ia}^\top p_{ib}$ is the prompt component of item $i$, $\widehat\theta_i$ is its corrected estimate, and $V_i=\Var(\widehat\theta_i)$ is its exact variance.

\begin{proposition}[Unbiased estimate of the exact variance]
\label{prop:nn-unbiased-variance}
Let $R\ge4$, and write $x^{(m)}=x(x-1)\cdots(x-m+1)$. Expand $\theta_i^2=\bigl(\sum_{a,b}W_{ab}\,p_{ia}^\top p_{ib}\bigr)^2$ into monomials $\prod_a\prod_c p_{ia}(c)^{m_{ac}}$, and let $\widehat Q_i$ be the result of replacing each monomial by $\prod_a\bigl[\prod_c n_{ia}(c)^{(m_{ac})}/R^{(m_a)}\bigr]$, where $n_{ia}(c)$ is the number of calls of cell $a$ that return $c$ and $m_a=\sum_cm_{ac}\le4$. Then $\widehat V_i=\widehat\theta_i^{\,2}-\widehat Q_i$ satisfies $\E\widehat V_i=V_i$ for every array of cell laws, it is the only unbiased estimate of $V_i$ that is a function of the cell counts, and no unbiased estimate of $V_i$ has a smaller variance. For $R=2$ and $R=3$, no unbiased estimate of $V_i$ exists.
\end{proposition}

\paragraph{Proof.}
Within a cell, $\E\prod_c n(c)^{(m_c)}=R^{(m)}\prod_c p(c)^{m_c}$ whenever $m=\sum_cm_c\le R$, and the cells are independent, so $\E\widehat Q_i=\theta_i^2$ and $\E\widehat V_i=\E\widehat\theta_i^{\,2}-\theta_i^2=V_i$. Uniqueness and minimum variance follow from the completeness of the counts, as in Proposition~\ref{prop:nn-umvu}. For $R<4$, move the law of one cell $a$ along $(x,1-x,0,\dots)$ with the other cells fixed. The expectation of any statistic is then a polynomial of degree at most $R$ in $x$, whereas $V_i$ has degree four, with leading coefficient $-8W_{aa}^2(2R-3)/[R(R-1)]\ne0$.

We write EV-U for the interval $\widehat\theta\pm z\,[\max(0,\sum_i\widehat V_i)]^{1/2}/N$. Table~\ref{tab:further-intervals} compares it with EV and with classical intervals for the cooling contrast of the prompt component, in simulations whose known cell laws are the pooled cell frequencies of each judge. Every interval is built around the same point estimate. EV-U has the coverage and the width of the interval that uses the true variance, and it is 4\% to 24\% narrower than EV, which keeps at least nominal coverage in every column. The classical intervals differ from one another. The per-category fixed-effects interval covers 85\% to 87\%, because it ignores the covariance between the category indicators. The per-category random-effects interval keeps at least nominal coverage in every column (98.8\% to 100.0\%) and is 1.3 to 3.7 times as wide as EV-U. The percentile bootstrap of calls almost never covers, because the resampled statistic is centered at the uncorrected plug-in dispersion. The centered bootstrap, which measures deviations from the plug-in dispersion of the resampled laws, covers 96.3\% and 96.7\% for Qwen and 92.9\% and 94.2\% for Gemma. The multivariate fixed-effects interval, which pools the within-cell covariance matrices, comes close to nominal coverage in these columns (93.7\% to 95.4\%). In the same simulations, the classical chi-square interval for the cooling contrast of the call component covers 96.0\% to 98.4\% for Qwen and 51.8\% to 53.5\% for Gemma ($R=2$, 4, and 8), because normal theory understates the variance of a within-cell variance for nearly deterministic cells. The Hoeffding interval covers in every dataset.

\begin{table}[!htbp]\centering\small
\setlength{\tabcolsep}{3pt}\renewcommand{\arraystretch}{1.05}
\caption{Interval methods for the prompt contrast in simulations with known cell laws: coverage (\%) and mean width ($\times10^3$) of nominal 95\% intervals, 10,000 datasets per column (4,000 for the bootstrap rows; post hoc). The centered bootstrap measures deviations from the plug-in dispersion of the resampled laws.}
\label{tab:further-intervals}
\begin{tabular*}{\linewidth}{@{\extracolsep{\fill}}lrrrr@{}}\toprule
Interval & Qwen, $R=4$ & Qwen, $R=8$ & Gemma, $R=4$ & Gemma, $R=8$\\\midrule
EV (Theorem~\ref{thm:nn-variance}) & 98.9 $\cdot$ 9.90 & 98.0 $\cdot$ 5.54 & 97.0 $\cdot$ 3.44 & 95.8 $\cdot$ 2.09\\
EV-U (Proposition~\ref{prop:nn-unbiased-variance}) & 94.6 $\cdot$ 7.52 & 94.9 $\cdot$ 4.58 & 94.3 $\cdot$ 3.04 & 94.5 $\cdot$ 1.99\\
True variance (reference) & 94.6 $\cdot$ 7.53 & 94.9 $\cdot$ 4.58 & 94.6 $\cdot$ 3.05 & 94.8 $\cdot$ 2.00\\
Fixed effects, per category & 84.7 $\cdot$ 5.54 & 86.0 $\cdot$ 3.47 & 85.2 $\cdot$ 2.30 & 87.3 $\cdot$ 1.56\\
Fixed effects, multivariate & 93.7 $\cdot$ 7.25 & 94.7 $\cdot$ 4.54 & 94.4 $\cdot$ 3.02 & 95.4 $\cdot$ 2.06\\
Random effects, per category & 98.8 $\cdot$ 9.87 & 100.0 $\cdot$ 8.83 & 100.0 $\cdot$ 7.54 & 100.0 $\cdot$ 7.34\\
Percentile bootstrap of calls & 0.0 $\cdot$ 8.25 & 0.0 $\cdot$ 5.15 & 0.0 $\cdot$ 2.88 & 0.1 $\cdot$ 1.95\\
Centered bootstrap of calls & 96.3 $\cdot$ 8.25 & 96.7 $\cdot$ 5.15 & 92.9 $\cdot$ 2.88 & 94.2 $\cdot$ 1.95\\
Cantelli (Theorem~\ref{thm:nn-cantelli}) & 100.0 $\cdot$ 86.22 & 100.0 $\cdot$ 49.55 & 100.0 $\cdot$ 79.43 & 100.0 $\cdot$ 43.70\\
Item-paired (IP) & 100.0 $\cdot$ 32.88 & 100.0 $\cdot$ 32.31 & 100.0 $\cdot$ 17.86 & 100.0 $\cdot$ 17.72\\\bottomrule
\end{tabular*}
\end{table}

Whether the multivariate interval keeps its coverage depends on which prompts are the noisiest. Its variance equals the exact variance when all cells share one covariance matrix. Otherwise the difference between the exact variance and the normal-theory variance has leading term $R^{-1}\sum_a d_{ia}^\top(\Sigma_{ia}-\bar\Sigma_i)d_{ia}$, which is positive when the prompts farthest from the item mean are the noisiest ($\bar\Sigma_i$ is the mean covariance matrix of the item's cells). When the cell profiles are rearranged within each item so that the farthest prompts are the noisiest, the multivariate interval covers 90.5\% to 93.8\% across the four columns, and 95.8\% to 98.8\% in the opposite arrangement, while EV-U covers 94.4\% to 95.3\% in both (5,000 datasets each; these arrangements are constructed, not observed).

Intervals built on an estimated variance need enough cells with disagreement. For $T=0$ taken alone, 98.8\% of Qwen's and 99.5\% of Gemma's simulated cells are deterministic. The three variance-based intervals then stay close to nominal for Qwen (EV 94.0\% to 94.6\%, EV-U 93.8\% to 94.3\%, and the multivariate fixed-effects interval 92.6\% to 93.2\%) and fall below it for Gemma (76.0\% to 83.2\%, 79.2\% to 87.4\%, and 80.1\% to 83.0\%). In a sweep of constructed laws, EV-U covers 94.5\% when at least 20 cells show any disagreement and 83.2\% otherwise. The limits of Proposition~\ref{prop:nn-exponential} need no variance estimate and remain valid there. The cooling contrasts reported in this paper are not in that regime: on the stored calls the cooled temperature carries 1\% to 8\% of their variance, and EV covers at least 95.8\% in every column of Table~\ref{tab:further-intervals}. On the original calls, EV-U narrows the Bonferroni-adjusted intervals of Table~\ref{tab:temperature-main} from $[.0363,.0489]$ to $[.0384,.0469]$ for Qwen and from $[.0120,.0164]$ to $[.0124,.0160]$ for Gemma (Table~\ref{tab:discussion-call-level}).

\paragraph{Strategy selection against simpler rules.}
Would a simpler rule choose between one call and voting as well as the forecast does? Table~\ref{tab:further-selection} compares the forecast with a pilot, with a forecast from one-order calls, with the two fixed strategies, and with threshold rules, over 1,000 draws of the development calls and the eight accuracy settings (Table~\ref{tab:discussion-regret} used 50 draws). The single-grouping pilot has higher regret than the forecast at every budget on both test sets, with every interval over draws above zero. A pilot averaged over 25 groupings reproduces the forecast: the variance of the pilot's accuracy estimate is 1.08 to 1.92 times the forecast's with one grouping and at most 1.04 times with 25, as Proposition~\ref{prop:nn-umvu} implies. A forecast from the same number of calls in one order, under the more favorable assumption above, has higher regret than the crossed forecast at every budget, with every interval over draws above zero, and stops improving after 16 calls. The gap comes from the condition with the most pure ties (Table~\ref{tab:further-ties}), and the one-order forecast understates pure ties. For Qwen at $T=0$ under the fair 2:2 rule, D is 2.43 and 1.68~pp more accurate than one call on the two test sets. There the crossed forecast chooses D in 67\%, 88\%, 98\%, 100\%, and 100\% of draws at the five budgets, and the one-order forecast in 9\%, 8\%, 3\%, 0\%, and 0\%. Over the eight settings, the choices fixed before E9 reach 79.54\% accuracy on the E9 items at 1.75 deployment calls per decision, against 78.66\% for always one call and 76.18\% for always D at four calls.

\begin{table}[!htbp]\centering\small
\setlength{\tabcolsep}{3pt}\renewcommand{\arraystretch}{1.05}
\caption{Mean regret (pp) on the 200 E2 and E3 items / on the 120 E9 items, over 1,000 draws of the development calls and the eight accuracy settings (post hoc). At 192 calls only six distinct draws exist. The one-order forecast uses $B/2$ prompts in one order with two calls each (all 48 with four calls at $B=192$) and assumes no position effect; its regret is averaged over the two orders.}
\label{tab:further-selection}
\begin{tabular*}{\linewidth}{@{\extracolsep{\fill}}lrrrrr@{}}\toprule
Calls per item and temperature ($B$) & 8 & 16 & 32 & 96 & 192\\\midrule
Forecast & .18 / .16 & .09 / .08 & .05 / .05 & .02 / .02 & .00 / .00\\
Pilot, one grouping & .22 / .20 & .12 / .10 & .09 / .08 & .04 / .04 & .03 / .03\\
Pilot, 25 groupings & .18 / .16 & .09 / .08 & .05 / .05 & .02 / .02 & .00 / .00\\
Forecast from one-order calls & .34 / .27 & .29 / .21 & .30 / .21 & .30 / .21 & .30 / .21\\
Forecast (Qwen, $T=0$, fair rule only) & .80 / .55 & .30 / .21 & .05 / .03 & .00 / .00 & .00 / .00\\
One-order forecast (same setting) & 2.21 / 1.54 & 2.25 / 1.56 & 2.35 / 1.63 & 2.43 / 1.68 & 2.43 / 1.68\\
\multicolumn{6}{@{}l}{\emph{Flip-rate threshold with a margin, fitted on}}\\
\quad all eight settings & .14 / .10 & .10 / .07 & .06 / .04 & .00 / .00 & .00 / .00\\
\quad the other judge & 1.37 / 1.02 & 1.37 / 1.02 & 1.37 / 1.02 & 1.37 / 1.02 & 1.37 / 1.02\\
\quad the other tie rule & 2.41 / 2.35 & 2.44 / 2.33 & 2.58 / 2.47 & 2.81 / 2.71 & 2.84 / 2.74\\
\multicolumn{6}{@{}l}{\emph{Repeat-disagreement threshold with a margin, fitted on}}\\
\quad all eight settings & .19 / .16 & .12 / .11 & .09 / .09 & .06 / .07 & .06 / .07\\
\quad the other judge & 1.37 / 1.02 & 1.37 / 1.02 & 1.37 / 1.02 & 1.37 / 1.02 & 1.37 / 1.02\\
\quad the other tie rule & 2.73 / 2.85 & 3.07 / 3.21 & 3.11 / 3.24 & 3.27 / 3.41 & 3.28 / 3.42\\
Always one call & \multicolumn{5}{c}{1.24 / .88}\\
Always D & \multicolumn{5}{c}{2.67 / 3.36}\\\bottomrule
\end{tabular*}
\end{table}

A threshold rule chooses D when a disagreement statistic of the development calls, the repeat disagreement or the paraphrase flip rate, exceeds a threshold. The threshold must be fitted on settings whose better strategy is already known, which we take from the census forecast. Without a margin, the threshold is the smallest value that classifies the largest number of fitting settings correctly; with a margin, it is placed at the midpoint of the gap between the settings in which D is better and those in which one call is better. Fitted with a margin on all eight settings, the flip-rate rule matches the forecast: its regret is within .06 pp of the forecast's at every budget. These fitted thresholds do not transfer from one judge or tie rule to the other. Fitted on one judge and applied to the other, or on one tie rule and applied to the other, every threshold rule, with or without the margin, has higher regret than the forecast at every budget, by .51 to 3.42 pp. The two failures have different causes. Within a judge the two temperatures share the same better strategy, so one judge's settings do not identify a threshold and the fitted rule is a constant choice. Neither statistic depends on the tie rule, so a threshold fitted under one rule cannot adapt to the other. If the better strategy of each census setting is instead read from the census deployments, one judge's fair-rule settings identify a threshold, and the leave-one-judge-out regret of the flip-rate rule with a margin is then within .01 pp of the forecast's at 8 calls and .06 to .14 pp above it from 16 calls on. The forecast needs no fitting.

Accuracy is the endpoint $c=1$ of the decision loss $L_c$ of Equation~\eqref{eq:position-abstention-cost}. The forecast also chooses well when an abstention costs less than an error. At each cost on the grid $c=0,.05,\dots,1$ the census forecast chooses the strategy with the lower forecast loss, and we average the loss over the grid and the four judge-by-temperature conditions under the original rule. Against always using one call, this choice has 3.02 pp $[2.39,3.65]$ less loss on the E9 items and 3.31 pp $[2.80,3.82]$ less on the 200 E2 and E3 items; at $c=0$ the differences are 9.57 and 9.67 pp. Against always voting, it has 1.45 pp $[.77,2.14]$ and 1.00 pp $[.53,1.48]$ less; at $c=1$ the differences are 6.44 and 5.06 pp. Its mean regret against the better strategy of each condition is .05 and .10 pp, and .13 and .16 pp when the choice uses 8 development calls per item and temperature.

In every condition the forecast chooses D below a break-even cost and one call above it. For Qwen at $T=1$ and $T=0$ and for Gemma at $T=1$ and $T=0$, the census break-evens are .69, .53, .55, and .54. In the same order, the observed break-evens are .79 $[.66,.96]$, .57 $[.53,.63]$, .48 $[.38,.60]$, and .48 $[.39,.60]$ on the E9 items, and .89 $[.77,1.03]$, .59 $[.55,.64]$, .53 $[.41,.67]$, and .52 $[.41,.63]$ on the 200 items.

The ten held-out settings of Table~\ref{tab:discussion-heldout} allow the same comparison beyond these four conditions (Qwen, Gemma, and GPT-OSS on LLMBar and JudgeBench, and four Qwen judging policies, all at $T=1$; post hoc). Each setting's strategy is chosen from the forecast on its 20 or 24 development items and evaluated on its 60 or 152 test items. Over the same grid of costs, the chosen strategy has on average 3.69 pp $[3.18,4.20]$ less loss than always using one call (10.99 pp at $c=0$) and 1.20~pp $[.88,1.52]$ less than always voting (5.92 pp at $c=1$), with a mean regret of .12 pp (item-paired intervals, stratified over the three item sets). The observed break-evens range from .43 to 1.23 across the ten settings.

\paragraph{What the pure ties consist of.}
A pure 2:2 tie of D has one of four vote patterns. In a first-position tie, all four calls choose the first-shown candidate; in a second-position tie, all four choose the second-shown candidate. In a prompt tie, each prompt chooses the same candidate in both orders and the two prompts disagree. In a mixed tie, one prompt follows the first-shown candidate and the other the second-shown. We call the first two patterns position ties. Table~\ref{tab:further-ties} counts the four patterns in the independent deployments of the census, E1, E2, E3, and E9. At $T=0$, 905 of Qwen's 929 pure ties are first-position ties, including all 124 on the census. Two calls of the same cell disagree in only .2\% to .6\% of pairs at $T=0$, so in these votes the verdict under both prompts reverses when the candidates are swapped. If the cell laws did not depend on the order, position ties and mixed ties would be equally frequent in expectation, whatever the prompt effects and the call variation; Qwen's counts at $T=0$ are 909 and 19.

\begin{table}[!htbp]\centering\small
\setlength{\tabcolsep}{3pt}\renewcommand{\arraystretch}{1.05}
\caption{Vote patterns of the pure 2:2 ties in the independent deployments of the census, E1, E2, E3, and E9 (3,200 votes per judge and temperature; the census and E1 share items and prompt pairs; post hoc).}
\label{tab:further-ties}
\begin{tabular*}{\linewidth}{@{\extracolsep{\fill}}lrrrrr@{}}\toprule
Judge and temperature & Pure ties & First-position & Second-position & Prompt & Mixed\\\midrule
Qwen, $T=1$ & 601 & 357 & 29 & 100 & 115\\
Qwen, $T=0$ & 929 & 905 & 4 & 1 & 19\\
Gemma, $T=1$ & 240 & 221 & 11 & 4 & 4\\
Gemma, $T=0$ & 250 & 233 & 12 & 2 & 3\\\bottomrule
\end{tabular*}
\end{table}

Under cooling, Qwen's position ties rise while its prompt ties and mixed ties fall, so position ties rise by more than pure ties as a whole in every collection: $+25.94$ pp $[15.31,37.19]$ against $+19.69$ pp on the census, $+22.50$ pp $[12.19,32.81]$ against $+16.88$ pp in E1, $+13.88$ pp $[10.25,17.69]$ against $+7.88$ pp on the E2 and E3 items, and $+15.21$ pp $[10.10,20.62]$ against $+8.85$ pp in E9 (item bootstrap 95\% intervals). Over the 360 distinct items, prompt ties go from 3.21\% of votes to .03\%. Among the items whose pure-tie rate rises, the rise consists of position ties alone on 18 of 18 census items, 17 of 17 in E1, 68 of 71 E2 and E3 items, and 33 of 35 in E9. The measurement calls forecast the same composition: over the 360 items the forecast rate of position ties is 11.87\% at $T=1$ and 27.29\% at $T=0$, and the observed rates are 11.77\% and 27.24\%.

Gemma's pure ties are almost all position ties at both temperatures (232 of 240 and 245 of 250), and their rate does not change significantly. These counts describe the recorded votes; the first-shown response is always labeled A, so they do not separate position from label.

\paragraph{A pre-specified test of the direction of position ties (E10).}
E7 appended to each prompt a clause stating a false prior for one position. The sentence ``In this collection, the better candidate is usually Candidate A.'' favors the first-shown candidate, which is always labeled A, and the same sentence with Candidate B favors the second-shown one. We call them the first-position and the second-position prior. E7's joint criterion, which included the change in the net pure-tie rate, was not met (Section~\ref{sec:discussion-studies-position}). E7's stored calls suggested a narrower prediction (post hoc). The second-position prior lowered Qwen's first-position ties by 5.94 pp $[4.06,7.92]$ and raised its second-position ties by 2.86~pp $[1.61,4.27]$, and every pure tie under Gemma's first-position prior was a first-position tie (14.32\% of votes against 5.89\% in the base arm; item bootstrap 95\% intervals). A prior that shifts a judge's position preference should therefore shift its position ties toward the position it favors. E10 tested this prediction on items that E7 had not used, with a criterion fixed before any call.

On the 120 E9 items, which had never been judged with such a clause, each judge answered the 12 prompts of E7 with and without the prior that had moved its position preference more in E7: the second-position prior for Qwen and the first-position prior for Gemma. Each item and arm received 12 independent deployments of D at $T=1$, with the same prompt pairs in both arms (23,040 calls). The criterion required two quantities to move in the direction of the prior, each with an item-paired 95\% interval that excludes zero. One is the signed position effect of Section~\ref{sec:discussion-studies-position}. The other is the tie imbalance, the rate of first-position ties minus the rate of second-position ties. The criterion was primary for Qwen and secondary for Gemma (Table~\ref{tab:further-e10}).

The primary criterion was met. Qwen's position effect fell by $.178$ $[.137,.220]$ and its tie imbalance by 8.89 pp $[6.09,11.69]$. First-position ties went from 10.97\% of votes to 4.17\%, and second-position ties from .97\% to 3.06\%. The estimated change in the tie imbalance is negative in each of the three benchmarks. The net pure-tie rate changed by $-2.64$ pp $[-5.61,.33]$ ($p=.08$, two-sided). Second-position ties ($+2.08$ pp) and other 2:2 patterns ($+2.08$ pp) replaced part of the first-position ties.

The secondary criterion was not met. Over the 120 items the prior raised Gemma's position effect by $.044$ $[.010,.079]$, against $.109$ in E7, and the change in its tie imbalance, $+3.13$ pp $[-.39,6.64]$, has the predicted sign, but its interval includes zero ($p=.08$). The breakdown by benchmark, which the plan lists as descriptive, places the change on JudgeBench. On the 24 JudgeBench items the prior raised Gemma's position effect by $.193$ $[.056,.329]$ and its tie imbalance by 14.93 pp $[1.13,28.73]$, close to the 60 JudgeBench items of E7 ($.189$ and 16.04 pp; post hoc). On the 96 LLMBar and RewardBench~2 items the position effect changed by $+.006$ and $+.008$ and the tie imbalance by $+.35$ and $+.12$ pp, with every interval including zero.

E10 used E7's clauses and prompts on different items, so the two studies together cover 240 distinct items, with 16 and 12 independent deployments of D per item and arm at $T=1$. Pooled post hoc with equal weight per item, Qwen's second-position prior changes the tie imbalance by $-8.85$ pp $[-10.74,-6.95]$ and the net pure-tie rate by $-2.20$ pp $[-4.12,-.29]$, and Gemma's first-position prior changes them by $+5.78$ pp $[3.12,8.44]$ and $+5.80$ pp $[3.13,8.46]$ (item-paired 95\% $t$ intervals).

\clearpage
\begin{table}[!htbp]\centering\small
\setlength{\tabcolsep}{2.5pt}\renewcommand{\arraystretch}{1.05}
\caption{E10: position-prior clauses on the 120 E9 items (12 prompts, 12 deployments of D per item and arm at $T=1$; pre-specified). Changes are clause minus base with item-paired 95\% intervals.}
\label{tab:further-e10}
\begin{tabular*}{\linewidth}{@{\extracolsep{\fill}}lrrlrrl@{}}\toprule
& \multicolumn{3}{c}{Qwen, second-position prior} & \multicolumn{3}{c}{Gemma, first-position prior}\\
\cmidrule(lr){2-4}\cmidrule(l){5-7}
Quantity & Base & Clause & Change & Base & Clause & Change\\\midrule
Signed position effect & $.212$ & $.034$ & $-.178$ $[-.220,-.137]$ & $.098$ & $.142$ & $+.044$ $[.010,.079]$\\
First-position ties (\% of votes) & $10.97$ & $4.17$ & $-6.81$ $[-9.36,-4.25]$ & $9.03$ & $12.22$ & $+3.19$ $[-.23,6.62]$\\
Second-position ties (\%) & $.97$ & $3.06$ & $+2.08$ $[.99,3.17]$ & $.83$ & $.90$ & $+.07$ $[-.72,.86]$\\
Tie imbalance (pp) & $10.00$ & $1.11$ & $-8.89$ $[-11.69,-6.09]$ & $8.19$ & $11.32$ & $+3.13$ $[-.39,6.64]$\\
All pure 2:2 ties (\%) & $17.71$ & $15.07$ & $-2.64$ $[-5.61,.33]$ & $9.86$ & $13.13$ & $+3.26$ $[-.25,6.78]$\\\bottomrule
\end{tabular*}
\end{table}

\end{document}